\documentclass[aps,pra,twocolumn,groupedaddress,floatfix]{revtex4-1}

\usepackage{graphicx}
\usepackage{dcolumn}
\usepackage{bm}
\usepackage{amsmath}
\usepackage{xcolor}
\usepackage{braket}

\usepackage{amsfonts}
\usepackage{blkarray}
\usepackage{amssymb}
\usepackage{multirow}
\usepackage{mathrsfs}

\DeclareMathOperator{\Tr}{Tr}

\begin{document}

\title{Interference, spectral momentum correlations, entanglement, and Bell inequality for a trapped 
interacting ultracold atomic dimer: Analogies with biphoton interferometry}

\author{Constantine Yannouleas}
\email{Constantine.Yannouleas@physics.gatech.edu}
\author{Benedikt B. Brandt}
\email{benbra@gatech.edu}
\author{Uzi Landman}
\email{Uzi.Landman@physics.gatech.edu}

\affiliation{School of Physics, Georgia Institute of Technology,
             Atlanta, Georgia 30332-0430}

\date{14 September 2018; Phys. Rev. A {\bf 99}, 013616 (2019)}

\begin{abstract}
Aiming at elucidating similarities and differences between quantum-optics biphoton interference phenomena and the 
quantum physics of quasi-one-dimensional double-well optically-trapped ultracold neutral bosonic or fermionic 
atoms, we show that the analog of the optical biphoton joint-coincidence spectral correlations, studied with 
massless non-interacting biphotons emanating from EPR-Bell-Bohm single-occupancy sources, corresponds to a 
distinct contribution in the total second-order momentum correlations of the massive, interacting, and 
time-evolving ultracold atoms. This single-occupancy contribution can be extracted from the total second-order 
momentum correlation function measured in time-of-flight experiments, which for the trapped atomic system 
contains, in general, a double-occupancy, NOON, component. The dynamics of the two-particle system are modeled by a
Hubbard Hamiltonian. The general form of this partial coincidence spectrum is a cosine-square quantum beating 
dependent on the difference of the momenta of the two particles, while the corresponding coincidence probability 
proper, familiar from its role in describing the Hong-Ou-Mandel coincidence dip of overlapping photons, results 
from an integration over the particle momenta. Because the second-order momentum correlations are mirrored in the 
time-of-flight spectra in space, our theoretical findings provide impetus for time-of-flight experimental protocols
for emulating with (massive) ultracold atoms venerable optical interferometries that use two space-time separated 
and entangled (massless) photons or double-slit optical sources. The implementation of such developments will 
facilitate testing of fundamental aspects and enable applications of quantum physics with trapped massive ultracold
atoms, that is, investigations of nonlocality and violation of Bell inequalities, entanglement, and quantum 
information science.
\end{abstract}

\maketitle

\section{Introduction} 
\label{intr}

The rapid advances in the experimental control of trapped ions and ultracold neutral atoms in optical 
lattices (whether bosonic or fermionic) is opening unprecedented opportunities for simulating phenomena that 
allow new vistas into issues of fundamental value pertaining to the foundation of the quantum description of 
nature, as well as for studying complex condensed-matter systems and exploring their many-body physics.
Accordingly, novel {\it in situ\/} \cite{geme09,grei09,grei15,hule15,joch15} 
and time-of-flight (TOF) \cite{foel05,foel06,bouc16,hodg17,schm17.1,schm17.2,berg18} experimental techniques and 
protocols that measure key quantities such as second- and higher-order many-body correlations of interacting 
particles in real (space coordinates) and momentum space, respectively, are being developed. In addition, the
importance of such quantities has been reflected in a growing number of theoretical studies 
\cite{roma04,baks07,pfan07,zinn14,poll18} which have mainly analyzed space correlations. In this sense it is 
notable that two recent theoretical publications \cite{bran17,bran18} have focused, with the use of exact 
diagonalization and the Hubbard model, on second-order momentum correlations for trapped, interacting ultracold 
atoms beyond earlier studies of first-order momentum correlations \cite{ming02,olsh03}.

Simultaneously with the many-body-physics-oriented experimental endeavors cited above, ultracold bosons in optical 
lattices in conjunction with site microscopy have been employed \cite{kauf14,isla15} to probe 
indistinguishability and mode entanglement using quantum-interference aspects in real space for two separated and
noninteracting bosonic atoms. In this respect, it is pertinent to note here two earlier experimental publications 
\cite{foel05,foel06} on ultracold atomic gases released from an optical lattice with a connection to the Hanbury 
Brown-Twiss interferometry. Such developments promise to generate a wealth of technological applications in 
the fabrication of quantum devices and for quantum information processing \cite{bloc06}.

\subsection{Quantum optics with massive interacting double-well-trapped ultracold atom dimer: 
momentum correlations, interference, entanglement and  Bell inequality  testing}
\label{qomip}

The aim of this paper is to elucidate comparisons between optical biphoton and trapped-dimer-atoms experiments, 
and serve as a resource and a guide for the analysis of current experiments, as well as for the design of future 
ones. Toward this goal we focus on a system made of a pair of interacting atoms trapped in a double-well optical 
potential. Indeed, with having control over the trapping potential [that is, the confining wells' frequencies 
along the axis $(x)$ connecting the two wells and in the transverse directions $(y,z)$, and the relative depth of 
the two wells],  distance $(D=2d)$ between the wells, the tunneling parameter $(J)$ between the two wells, and the 
strength $({\cal U})$ of the contact interaction between the trapped 
ultracold atoms (via the use of the Feshbach resonance technique), makes this system well suited for comparing, 
elucidating, and placing in context the results of specific measurements employing ultracold (massive and 
interacting) atoms by invoking analogy to the large body of work done in the past several decades in quantum optics
(using massless non-interacting photons). 
Furthermore, the double-well trapped ultracold atom system can serve as a 
test-bed for assessing concepts and formalisms addressing many-body interacting and highly-correlated systems, 
as well as for testing fundamental quantum behavior (addressing entanglement and violation of Bell’s inequalities)
with implications for quantum information and quantum simulations.

\textcolor{black}{
In the double-well-trapped interacting-atoms experiment, 
one prepares first, an eigenstate of the system (modeled here by the 
corresponding Fermi- or Bose-Hubbard model Hamiltonian), or a non-stationary state made of a superposition of 
eigenstates; in this work, we restrict ourselves to a quasi-one-dimensional double-well confinement. There are two
ways for probing the system after preparation: (i) {\it in-situ\/} detection (imaging) of individual particles' 
positions, which may be combined with resolving the particle's hyperfine (spin) state, and (ii) release of the 
optical confinement, with the resulting time-of-flight measurement allowing determination of the momentum 
wave-function with single-atom resolution, as well as the construction of two-body (second-order) momentum 
correlation maps. In this paper, we address mainly the second (ii) measurement protocol.
}

\textcolor{black}{ 
In making the analogy between the atom-based (matter-wave) interferometry and the optical one, the two wells 
(making up the double-well) are regarded as the sources [left ($L$) and right ($R$)] from which the particles 
emanate after the confinement is relieved (trap-release), arriving upon free-flight to the far-field detectors 
where their momenta are measured. As we discuss at some length below (see in particular Sect.\ \ref{simi}), the 
development and detailed characterization of the photon sources in quantum-optics interferometry measurements 
played an important role, with the primary source produced via spontaneous parametric down conversion generating 
twin-pairs of spatially separated photons in an entangled (EPR-Bell-Bohm state \cite{shih03}); non-entangled 
photon pairs from independent primary sources have also been produced \cite{sant02}. 
}

\textcolor{black}{
The key difference of the 
double-well (two-atom) source from the quantum-optics photon sources described above is that in general (for an 
arbitrary value of the interparticle interaction ${\cal U}$), the state of the two trapped atoms contains a 
superposition of both singly-occupied [with one particle in each well, EPR-Bell-Bohm state($\pm$)=
$(|1_L,1_R\rangle \pm |1_R,1_L\rangle)/\sqrt{2}$] and doubly-occupied [with both particles in one of the wells,
NOON($\pm$) $= (|2,0\rangle \pm |0,2\rangle)/\sqrt{2}$] components; only for infinitely strong interparticle 
repulsion, does one obtain a pure single-occupancy two-particle state. The two-particle momentum correlation 
function measured in the TOF experiments gets contributions from all components of the two-particle wave function 
of the double-well-trapped system. In order to make direct contact with the majority of quantum optics coincidence 
interferometry measurements that use single-occupancy (referred to also as twin-pair) sources, it is imperative  
that a method be developed for extraction of the contribution of the single-occupancy component (soc) from the  
total momentum correlation function determined by the total two-atom wavefunction. Indeed such methodology is 
developed in this work. In the following, the extracted contribution associated with the single-occupancy 
wavefunction component (the EPR-Bell-Bohm component), is called the single-occupancy-component partial 
joint-coincidence probability spectral map, $p_{\rm soc} (k_1,k_2)$, where $k_1$, $k_2$ are the TOF measured 
momenta of the two particles. Integration over the $k_1$ and $k_2$ momenta, yields a scalar joint-coincidence
probability, $P_{11} = \int \int p_{\rm soc} (k_1,k_2) dk_1 dk_2$, which can be determined by {\it in situ\/} 
two-atom double-well measurements \cite{kauf14,bran18}; $P_{11}$ is directly analogous to the joint-coincidence 
probabilty measured in the celebrated Hong-Ou-Mandel quantum-optics experiment \cite{hong87}. 
}

\textcolor{black}{
The methodology developed in this paper, in conjunction with the incorporation of interaction effects via the 
Hubbard model, opens the way for gaining deep insights into the structure of the many-body wave-function and its 
evolution as a function of the strength of interatomic interaction (both repulsive and attractive). Subsequently, 
it is used here in a discussion of time-evolution when starting from a non-stationary state in the double-well, 
exploration of entanglement aspects, and for demonstration of the violation of Bell's inequalities with trapped 
ultracold atoms. A more detailed plan of the paper is offered in the last sub-section of this Introduction.
}

More precisely, motivated by the potential of time-of-flight measurements, we develop in this paper a theory of 
second-order interference (referred to as fourth-order in quantum optics \cite{mand99}) in momentum space. 
This development, relating the field of quantum optics, employing massless non-interacting photons, to that of the
physics of (massive and interacting) ultracold atoms, is indeed most 
desirable and appears natural when one realizes the following
correspondences: (i) $\omega/c \rightarrow k$, and (ii)  $c \Delta \tau \rightarrow D$, where the quantities to 
the left of the arrow, $\omega$, $c$, and $\Delta \tau$,  are, respectively, the frequency, speed of light, and 
time delay between individual photons, and those to the right of the arrows are the atom momentum ($k$) and the 
interatomic distance ($D$, distance between the optical-lattice microtraps, or optical tweezer trapping sites).
Our approach allows us to explore analogies with biphoton (nonlocal two-photon) interference 
\cite{mand99,oubook,shihbook} beyond the example of the Hong-Ou-Mandel (HOM) dip \cite{hong87}. 
\textcolor{black}{
We wish to clarify at this point that the measurements at the time-of-flight far-field image are actually 
performed in space and that the space coordinates $X_1$ and $X_2$ for the positions of the two particles at the 
far field are related to the single-particle momenta $k_1$ and $k_2$ at the source (i.e., the double-well trap) as 
$X_j=\hbar k_j t_{\rm TOF})/M$, $j=1,2$ \cite{altm04}, where $M$ is 
the mass of the atom and $t_{\rm TOF}$ is the time of flight \cite{note2}.
}

\subsection{Hong-Ou-Mandel and other interference phenomena}
\label{dip}

\textcolor{black}{ 
In the original HOM experiment \cite{hong87}, two photodetectors were used to monitor the output modes of a beam 
splitter on which two photons impinged. The coincidence count of the detectors, $P_{11}$, was found to drop to zero
(total destructive interference) when the identical input photons overlap perfectly in both time and space on a 
50\%-50\% beamsplitter. This is called the Hong-Ou-Mandel dip, or HOM dip.  The HOM dip as a function of the time 
delay $\Delta \tau$ between the two photons has a characteristic shape resembling an inverted Gaussian flanked by 
two shoulders, e.g., in the simplest case, $P_{11}(\Delta \tau )
\propto 1-\exp[-(\Delta \tau)^2(\Delta \omega)^2/2]$, where $\Delta \omega$ is the frequency
bandwidth of the downconverted photons \cite{hong87,mand99,branc17}.
}

\textcolor{black}{ 
The HOM dip has been reproduced using two beams of traveling ultracold He atoms 
\cite{aspe15}, as well as beams of free electrons \cite{liu98,jonc12,bocq13,burk07}. 
The parameter underlying the HOM coincidence dip is the extent of the overlap of the wave packets of two atoms in 
the beams, which is controlled by the time delay $\Delta \tau$ [or equivalently relative distance $D$, see (ii) 
above] between the atoms. These accomplishments with free-space particle beams are in the spirit of the original 
HOM dip \cite{hong87,branc17}. In this context, we note a proposal \cite{kher14} to reproduce the HOM dip 
with two colliding Bose-Einstein condensates.  
In the case of trapped noninteracting ultracold atoms, the demonstrated 
\cite{kauf14,isla15} joint-coincidence variation is due to the tunneling -- described by the tunneling parameter 
$J$ in the Hubbard Hamiltonian [see Eqs.\ (\ref{hbspl})-(\ref{hf}) in Sec.\ \ref{hubs1}] -- 
between the two separated wells (near-vanishing wave packet overlap). The ensuing time evolution generates (as a 
function of time $t$) a sinusoidal pattern for $P_{11}(t)$, resulting from the combinations 
$\exp(i\phi) \pm \exp(-i\phi)$ associated with the accumulated phase $\phi \propto J t/\hbar$.  
}

\textcolor{black}{ 
Constructive and destructive interference effects in quantum optics have been shown to occur 
\cite{ghos87,mand88,shih88,fran89,mand90,kwia90,rari90,shih96,remp04} also without mixing of the photon beams (as it
takes place on the beam-splitter in the HOM experiments \cite{hong87,liu98,jonc12,bocq13,burk07,aspe15}). In these 
non-mixing cases, the association of the resulting sinusoidal coincidence probability curves is affected by 
accumulation of phases, brought about by a variety of experimental techniques that are used to control the optical 
path lengths, or with the use of phase-shifting devices. These phase-based interference phenomena opened further 
perspectives for fundamental quantum physics investigations and applications, in particular nonlocality, 
entanglement, and testing of the Bell inequalities 
\cite{mand99,oubook,shihbook,ghos87,mand88,shih88,fran89,mand90,kwia90,rari90,shih96,remp04} using EPR-Bell-Bohm
biphoton states generated via spontaneous parametric down conversion; the non-locality mentioned above, reflects 
the separation of the two photons in these non-mixing interference phenomena. More recently, 
double-slit biphoton quantum-optics experiments have been performed \cite{neve07,bobr14,exte09,wang17} where, 
in addition to the EPR-Bell-Bohm component, a double-occupancy NOON component is included in the prepared biphoton 
state.
}

\textcolor{black}{ 
Here we demonstrate an extensive correspondence between the dynamical 
evolution of two {\it interacting} ultracold fermionic or bosonic atoms trapped in a double well with the 
physics underlying the biphoton nonlocal quantum interference \cite{mand99,oubook,shihbook,exte09}.
We show that this analogy extends beyond, and carries deeper consequences, than just the pattern of 
the integrated scalar coincidence probability $P_{11}$ (see Sec.\ \ref{qomip} above and Sec.\ \ref{jop} below) 
associated with the EPR-Bell-Bohm-state component, to include an analogy between the underlying frequency 
interferograms (optical spectral frequency correlation maps over the frequencies ($\omega_1,\omega_2$) of the
two massless photons \cite{gerr15.1,gerr15.2,kwia18,exte09}) and the partial joint-coincidence probability maps,
$p_{\rm soc}(k_1,k_2)$ (see Sec.\ \ref{qomip} above and Sec.\ \ref{jop} below), for the two trapped massive atoms.
}

\textcolor{black}{
The $p_{\rm soc}(k_1,k_2)$ map constitutes a distinct contribution to 
the total second-order momentum correlation maps,
exhibiting a general form of a cosine-square quantum beat on the {\it difference\/} of the two momenta with fringes 
parallel to the main diagonal of the maps. Furthermore, we demonstrate that the NOON component of the two-atom
wave function generates another distinct contribution to the total second-order momentum correlation maps
exhibiting a general form of a sinusoidal quantum beat on the {\it sum\/} of the two particle momenta with fringes
parallel to the antidiagonal of the maps; this behavior is in agreement with the findings from recent 
double-slit biphoton quantum-optics experiments \cite{neve07,bobr14,exte09,wang17}.
}

\textcolor{black}{ 
Our findings will enable experimental extraction of the massive-particle single-occupancy- and NOON-component 
interference contribution terms from time-of-flight measurements which mirror \cite{altm04} the total second-order 
momentum correlations of the trapped ultracold particles. In this context, we note ongoing efforts in the 
experimental community to explicitly measure \cite{berg18} the total second-order 
momentum correlations of two interacting double-well trapped fermions or to devise protocols based on such 
correlations for the characterization of entanglement of two noninteracting distinguishable bosons 
\cite{schm17.2}.  
}

\subsection{Theoretical methodology}

The theoretical model employed in these investigation is the Hubbard model, formulated and implemented for three 
case studies of a double-well-trapped ultracold atom dimer: (i) two spinless bosons, (ii) two spin-1/2 bosons, and
(iii) two spin-1/2 fermions. The use of the Hubbard model allowed us to employ efficiently and effectively a 
unifying theoretical methodology to systems of varied characteristics, e.g., quantum statistics (bosons, fermions) 
and spin functions, and across the entire range of interparticle contact interaction range (from strong attraction 
to the high repulsion limit). Furthermore, along with the numerical results we provide a wealth of analytical 
solutions that we expect to aid the design of future experiments, as well as guide the analysis of current and 
forthcoming investigations. 

It is pertinent to add here that in past publications 
\cite{bran15,bran16,bran17,bran17.2,bran18}, we have employed the Hubbard model in conjunction with exact 
diagonalization (EXD) of the microscopic Hamiltonian, through the use of extensive, large-scale, convergent 
configuration interaction (CI) calculations. These comparative studies have verified that, with proper 
parametrization of the Hubbard model Hamiltonian (via use of the microscopic EXD calculations), the results of the 
Hubbard model agree well with those obtained from the ab-initio microscopic EXD calculations, as well as with 
experimental results, when available; see, e.g., \cite{joch15,bran15,bran18}.

Beyond uncovering valuable analytical and numerical solutions to complex physical problems, the results obtained 
in this work illustrate the outstanding merits of the Hubbard-model framework used here, in enabling and aiding 
theoretical research into fundamental problems, as entanglement and its dependence on interparticle interactions 
and on time, as well as through illustrating the suitability and applicability of the model to 
Bell-inequality-testing of the non-local nature of the phenomena studied here.

\subsection{Plan of the paper}

In light of the broad scope of this paper, as well as the varied audiences targeted by it (including researchers in
the fields of ultracold atoms, quantum optics, and condensed-matter physics), we provide here a most detailed plan
of the manuscript, aiming at facilitating easy navigation through this rather expansive work.  
The paper is organized as follows. In Sec.\ \ref{hubb}, the Hubbard model is briefly reviewed in a compact way 
applicable to all three cases of a pair of particle considered here [see (i)-(iii), above]. The Hubbard model 
eigenstates are given in analytical forms for all three cases in Sec.\ \ref{hubs2}. 
Sec.\ \ref{pint} addresses the topic of particle coincidence interferograms and the second-order momentum 
correlations. The discussion is divided into the following subsections: III A: Two-particle wave functions in 
momentum space; III B: The total second-order momentum correlations; 
\textcolor{black}{
III C: The two-particle coincidence 
interferogram, where a special role is played in the interpretation of the time-of-flight experiments by the 
extracted partial joint-coincidence probability spectrum $p_{\rm soc}(k_1, k_2)$ for detecting a pair of particles 
in the time-of-flight expansion image (far field), with the simultneously detected particles being associated with 
the single-occupancy-in-each-well component of the total two-atom wave function.} This partial joint-coincidence 
probability is of particular significance here, because unlike the primary photon optics sources used in the 
biphoton HOM experiments where the twin-pair of photons are generated in an EPR-Bell-Bohm entangled state (pure, 
with no double-occupancy contribution), our source, namely an ultracold atom dimer trapped in a double well, 
contains an entangled double-occupancy [NOON($\pm$)] component. This partial joint coincidence probability is 
related to the part of the momentum total wave function that involves exclusively the symmetrized or 
antisymmetrized cross products of both the left and right single-particle orbitals, respectively. Finally, Sec.\ 
\ref{illu} contains illustrative special cases, where selected cases are illustrated in the context of 
time-dependent evolution of a prepared wave-packet. 
\textcolor{black}{
Sec.\ III C discusses also the complementary double-occupancy spectral correlation maps, 
$p_{20+02}(k_1,k_2)$, associated with the NOON($\pm$) component of the two-atom wave function.}

Sec.\ \ref{simi} is devoted to a discussion of the similarities and differences of 
the double-well-trapped ultracold-atom-dimer with the biphoton used in quantum optics experiments. First we discuss
in IV A the double well as a different type of source producing a larger variety of pairs of entangled particles, 
and in IV B we present a detailed mathematical analysis of the correspondence between the joint-probabilities in 
quantum optics and the 2nd-order momentum correlations of two double-well-trapped ultracold atoms [expanding on the
comment we made above in connection to $p_{\rm soc}(k_1, k_2)$]. A connection between the the ultracold-atoms 
interference results and the Hanbury Brown-Twiss interferometry is discussed in Sec.\ \ref{hbt}.  

Sec.\ \ref{exp} is devoted to a demonstration, using the Hubbard model, 
that in the context of two ultracold atoms 
confined in a double-well trap, our theoretical extraction of the partial coincidence probability 
$p_{\rm soc}(k_1, k_2)$ from the total joint coincidence probability, allows for the use of interacting massive 
trapped particles to experimentally test the Bell inequalities, in close analogy with previous quantum-optics 
experiments \cite{rari90,aspe82,aspe82.2,ou88} that used twin pairs of entangled, but separated, photons. 
Entanglement aspects are discussed and illustrated, again with the use of the Hubbard model, in Secs.\ VI and 
VII. These show analytical and numerical results from investigations of entanglement properties of the double-well
trapped dimer, and the effects of interparticle interaction on the entanglement characteristics. 
In Sec.\ \ref{vn}, we discuss the von Neumann entropy for mode entanglement for the Hubbard-dimer eigenstates 
(Sec.\ \ref{vneig}) and for time-dependent wave packets made from these states. Entanglement concurrencies 
for two particles are displayed in Sec.\ \ref{conc} for both the eigenstates (Sec.\ \ref{chde}) and for 
wave packets made of these states (Sec.\ \ref{ctdp}). We summarize the paper, including a discussion of 
recent work on double photoionization of molecular hydrogen in Sec.\ \ref{summ}. 

Appendix \ref{a1} contains results pertaining to the diagonalization of the Hubbard-dimer Hamiltonians for the 
three cases discussed in this paper. 
\textcolor{black}{Because of the interdisciplinary nature of this work, and to allow a quick 
reference and recall of the definitions of quantities and symbols used throughout the paper, we also provide in 
this arXiv version in Appendix \ref{a2} a glossary of terms appearing in the following sections; this glossary
is removed from the published PRA version.}

\section{Hubbard model for two interacting particles in a double well} 
\label{hubb}

\subsection{Historical introduction and current effort}
\label{shubb}

\textcolor{black}{
We begin with a short summary of the history of the Hubbard model and a summary of our current efforts employing 
this model for studies of trapped finite ultracold atom systems; readers familiar with the Hubbard model may skip 
this introductory material.} 

The Hubbard model, independently conceived in several papers in 1963 
(see an editorial \cite{hubb13} on the occasion
of half a century of the Hubbard model), all aiming at treating correlated electrons in solids, is one of the most
successful quantum mechanical model Hamiltonians in condensed matter physics. From a technical point of view the 
model is an extension of the so-called tight-binding model, where particles can hop (tunnel) between lattice 
sites; in most applications only nearest-neighbor sites are included, and all hopping events have the same kinetic
energy, denoted in this paper as $-J$ instead of the more common notation $-t$ (in this paper, $t$ denotes time). 
The interaction between particles is limited in the simplest Hubbard 
Hamiltonian \cite{hubb63,hubb64} to that between particles occupying the same site, represented by an energy $U$. 

Applications of the Hubbard model to ultracold atoms trapped in optical-lattice potentials have been discussed 
since the late 1990’s \cite{jaks98,jaks05,essl10} and they have been shown to be a most useful and versatile tool 
in this field. It should be noted here that the simplicity of the Hubbard model is rather deceptive.
Indeed it has been found to be a ``mathematically hard'' problem, and an exact solution has been obtained only 
for the one-dimensional case. However, with increasing computer power it is possible to solve more complete Hubbard 
models -- that is, extended Hubbard Hamiltonians \cite{dutt15} that may include hopping processes beyond 
nearest-neighboring sites, consideration of interparticle inter-site interactions (beyond the on-site Hubbard 
$U$), and multistate Hubbard models. 

Furthermore, the Hubbard model, originally written for fermions (electrons), has been adapted to treat bosonic 
systems \cite{fish89,rama93}. Indeed occasions where experimental findings boosted the popularity of 
the model include the experimental demonstrations of transitions from a superfluid to a Mott insulator, found 
first for an optical lattice of ultracold bosonic atoms \cite{grei02} and later for fermionic ones \cite{joer08}.
Applications of the Bose-Hubbard model to two coupled Bose-Einstein condensates trapped in a double well 
confinement \cite{poll10,penn17} have further contributed to this popularity.
Another surge in the popularity of the model has been marked by the introduction of an adaptation of the Hubbard 
model, the so called $t$-$J$ model \cite{bran16,bran17.2,ande04}, as a candidate model for the emergence of a 
superconducting state, developed in the context of searching for a theory of high-T$_c$ superconductivity. From a 
practical perspective, increasing computer power and new computational platforms allow the numerical solution of ever
larger and more complex Hubbard model Hamiltonians.

In the past several years, we have developed and applied Fermi and Bose Hubbard models for the treatment of finite 
ultracold atom systems trapped in optical lattices of variable size \cite{bran15,bran16,bran17.2,bran17,bran18}. 
In these studies, we have treated varied systems, from Fermi dimers trapped in quasi-one-dimensional double wells 
as the elementary building blocks of the Hubbard model \cite{bran15}, to trapped Fermi dimers and trimers and 
finite spin chains \cite{bran16}, and  to investigations 
on eight fermions in coupled 4-site plaquettes as basic units, 
aiming at emulation and development of effective models for uncovering hole-pairing in high-T$_c$ 
superconductivity \cite{bran17.2}. 

More recently we have formulated and explored the properties 
of double-well trapped interacting ultracold atom systems (fermions or bosons) via investigations of two-particle
(second-order) density matrix and second order momentum correlations, exhibiting quantum bi-particle interference 
behavior, thus extending earlier (massless and non-interacting) bi-photon fundamental quantum behavior to the 
domain of massive and interacting quantum systems \cite{bran17,bran18}. 

In all these previous studies,
we have performed Hubbard model calculations in conjunction with exact diagonalization (EXD) of the corresponding 
microscopic Hamiltonian through large-scale, convergent, configuration-interaction (CI) calculations. These 
calculations allowed us to determine the appropriate Hubbard Hamiltonian parameters via fitting the Hubbard model 
results to the corresponding results from the EXD calculations. In all of these studies, the Hubbard modeling 
provided a faithful description of the EXD results. Particularly relevant to our current paper are our 
comparative results for the two-particle density matrix and two-particle (second-order) momentum correlation maps 
for bosonic and fermionic ultracold atoms calculated for the entire range (repulsive and attractive) of 
interatomic contact interactions \cite{bran17,bran18}.

\subsection{Two-site Hubbard-model Hamiltonians} 
\label{hubs1}

{\it (i) Two spinless bosons:\/} In this case the two-site Hubbard Hamiltonian has the following form
in second quantization:
\begin{align}
H_{spinless}^B = -J \sum_{i \neq j=1}^2 \hat{b}_i^\dagger \hat{b}_j  + 
\frac{U}{2} \sum_{i=1}^2 \hat{n}_i(\hat{n}_i-1),
\label{hbspl}
\end{align}
where $\hat{b}_i^\dagger$ and $\hat{b}_j$ are bosonic operators and 
$\hat{n}_i=\hat{b}_i^\dagger \hat{b}_i$ is the number 
operator at each site $i=1,2$. $J$ is the tunneling parameter between the two wells and $U$ is the onsite
Hubbard parameter. $U$ can be either positive (repulsive interaction) or negative (attractive interaction).  
\textcolor{black}{
Note that in this work, depending on context, we designate the two wells (sites) as ``1'' and ``2'', or
alternatively as $L$ (left) and $R$ (right).}\\
~~~~\\

{\it (ii) Two spin-1/2 bosons:\/} In this case the two-site Hubbard Hamiltonian has the following form
in second quantization:
\begin{align}
H_{spin-1/2}^B = -J \sum_{i \neq j=1,\sigma}^2 \hat{b}_{i,\sigma}^\dagger \hat{b}_{j,\sigma}  + 
\frac{U}{2} \sum_{i=1}^2 \hat{N}_i(\hat{N}_i-1),
\label{hbspf}
\end{align}
where $\hat{b}_{i,\sigma}^\dagger$ and $\hat{b}_{j,\sigma}$ are bosonic operators and
$\hat{N}_i=\sum_\sigma \hat{b}_{i,\sigma}^\dagger \hat{b}_{i,\sigma}$, 
with $\sigma$ denoting the up ($\uparrow$) or down ($\downarrow$) spin;
$\hat{N}_i$ is the number operator at each site $i$ including spin. \\
~~~~\\
{\it (iii) Two spin-1/2 fermions:\/} In this case the two-site Hubbard Hamiltonian has the following form
in second quantization:
\begin{align}
H^F = -J \sum_{i \neq j=1,\sigma}^2 \hat{a}_{i,\sigma}^\dagger \hat{a}_{j,\sigma}  + 
U \sum_{i=1}^2 \hat{n}^f_{i\uparrow}\hat{n}^f_{i\downarrow},
\label{hf}
\end{align}
where $\hat{a}_{i,\sigma}^\dagger$ and $\hat{a}_{j,\sigma}$ are fermionic operators and
$\hat{n}^f_{i,\sigma}=\hat{a}_{i,\sigma}^\dagger \hat{a}_{i,\sigma}$;
$\hat{n}^f_{i,\sigma}$ is the number operator at each site $i$ for a given spin.

The solutions of the above three Hubbard Hamiltonians are obtained by diagonalizing the associated
matrix Hamiltonians as described in Appendix \ref{a1}.

\subsection{Hubbard-model eigenstates} 
\label{hubs2}

Following Refs.\ \cite{bran17,bran18}, we assign space orbitals (Wannier-type single-particle wave functions)
to the trapped particles in the double well. In particular, we assume that the pair of single-particle 
$\psi_L(x)$ and $\psi_R(x)$ ground-state orbitals for each corresponding left or right well
are sufficient for defining the relevant many-body Hilbert space (case of well separated wells). Then the 
two-particle wave functions (see Hubbard solutions in Appendix \ref{a1}) associated with the corresponding 
two-site Hubbard model can be summarized in a compact way for all three cases examined here; namely, 
(i) two spinless bosons, (ii) two spin-1/2 bosons, and (iii) two spin-1/2 fermions.

For all three cases of the Hubbard dimer considered here, the two-particle Hubbard eigenfunctions are written 
as a product of a space part $\Psi(x_1,x_2)$ and a spin part $\chi(S,S_z)$ where $S$ is the total spin and 
$S_z$ is its projection. For spinless particles, obviously one has $\chi(S,S_z) =1$. For two spin-1/2 particles,
one has two spin eigenfunctions with spin projection $S_z=0$,
\begin{equation}
\sqrt{2}\chi(0,0) = \alpha(1)\beta(2)-\beta(1)\alpha(2),
\label{chi0} 
\end{equation}
a spin singlet state, which is antisymmetric under an exchange of particle indices, and
\begin{equation}
\sqrt{2}\chi(1,0) = \alpha(1)\beta(2)+\beta(1)\alpha(2),
\label{chi1} 
\end{equation}
a spin triplet state, which is symmetric under an exchange of particle indices. Similarly, for all three cases, 
the space part is a superposition of either symmetric or antisymmetric combinations of the $L$, $R$ space 
orbitals. For the symmetric combinations, one has three possibilities: 
\begin{equation}
\sqrt{2}n_+ \Phi_{S1}(x_1,x_2)=\psi_L(x_1)\psi_R(x_2)+\psi_R(x_1)\psi_L(x_2),
\label{phis1}
\end{equation}
or
\begin{equation}
\sqrt{2}n_- \Phi_{S2}(x_1,x_2)=\psi_L(x_1)\psi_L(x_2)-\psi_R(x_1)\psi_R(x_2),
\label{phis2}
\end{equation}
or
\begin{equation}
\sqrt{2}n_+ \Phi_{S3}(x_1,x_2)=\psi_L(x_1)\psi_L(x_2)+\psi_R(x_1)\psi_R(x_2).
\label{phis3}
\end{equation}
For the antisymmetric combination, there is a single possibility:
\begin{equation}
\sqrt{2}n_- \Phi_A(x_1,x_2)=\psi_L(x_1)\psi_R(x_2)-\psi_R(x_1)\psi_L(x_2).
\label{phia}
\end{equation}
The above [Eqs.\ (\ref{phis1})-(\ref{phia})] applies for cases where the overlap 
$S=\int \psi_L(x)\psi_R(x)dx$ is small. In such cases $n_{\pm}^2=1 \pm S^2 \approx 1$.

Taking into account that the total wave functions $\varphi$'s are symmetric (antisymmetric) for bosons 
(fermions) under interchange of particle indices, one has\\
~~~\\
{\it (i) for two spinless bosons (that is, two bosonic atoms with the same hyperfine state):\/}
\begin{align}
\begin{split}
\varphi_{1} &= \Xi_{S1}(x_1,x_2) \\
\varphi_{2} &= \Xi_{S2}(x_1,x_2) \\
\varphi_{3} &= \Xi_{S3}(x_1,x_2),
\end{split}
\label{vphi_i}
\end{align}

{\it (ii) for two spin-1/2 bosons (that is two bosonic atoms in a pair of hyperfine states):\/}
\begin{align}
\begin{split}
\varphi_{1} &= \Xi_{S1}(x_1,x_2) \chi(1,0) \\
\varphi_{2} &= \Xi_{S2}(x_1,x_2) \chi(1,0) \\
\varphi_{3} &= \Xi_{S3}(x_1,x_2) \chi(1,0) \\
\varphi_4  &= \Xi_A (x_1,x_2) \chi(0,0),
\end{split}
\label{vphi_ii}
\end{align}

{\it (iii) for two spin-1/2 fermions:\/}
\begin{align}
\begin{split}
\varphi_{1} &= \Xi_{S1}(x_1,x_2) \chi(0,0) \\
\varphi_{2} &= \Xi_{S2}(x_1,x_2) \chi(0,0) \\
\varphi_{3} &= \Xi_{S3}(x_1,x_2) \chi(0,0) \\
\varphi_4 &= \Xi_{A}(x_1,x_2) \chi(1,0),
\end{split}
\label{vphi_iii}
\end{align}
where 
\begin{align}
\begin{split}
\Xi_{S1}(x_1,x_2)& = {\cal A}({\cal U}) \Phi_{S1}(x_1,x_2) + {\cal B} ({\cal U}) \Phi_{S3} (x_1,x_2)\\
\Xi_{S2}(x_1,x_2)& = \Phi_{S2}(x_1,x_2) \\
\Xi_{S3}(x_1,x_2)& = {\cal D}({\cal U}) \Phi_{S1}(x_1,x_2) + {\cal E} ({\cal U}) \Phi_{S3} (x_1,x_2)  \\
\Xi_A (x_1,x_2) & = \Phi_A (x_1,x_2),
\end{split}
\label{xi}
\end{align}
and the coefficients ${\cal A}$, ${\cal B}$, ${\cal D}$, and ${\cal E}$ are given by
\begin{align}
\begin{split}
{\cal A}({\cal U}) & =
\frac{{\cal U}+\sqrt{{\cal U}^2+16}}{\sqrt{2}\sqrt{ {\cal U}^2 +{\cal U} \sqrt{{\cal U}^2+16}+16}}, \\
{\cal B}({\cal U}) & =
\frac{4}{\sqrt{2}\sqrt{ {\cal U}^2 +{\cal U} \sqrt{{\cal U}^2+16}+16}}, \\
{\cal D}({\cal U}) & = -{\cal A}(-{\cal U}), \\
{\cal E}({\cal U}) & = {\cal B}(-{\cal U}),
\end{split}
\label{abde}
\end{align}
and ${\cal U}=U/J$, where $U$ and $J$ are the Hubbard parameters for on-site interaction and intersite
tunneling, respectively.

\begin{figure}[t]
\includegraphics[width=7.5cm]{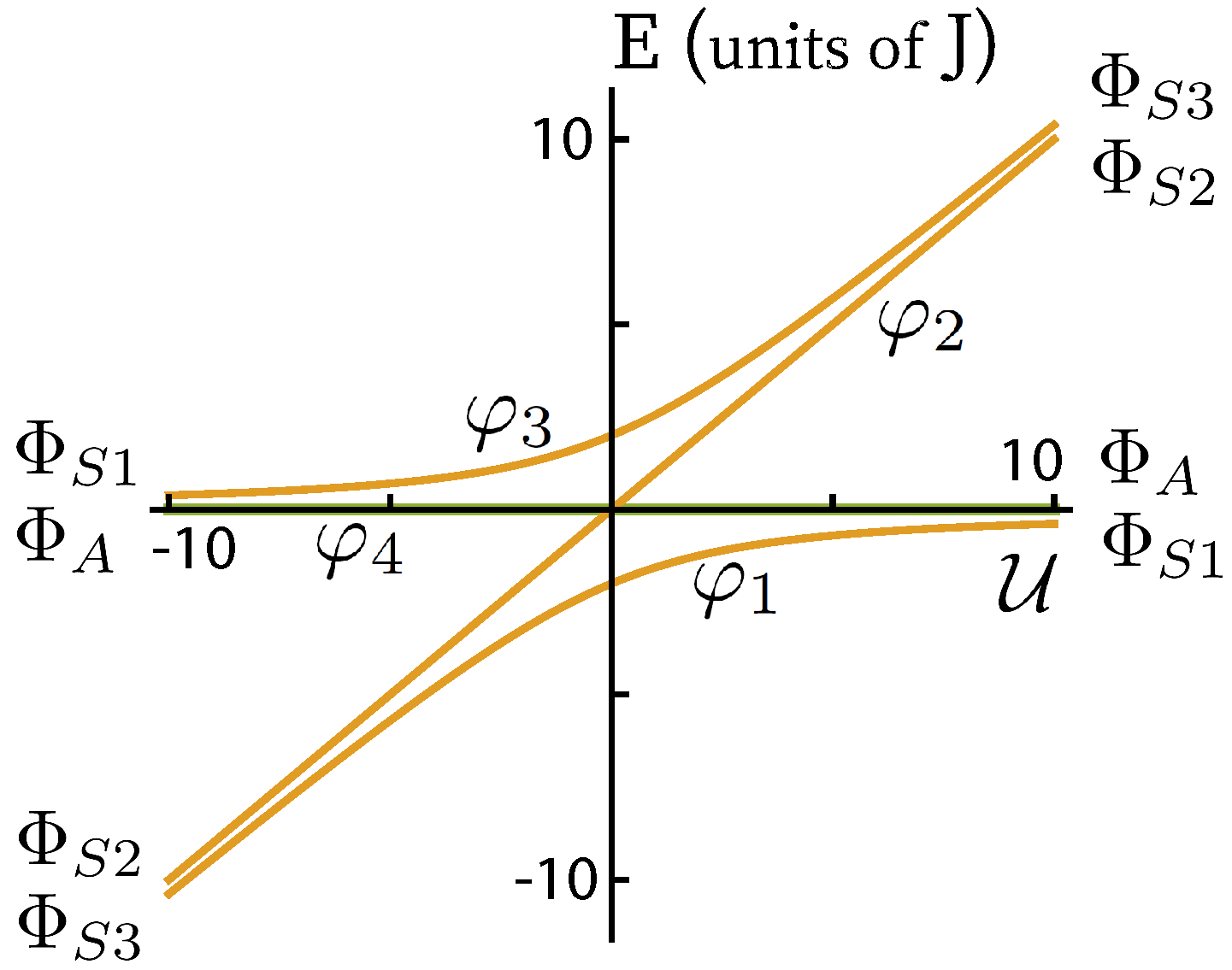}
\caption{The Hubbard-dimer energy levels for all three cases of (i) spinless bosons, (ii) spin-1/2 bosons, and
(iii) spin-1/2 fermions given by Eq.\ (\ref{ener}). See text for an explanation of the symbols  $\varphi$'s 
denoting the four Hubbard stationary wave functions, as well as their limiting $\Phi$ forms at 
${\cal U} \rightarrow \pm \infty$.}
\label{fig1}
\end{figure}

The Hubbard eigenenergies corresponding to these eigenfunctions are independent of the bosonic or fermionic 
nature of the two particles. They are given by
\begin{align}
\begin{split}
E_1& = \frac{J}{2} ( {\cal U} - \sqrt{{\cal U}^2 +16}) \\
E_2& = J {\cal U} = U\\
E_3& = \frac{J}{2} ( {\cal U} + \sqrt{{\cal U}^2 +16})\\
E_4 & = 0.
\end{split}
\label{ener}
\end{align}

These energies are plotted in Fig.\ \ref{fig1}. For spinless bosons, the total antisymmetric wave function
$\varphi_4=\Phi_A$ and corresponding energy $E_4$ are apparently missing. 

\section{Particle coincidence interferogram and total second-order momentum correlations} 
\label{pint}

\subsection{Two-particle wave functions in momentum space} 
\label{tpwfm}

The space orbital of a particle trapped within the $j$th well can be approximated by a displaced Gaussian 
function [localized about the position $x=d_j$, $j=1,2$ or, equivalently $L$ (left) or $R$ (right) in our 
one-dimensional trap], given by 
\begin{equation}
\psi_{j}(x)=\frac{1}{(2\pi)^{1/4}\sqrt{s}}\exp \left(-\frac{(x-d_j)^2}{4s^2}\right),
\label{psix}
\end{equation}
where $s$ denotes the width of the Gaussian function. The single-particle orbital $\psi_j (k)$ in the 
momentum Hilbert space is given by the Fourier transform of $\psi_j (x)$, namely 
$\psi_j (k) = (1/\sqrt{2\pi})\int_{-\infty}^{\infty}\psi_j (x)\exp (ikx) dx$. Performing this Fourier 
transform, one finds
\begin{equation}
\psi_{j}(k)=\frac{2^{1/4}\sqrt{s}}{\pi^{1/4}}\exp (-k^2s^2)\exp (id_j k).
\label{psik}
\end{equation}
Below we will also use the notation $L$ (left), $R$ (right) to denote the $j=1$, $j=2$ wells, respectively.

The Fourier transformed quantities $\Phi_{Si}(k_1,k_2)$ and $\Phi_A(k_1,k_2)$ preserve the same symmetry under
particle exchange [see Eqs.\ (\ref{phis1})-(\ref{phia}) for the coordinate-space orbitals]
and are easily obtained by replacing $\psi_j(x_i)$ by $\psi_j(k_i)$ [see Eq.\ (\ref{psik})].
Taking for simplicity the position of the left (right) well to be at $d_1=-d$ ($d_2=d$), this yields:  
\begin{equation}
\Phi_{S1}(k_1,k_2)=\frac{2s}{\sqrt{\pi}} e^{-s^2(k_1^2+k_2^2)}\cos[d(k_1-k_2)],
\label{phisk1}
\end{equation}
and
\begin{equation}
\Phi_{S2}(k_1,k_2)=-i \frac{2s}{\sqrt{\pi}} e^{-s^2(k_1^2+k_2^2)}\sin[d(k_1+k_2)],
\label{phisk2}
\end{equation}
and
\begin{equation}
\Phi_{S3}(k_1,k_2)=\frac{2s}{\sqrt{\pi}} e^{-s^2(k_1^2+k_2^2)}\cos[d(k_1+k_2)],
\label{phisk3}
\end{equation}
and
\begin{equation}
\Phi_A(k_1,k_2)=-i \frac{2s}{\sqrt{\pi}} e^{-s^2(k_1^2+k_2^2)}\sin[d(k_1-k_2)].
\label{phiak}
\end{equation}

The total spin-space wave function is symmetric for bosons and antisymmetric for fermions. In addition, for
$N=2$ massive particles and for eigenstates and wave packets that conserve the total spin, the spin and space 
degrees of freedom separate, i.e., in all cases the total wave function is a product of a pure spin eigenfunction 
and a pure space component. As a result, the corresponding momentum part of the total many-body wave function, in 
the general case of a nonstationary superposition of the Hubbard-dimer eigenstates (referred to also as wave 
packets below) that also conserves the total spin, can be written as:\\
~~~~~\\
(A) a symmetric superposition of the form
\begin{equation} 
\Psi_S(k_1,k_2)=\sum_{i=1}^3 {\cal C}_i({\cal U},t)\Phi_{Si}(k_1,k_2)
\label{psis}
\end{equation}
for the following three cases: (A1) two spinless bosons, (A2) two spin-1/2 bosons in the {\it triplet\/} spin 
state, and (A3) for two spin-1/2 fermions in the {\it singlet\/} spin state. Note the time argument, $t$, in the 
above equation; for  examples of ${\cal C}_i ({\cal U},t)$, see Sec.\ \ref{illu}.\\
~~~~~\\
(B) a single term 
\begin{equation}
\Psi_A(k_1,k_2)=\Phi_A(k_1,k_2)
\label{psia}
\end{equation}
in the following cases: (B1) two spin-1/2 bosons in the {\it singlet\/} spin state, or (B2) two spin-1/2 
fermions in the {\it triplet\/} spin state.\\ 
~~~~~\\
These results hold for both the ground and excited stationary eigenstates, as well as for the time-evolving 
wave packets of the two-site Hubbard model. The coefficients ${\cal C}_i({\cal U},t)$'s admit analytic 
dependence on the Hubbard parameter ${\cal U}$ for eigenstates and on both the interaction 
parameter ${\cal U}$ and time $t$ for wave 
packets; explicit examples will be discussed below. For eigenstates, these coefficients are real numbers, but
they are complex numbers for time-evolving wave packets. 

\subsection{The total second-order momentum correlations} 
\label{tsomc}

\textcolor{black}{
Generally, the second-order (two-particle) space density $\rho(x_1,x_1^\prime,x_2,x_2^\prime)$ 
for an $N$-particle system, is defined as an integral over the product of the many-body wave function 
$\Psi(x_1, x_2, \ldots,x_N)$ and its complex conjugate $\Psi^*(x_1^\prime, x_2^\prime, \ldots,x_N)$, 
taken over the coordinates $x_3,\ldots,x_N$ of $N-2$ particles. To obtain the second-order space correlation 
function, ${\cal G}(x_1,x_2)$, one sets $x_1^\prime=x_1$ and $x_2^\prime=x_2$. The second-order momentum 
correlation function ${\cal G}(k_1,k_2)$ is obtained via a Fourier transform (from real space to momentum space) 
of the two-particle space density  $\rho(x_1,x_1^\prime,x_2,x_2^\prime)$ \cite{bran17,bran18}. In the case
of $N=2$, the above general definition reduces to a simple expression for the two-particle correlation 
functions, as the modulus square of the two-particle wave function itself; this applies in both cases whether 
the two-particle wave function is written in space or in momentum coordinates. Consequently, the total 
second-order momentum correlations for the above-noted (A1)-(A3) cases [wave function with
space part symmetric under particle exchange, see Eq.\ (\ref{psis})] is given by
}
\begin{align}
\begin{split}
& {\cal G}_S(k_1,k_2) = |\Psi_S(k_1,k_2)|^2 = \\
& (4s^2/\pi) e^{-2 s^2(k_1^2+k_2^2)} \Big( {\cal C}_1^*{\cal C}_1 \cos^2[d(k_1-k_2)] + \\ 
& {\cal C}_2^*{\cal C}_2 \sin^2[d(k_1+k_2)] + {\cal C}_3^*{\cal C}_3 \cos^2[d(k_1+k_2)] + \\
& Re(-i{\cal C}_1^*{\cal C}_2) (\sin(2dk_1)+ \sin(2dk_2)) + \\ 
& Re({\cal C}_1^*{\cal C}_3) (\cos(2dk_1)+ \cos(2dk_2)) + \\
& Re(i{\cal C}_2^*{\cal C}_3) \sin[2d(k_1+k_2)] \Big).
\end{split}
\label{gsm}
\end{align}

The total second-order momentum correlations for the above (B1) and (B2) cases (wave function with a space part 
that is antisymmetric under particle exchange) are
\begin{align}
\begin{split}
 {\cal G}_A&(k_1,k_2) = |\Psi_A(k_1,k_2)|^2 = \\
& (4s^2/\pi) e^{-2 s^2(k_1^2+k_2^2)} \sin^2[d(k_1-k_2)]. 
\end{split}
\label{qa}
\end{align}

The specific cases of ${\cal G}_S(k_1,k_2)$ or ${\cal G}_A(k_1,k_2)$ for the four Hubbard eigenstates of two 
interacting spin-1/2 ultracold fermions were investigated in a recent publication \cite{bran18}; see Eqs.\ 
(5)-(6) therein. To facilitate the comparison for the three spin-singlet states in this case of two fermions, we 
note that ${\cal C}_2=0$ for the ground and 3rd excited states, and ${\cal C}_1={\cal C}_3=0$ for the 2nd 
excited state. We further remind the reader about the trigonometric identity $\cos^2(x)=(1+\cos(2x))/2$.  

\subsection{The two-particle coincidence interferogram} 
\label{jop}

\textcolor{black}{
The partial joint-coincidence probability spectrum $p_{\rm soc}(k_1,k_2)$ for detecting a pair of particles in 
the time-of-flight expansion image (far field) with the double-well-trapped particles belonging to the 
single-occupancy component of the two-atom wave function, that is with each of the particles originating from a 
{\it different\/} well (single-occupancy at each one of the two wells) is related (see also Ref.\ 
\cite{wang06}) to the part of the momentum total wave function that involves exclusively the symmetrized}
\begin{align}
\Phi_{S1}(k_1,k_2) = \psi_L(k_1)\psi_R(k_2)+\psi_R(k_1)\psi_L(k_2),
\label{cros}
\end{align}
or antisymmetrized 
\begin{align}
\Phi_A(k_1,k_2) = \psi_L(k_1)\psi_R(k_2)-\psi_R(k_1)\psi_L(k_2)
\label{croa}
\end{align}
cross products of {\it both\/} the left (indexed by a subscript $L$) and right (indexed by a subscript $R$) 
single-particle orbitals, respectively. As noted above, below we will use the subscript ``soc'' (single-occupancy 
component) to label this partial coincidence probability at the far field.

Taking into consideration the above and the expressions in Eq.\ (\ref{phisk1}) and Eq.\ (\ref{phiak}) [for 
$\Phi_{S1}(k_1,k_2)$ and $\Phi_A(k_1,k_2)$, respectively], the partial joint-coincidence probability spectrum, for 
detecting a pair of particles in the time-of-flight expansion image (far-field) with the double-well trapped 
particles belonging to the single-occupancy component of the two-atom wave function, is given by
\begin{align}
\begin{split}
p_{\rm soc}^S&(k_1,k_2)= {\cal C}_1^* {\cal C}_1 |\Phi_{S1}(k_1,k_2)|^2= \\
& {\cal C}_1^* {\cal C}_1 
\frac{4s^2}{\pi} e^{-2 s^2(k_1^2+k_2^2)} \cos^2[d(k_1-k_2)],
\end{split}
\label{p11s}
\end{align}  
for the (A1)-(A3) cases [wave function with momentum (or space) part symmetric under particle exchange, see Eq.\ 
(\ref{psis})], and by
\begin{align}
\begin{split}
p_{\rm soc}^A&(k_1,k_2)= |\Phi_A(k_1,k_2)|^2 = \\
& \frac{4s^2}{\pi} e^{-2 s^2(k_1^2+k_2^2)} \sin^2[d(k_1-k_2)],
\end{split}
\label{p11a}
\end{align}
for the (B1) and (B2) cases [wave function with a momentum (or space) part that is antisymmetric under particle 
exchange].

Likewise, the complementary double-occupancy probability spectrum, $p_{20+02}^S(k_1,k_2)$, or
$p_{20+02}^A(k_1,k_2)$, for detecting a pair of 
particles in the time-of-flight expansion image with both particles originating from the {\it same\/} well 
(whether the left or right one) is related to the part of the momentum total wave function that involves 
exclusively the product of left-left and right-right orbitals, namely
\begin{align}
\begin{split}
&p_{20+02}^S(k_1,k_2)= \\
&{\cal C}_2^* {\cal C}_2 |\Phi_{S2}(k_1,k_2)|^2 + {\cal C}_3^* {\cal C}_3 |\Phi_{S3}(k_1,k_2)|^2 +\\
& {\cal C}_2^*{\cal C}_3 \Phi^*_{S2}(k_1,k_2) \Phi_{S3}(k_1,k_2) +
{\cal C}_2 {\cal C}_3^* \Phi_{S2}(k_1,k_2)\Phi_{S3}^*(k_1,k_2) = \\
& \frac{4s^2}{\pi} e^{-2 s^2(k_1^2+k_2^2)} \Big( {\cal C}_2^* {\cal C}_2 \cos^2[d(k_1+k_2)] + \\
&{\cal C}_3^* {\cal C}_3 \sin^2[d(k_1+k_2)] + Re(i{\cal C}_2^*{\cal C}_3) \sin[2d(k_1+k_2)] \Big).
\end{split}
\label{p20s}
\end{align}  
for the cases with a symmetric space (or momentum) part, and
\begin{align}
p_{20+02}^A(k_1,k_2)= 0
\label{p20a}
\end{align}
for the cases with an antisymmetric space (or momentum) part.

Furthermore, the {\it in situ\/} (integrated) joint single-occupancy probability, $P_{11}$, and the {\it in situ\/} 
(integrated) double-occupancy probability, $P_{20+02}$, associated with destructive and constructive interference, respectively, are obtained by an integration over the momenta $k_1$ and $k_2$. One gets,
\begin{align}
\begin{split}
P_{11}^S= 1-P_{20+02}^S = & \int \int p_{\rm soc}^S(k_1,k_2) dk_1 dk_2 = \\
& {\cal C}_1^*({\cal U},t) {\cal C}_1({\cal U},t).
\end{split}
\label{pp11s}
\end{align}   
for the cases with a symmetric space (or momentum) part, and
\begin{align}
P_{11}^A=1,
\end{align}
for the cases with an anti-symmetric space (or momentum) part. 

\subsection{Illustrative specific cases} 
\label{illu}

\begin{figure*}[t]
\includegraphics[width=14.5cm]{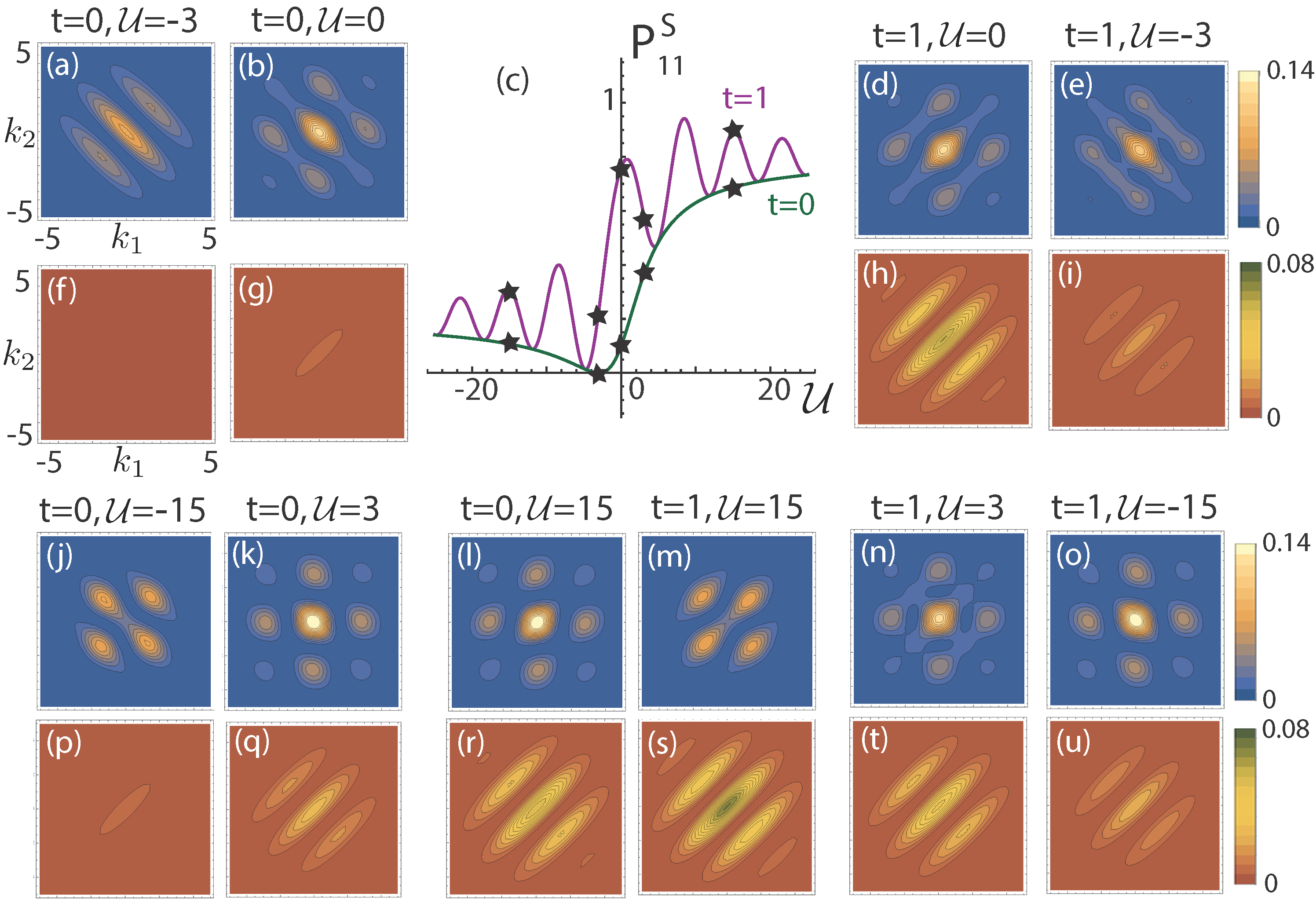}
\caption{Total momentum correlation ${\cal G}_S(k_1,k_2)$ maps [top panels with blue background denoted as 
(a,b,d,e,j-o)] and left-right joint-coincidence interferograms $p^S_{\rm soc}(k_1,k_2)$ [bottom panels with brown 
background denoted as (f,g,h,i,p-u)] for the space-symmetric 
two-particle wave packet defined in Eqs.\ (\ref{sup1}) and (\ref{wps}) for $\gamma=1/2$ and 
for two different times $t=0$ and $t=1$ $\hbar/J$ at five different values of the 
Hubbard interaction parameter ${\cal U}=0, \pm 3,\pm 15$. (c) Integrated coincidence probability 
$P^S_{11}$ as a function of ${\cal U}$ for two values of time $t=0$ and $t=1$ $\hbar/J$. 
Here and in all figures (unless explicitly stated otherwise): interwell distance $2d=2$ $\mu$m and
width $s=0.25$ $\mu$m. Momenta $k_1$ and $k_2$ in units of $1/\mu$m. ${\cal G}_S(k_1,k_2)$ and 
$p^S_{\rm soc}(k_1,k_2)$ in units of $\mu$m$^2$. See text for details.
}
\label{fig2}
\end{figure*}

One can generate a variety of time-evolving two-particle wave packets by considering an initial superposition of 
all four eigenstates $\varphi_i$, $i=1,\ldots,4$ of the Hubbard dimer 
[see Eqs.\ (\ref{vphi_i})-(\ref{vphi_iii})]. 
As an illustrative example, we will consider in this section an initial 
state that is a superposition of the lowest and highest in energy pair of eigenstates $\varphi_1$ and 
$\varphi_3$, i.e.,
\begin{align}
\Omega(t)=\frac{ \varphi_1 e^{ -i E_1 t/\hbar } + \gamma \varphi_3 e^{ -i E_3 t/\hbar } }
{ \sqrt{1+\gamma^2} }, 
\label{sup1}
\end{align} 
where $\gamma$ is a mixing parameter that can take both positive and negative values. 
\textcolor{black}{
In the noninteracting case
(${\cal U}=0$), the initial (at $t=0$) wave function in Eq.\ (\ref{sup1}) describes a single particle in each well,
i.e., an EPR-Bell-Bohm state, when the mixing parameter takes the value $\gamma=-1$. In the interacting case
(${\cal U} \neq 0$), the initial wave function in Eq.\ (\ref{sup1}) is not in general a pure EPR-Bell-Bohm state, 
but a mixed one comprising a NOON component as well; see below the case with constant $\gamma=1/2$ and 
independently variable ${\cal U}$. However, the initial wave function in Eq.\ (\ref{sup1}) can be forced to
be a pure EPR-Bell-Bohm state if the mixing parameter is chosen to depend on ${\cal U}$ as in Eq.\ (\ref{gamu})
below. Both of these two initial-state cases are investigated below.}   

The two eigenstates $\varphi_1$ and $\varphi_3$ have the same spin function $\chi(S,S_z)$, which factorizes. Thus, 
for investigating $\Omega(t)$, we can focus only on the time evolution of its space part, which has the form of 
Eq.\ (\ref{psis}) with the following specific coefficients
\begin{align}
\begin{split}
&{\cal C}_1({\cal U},t)= ( {\cal A}({\cal U}) e^{-i E_1 t/\hbar} + 
\gamma {\cal D}({\cal U}) e^{-i E_3 t/\hbar} )/
\sqrt{\gamma^2+1},\\
&{\cal C}_2({\cal U},t)=0,\\
&{\cal C}_3({\cal U},t)= ( {\cal B}({\cal U}) e^{-i E_1 t/\hbar} + 
\gamma {\cal E}({\cal U}) e^{-i E_3 t/\hbar} )/
\sqrt{\gamma^2+1}.
\end{split}
\label{wps}
\end{align}

For the wave packet specified by the coefficients in Eq.\ (\ref{wps}), the left-right joint-coincidence 
interferogram is given by
\begin{align}
\begin{split}
p_{\rm soc}^S(k_1,& k_2) = 
\frac{4 s^2 e^{-2 s^2 \left(k_1^2+k_2^2\right)} \cos ^2(d (k_1-k_2))}{\pi} \times \\
& \left(\frac{1}{2}+\frac{(1-\gamma^2) {\cal U} -8 \gamma \cos \left( tJ \sqrt{ {\cal U}^2+16}/\hbar\right)}
{2 \left(1+\gamma^2\right) \sqrt{{\cal U}^2+16}}\right).
\end{split}
\label{p11mapxmpl1}
\end{align}

Following the discussion in the previous section, summation over the momenta $k_1$ and $k_2$ yields the integrated
joint-coincidence probability
\begin{align}
\begin{split}
P_{11}^S =
\frac{1}{2}+\frac{(1-\gamma^2) {\cal U} -8 \gamma \cos \left( tJ \sqrt{ {\cal U}^2+16}/\hbar\right)}
{2 \left(1+\gamma^2\right) \sqrt{{\cal U}^2+16}}.
\end{split}
\label{p11xmpl1}
\end{align}

We note that this wave packet allows for the occurrence of total destructive interference. Indeed, from Eq.\ 
(\ref{p11xmpl1}), one sees that $P_{11}^S=0$ when $t=0$ and ${\cal U}=2(\gamma^2-1)/\gamma$.\\
~~~~\\
\textcolor{black}{
{\it Case of $\Omega(t)$ with constant $\gamma=1/2$ and variable ${\cal U}$ 
(Initial state contains both EPR-Bell-Bohm and NOON components):\/}} In this case, Eq.\ (\ref{p11xmpl1}) yields a 
vanishing integrated joint-coincidence probability, $P_{11}^S=0$, for $t=0$ and ${\cal U}=-3$. This vanishing value
is indicated by a star in the full curve of $P_{11}^S$ [specified by Eq.\ (\ref{p11xmpl1})] when plotted as 
a function of ${\cal U}$ while $t$ is kept constant at $t=0$; see Fig.\ \ref{fig2}(c). The corresponding curve 
for setting $t=1$ $\hbar/J$ in Eq.\ (\ref{p11xmpl1}) is also displayed for comparison in the same 
frame [Fig.\ \ref{fig2}(c)]. Unlike the $t=0$ curve, the $t=1$ curve does not reach a vanishing value at any 
point ${\cal U}$; in addition, it exhibits an oscillatory behavior with varying ${\cal U}$, in contrast to the 
$t=0$ curve. 

For a $2d=2$ $\mu$m interwell separation and a Wannier space-orbital width $s=0.25$ $\mu$m, 
total second-order momentum correlation maps 
${\cal G}_S(k_1,k_2)$ [top-row, blue-background frames) are 
displayed in Figs.\ \ref{fig2}(a,b,d,e,j-o) along with the corresponding joint-coincidence spectral maps 
$p_{\rm soc}^S(k_1,k_2)$ [bottom-row, brown-background frames] in Figs.\ \ref{fig2}(f,g,h,i,p-u) for the pairs 
of $(t,{\cal U})$ values indicated by black stars on the two $P^S_{11}$ curves displayed in Fig.\ 
\ref{fig2}(c). The topology of the patterns in the ${\cal G}_S(k_1,k_2)$ maps illustrate the fact that the total
2nd-order momentum maps result from the interference of several components that vary sinusoidally as a function 
of the single momenta $k_1$ and $k_2$, as well as their sum $k_1+k_2$ and difference $k_1-k_2$ [see Eq.\ 
(\ref{gsm})]. As a result this topology varies significantly between different pairs $(t,{\cal U})$
of time and interaction-strength values. We note that in all subsequent figures (unless explicitly stated 
otherwise), we will use the same interwell distance $2d=2$ $\mu$m and width $s=0.25$ $\mu$m, when such
parameters are relevant.

In contrast, the topology of the $p_{\rm soc}(k_1,k_2)$ maps remains unchanged, exhibiting a number of fringes 
parallel to the main diagonal. This reflects the fact that only one sinusoidal component dependent on the 
difference of the momenta $k_1-k_2$ contributes [see Eq.\ (\ref{p11s})] to the joint-coincidence correlation 
spectrum. The number and amplitude (visibility) of these fringes depend on the value of $P^S_{11}$ [see Eq.\ 
(\ref{pp11s})]; naturally for the special values $(t=0, {\cal U}=-3)$ (where $P^S_{11}=0$), no fringe 
structure is present [see Fig.\ \ref{fig2}(f)]. We note that the uniformity of the topology of fringes, as well 
as the dependencies on the difference of the single-photon momenta $k_1-k_2$ (or frequencies $\omega_1-\omega_2$)
and on the magnitude of the integrated joint-coincidence $P^S_{11}$ are also characteristic properties of the 
optical spectral correlation maps; see, e.g., Fig.\ 3 in Ref.\ \cite{gerr15.1} and Fig.\ 1 in 
Ref.\ \cite{gerr15.2}. 

\begin{figure}[t]
\includegraphics[width=7.5cm]{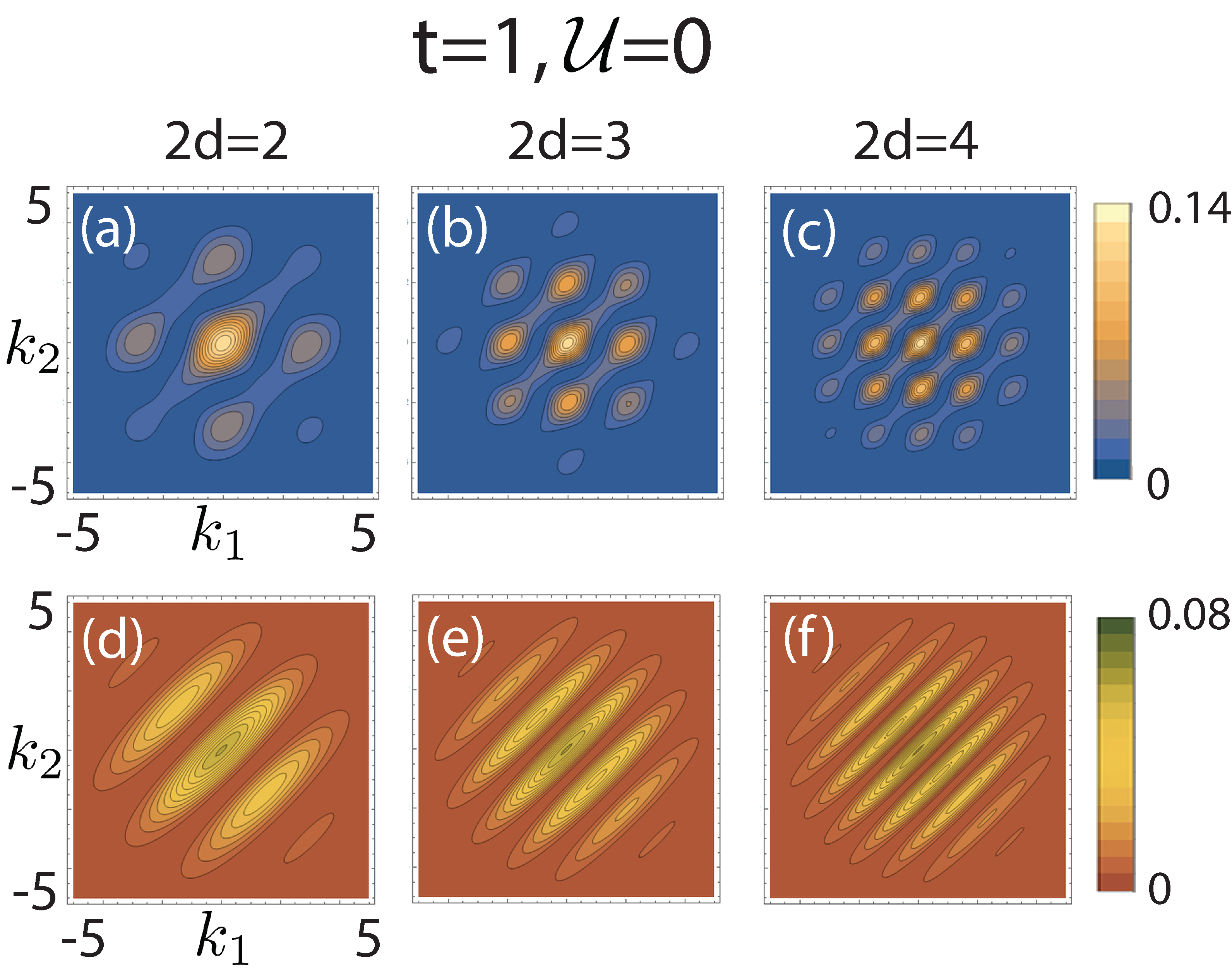}
\caption{2nd-order total momentum correlation maps [${\cal G}_S(k_1,k_2)$, upper row, blue panels] and 
corresponding joint-coincidence spectral maps [$p^S_{\rm soc}(k_1,k_2)$, lower row, brown panels] for 
the space-symmetric two-particle wave packet defined in Eqs.\ (\ref{sup1}) and (\ref{wps}) and for $\gamma=1/2$. 
Three different interwell separations, (a,d) $2d=2$ $\mu$m, (b,e) $2d=3$ $\mu$m, and (c,f) $2d=4$ $\mu$m, 
are considered. Note that the number of visible fringes increases with increasing $2d$. Remaining parameters 
$t=1$ $\hbar/J$, and ${\cal U}=0$. Momenta $k_1$ and $k_2$ in units of $1/\mu$m. ${\cal G}_S(k_1,k_2)$ and 
$p^S_{\rm soc}(k_1,k_2)$ in units of $\mu$m$^2$.}
\label{fig3}
\end{figure}

From Eq.\ (\ref{p11xmpl1}), it is seen that the integrated joint-coincidence probability 
$P_{11}^S$ is independent of the interwell separation $2d$; this is a consequence of the large interwell 
separation [$d >> s$, where $s$ is the Gaussian-width parameter in Eq.\ (\ref{psix})] which yields an
exponentially small overlap between the space orbitals of the two trapped particles. Because the interwell 
separation in our two-particle case corresponds to the time delay $\Delta \tau$ between the two photons that 
impinge on a beamsplitter in a HOM-like experimental arrangement (see also the electronic HOM
\cite{liu98,jonc12,bocq13} and Ref.\ \cite{aspe15}), it is apparent that our maps correspond to 
points on the shoulders of the HOM dip when compared to the spectral maps in Refs.\ \cite{gerr15.1,gerr15.2}.  
Nevertheless, a dependence on the interwell separation is evidenced by the $p_{\rm soc}(k_1,k_2)$ spectral maps 
themselves, because the distance between fringes equals $1/d$ [see the argument of the cosine term in 
Eq.\ (\ref{p11s})]. Thus a larger interwell separation yields a larger number of fringes within the
visible window allowed by the damping factor $\exp[-2s^2(k_1^2+k_2^2)]$; this behavior is illustrated in Fig.\ 
\ref{fig3}. To further stress the analogy with biphoton quantum optics, we note that a similar behavior is also 
present in the recently measured instances of optical joint-coincidence intensity spectra (see Fig.\ 3 in Ref.\ 
\cite{gerr15.1} and Fig.\ 1 in Ref.\ \cite{gerr15.2}).

\begin{figure}[t]
\includegraphics[width=8.0cm]{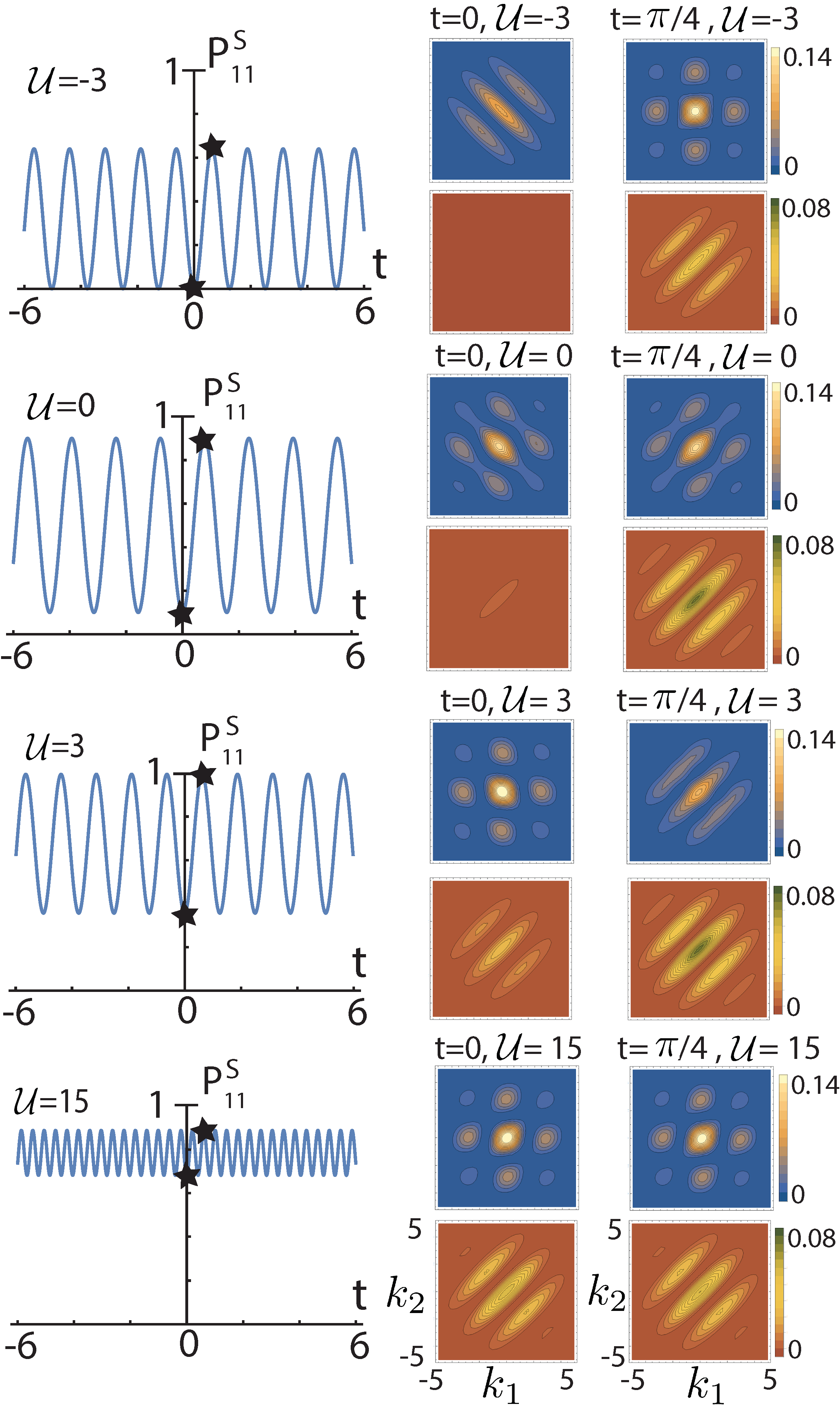}
\caption{Total momentum correlation ${\cal G}_S(k_1,k_2)$ maps (top panels, blue background) and left-right 
joint-coincidence interferograms $p_{\rm soc}(k_1,k_2)$ (bottom panels, brown background) for the space-symmetric
wave packet defined in Eqs.\ (\ref{sup1}) and (\ref{wps}) for $\gamma=1/2$ and for two 
different times $t=0$ and $t=\pi/4$ $\hbar/J$ at four different values of the Hubbard parameter 
${\cal U}=-3,0,3,15$ (from top to bottom). The frames on the left display the integrated joint-coincidence 
probability $P^S_{11}$ as a function of time $t$ for these four values of ${\cal U}$. 
Momenta $k_1$ and $k_2$ in units of $1/\mu$m. ${\cal G}_S(k_1,k_2)$ and $p^S_{\rm soc}(k_1,k_2)$
in units of $\mu$m$^2$. See text for details.
}
\label{fig4}
\end{figure}

Fig.\ \ref{fig4} (left column) displays a complimentary aspect of the joint-coincidence probability $P_{11}^S$, 
that is the behavior of the r.h.s. of Eq.\ (\ref{p11xmpl1}) as a function of time $t$ for constant ${\cal U}$ 
and $\gamma=1/2$; the four frames correspond (from top to bottom) to four different values 
${\cal U}=-3$, 0, 3, and 15. In all instances this time evolution is oscillatory and the period of oscillations 
$T$ decreases with increasing $|{\cal U}|$. Indeed from the argument of the cosine term in Eq.\ (\ref{p11xmpl1}), 
one has $T=2\pi\hbar/( J\sqrt{{\cal U}^2+16} )$; in addition, 
the amplitude of the oscillations decreases with increasing $|{\cal U}|$. For ${\cal U}=-3$ the
minima of the oscillations reach vanishing values. However, as mentioned previously, this vanishing of
$P_{11}^S$ does not correspond to the minimum value of an HOM dip because the overlap of the space orbitals 
of the two trapped atoms remains exponentially small ($d>>s$). 

For completeness, we present on the right side of Fig.\ \ref{fig4} the underlying joint-coincidence spectral 
decomposition maps at pair of points $(t,{\cal U})$ marked by a black star on each of the four $P_{11}^S$
curves. The corresponding total 2nd-order momentum correlation maps are also displayed for comparison. It is 
rewarding to observe that these maps offer further confirmation of the properties discussed in connection to Fig.\ 
\ref{fig2}, in particular the invariability of the topology of the fringes in the $p_{\rm soc}^S(k_1,k_2)$ 
interferograms (frames with brown background) and the dependence of the fringe intensity on the value of the 
integrated $P_{11}^S$. 

The oscillatory behavior of the joint-coincidence probability $P_{11}^S$ exhibited in Figs.\ 
\ref{fig2} and \ref{fig4} is analogous to that found in many experiments 
\cite{mand88,shih88,fran89,mand90,kwia90,rari90,shih96,remp04} of quantum optics employing time-delayed 
or distance-separated photons in biphoton-state configurations \cite{oubook,shihbook} which confirmed
important aspects of quantal entanglement and quantum nonlocality.\\
~~~~\\
\begin{figure*}[t]
\includegraphics[width=14.5cm]{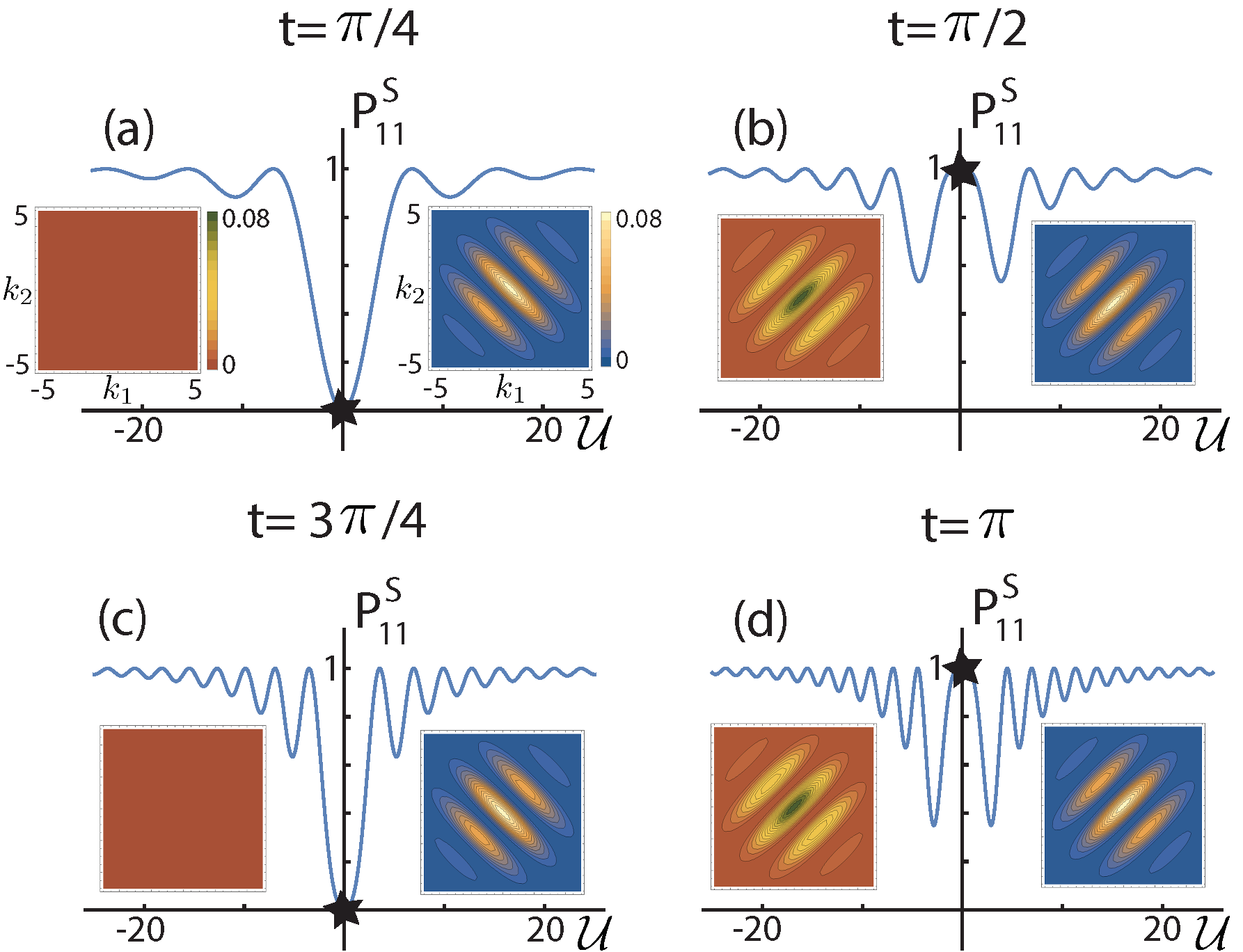}
\caption{
Integrated joint-coincidence probability curves $P^S_{11}$ [see Eq.\ (\ref{p11xmpl2})] as a function of ${\cal U}$ 
for four characteristic values of time, (a) $t=\pi/4$ $\hbar/J$, (b) 
$t=\pi/2$ $\hbar/J$, (c) $t=3\pi/4$ $\hbar/J$, and (d) $t=\pi$  $\hbar/J$.
This case corresponds to the space-symmetric wave packet defined in Eqs.\ (\ref{sup1}) and (\ref{wps}),
but for a variable $\gamma({\cal U})$ given by Eq.\ (\ref{gamu}). 
The insets display corresponding total 2nd-order momentum correlation ${\cal G}_S(k_1,k_2)$ maps (right panels, 
blue background) and left-right joint-coincidence interferograms 
$p_{\rm soc}(k_1,k_2)$ (left panels, brown background). 
Momenta $k_1$ and $k_2$ in units of $1/\mu$m. ${\cal G}_S(k_1,k_2)$ and $p^S_{\rm soc}(k_1,k_2)$
in units of $\mu$m$^2$. See text for details.
}
\label{fig5}
\end{figure*}

\textcolor{black}{
{\it Initial wave packet $\Omega(t=0)$ with one particle in each well for any value of ${\cal U}$:\/}} In general,
as mentioned previously, the initial wave packet in Eq.\ (\ref{sup1}) does not describe a state with one particle 
in each well. For this to happen, one must have 
${\cal C}_3({\cal U},0)={\cal B}({\cal U}) + \gamma {\cal E}({\cal U}) =0 $ 
[see Eq.\ (\ref{wps})], which yields a ${\cal U}$-dependent $\gamma$, i.e.,
\begin{align} 
\gamma({\cal U}) = \frac{1}{4}({\cal U}-\sqrt{{\cal U}^2+16}). 
\label{gamu}
\end{align}

In this case, Eqs.\ (\ref{p11mapxmpl1}) and (\ref{p11xmpl1}) yield the following two expressions
\begin{align}
\begin{split}
p^S_{\rm soc}(k_1,& k_2)= \frac{4 s^2 e^{-2 s^2 \left(k_1^2+k_2^2\right)} \cos^2[d (k_1-k_2)]}{\pi} \times \\
& \left(1-\frac{16}{{\cal U}^2+16}\sin^2\left(\frac{tJ\sqrt{{\cal U}^2+16}}{2\hbar}\right)\right),
\end{split}
\label{p11mapxmpl2}
\end{align}
and
\begin{align}
P^S_{11} =
1-\frac{16}{{\cal U}^2+16}\sin^2\left(\frac{tJ\sqrt{{\cal U}^2+16}}{2\hbar}\right).
\label{p11xmpl2}
\end{align}

From an inspection of Eq.\ (\ref{p11xmpl2}), it is seen that the integrated 
joint-coincidence probability now reaches
a vanishing value only for the non-interacting case (${\cal U}=0$). This defines a variant behavior compared to 
that discussed in the previous paragraph when $\gamma=1/2$ and was taken to be independent of the controlling 
parameter ${\cal U}$.

Fig.\ \ref{fig5} displays the time-oscillatory behavior as a function 
of ${\cal U}$ of the integrated joint-coincidence $P^S_{11}$ 
specified by Eq.\ (\ref{p11xmpl2}) at four characteristic time values $t=j\pi/4$ $\hbar/J$, $j=1,2,3,4$. 
For $t=\pi/4$ $\hbar/J$ and $t=3\pi/4$ $\hbar/J$, $P^S_{11}({\cal U}=0)$ equals zero, while for $t=\pi/2$ $\hbar/J$
and $t=\pi$ $\hbar/J$, $P^S_{11}({\cal U}=0)$ equals unity. In both cases, however, $P^S_{11}$ exhibits an
oscillatory behavior with diminishing amplitude and it approaches rather rapidly unity for 
${\cal U} \rightarrow \pm \infty$. The period $T_{\cal U}$ of the oscillations in ${\cal U}$ decreases
with increasing $t$ values according to ${\cal U}_P=\sqrt{4\hbar^2/(tJ)^2-16}$. 
\textcolor{black}{We note that similar ``quantum-beating'' patterns with variable amplitude, 
distinct from the HOM dip, have been reported (as early as 1988) in many experimental 
(see, e.g., Refs.\ \cite{mand88,remp04,oubook,shihbook}) or theoretical \cite{wang06} studies concerning
biphoton interference. 
}

For completeness, Fig.\ \ref{fig5} displays also total 2nd-order momentum ${\cal G}_S(k_1,k_2)$
correlation maps and corresponding $p^S_{\rm soc}(k_1,k_2)$ joint-coincidence spectral maps at the values 
marked by a black star on the $P^S_{11}$ curves. Again, in contrast to the ${\cal G}_S(k_1,k_2)$ maps (blue
background), one notices the invariability of the topology of the fringe pattern in the $p^S_{\rm soc}(k_1,k_2)$
maps (brown background). Apparently, in the case of the joint-coincidence spectral maps, there is no fringe 
structure when $P^S_{11}=0$ [maps with brown background in (a) and (c)], while the fringe visibility is maximum 
when $P^S_{11}=1$ [maps with brown background in (b) and (d)]. Furthermore the fringe structure of the
total 2nd-order momentum correlation maps (blue background) in (a) and (c) reflect the fact that the 
biphoton state is a pure NOON, $(|2,0>+|0,2>)/\sqrt{2}$, state (see also Ref.\ \cite{bran18}), while
in (b) and (d), they coincide with the $p^S_{11}(k_1,k_2)$ maps (brown background), which reflects the
fact that the biphoton is a pure maximally entangled Bell state, $(|1_L,1_R>+|1_R,1_L>)/\sqrt{2}$.
\textcolor{black}{
A measurement of the joint-coincidence probability [Eq.\ (\ref{p11xmpl2})] for ${\cal U} =0$, in a system of 
trapped bosonic atoms, has been reported recently \cite{rega18}.} 

\begin{figure}[t]
\includegraphics[width=7.5cm]{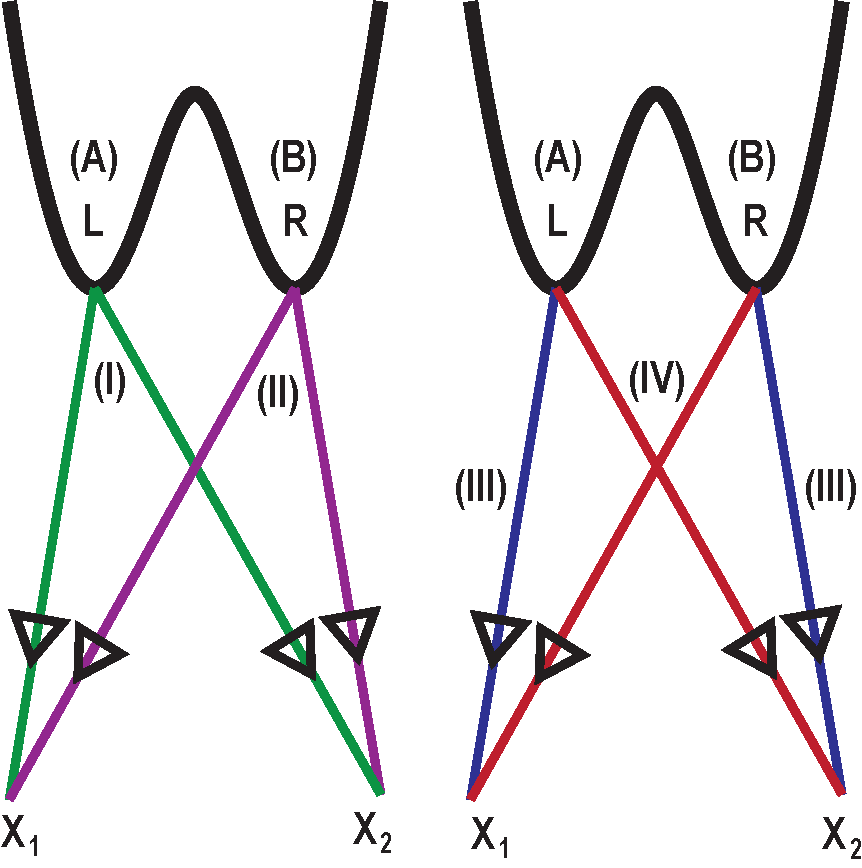}
\caption{Diagrams of the four different amplitudes that contribute to the total 2nd-order momentum correlations 
${\cal G}(k_1,k_2)$ [the total 4th-order coincidence $\widetilde{P}_2(x_1,x_2)$ 
in Ref.\ \cite{mand99}; see Eq.\ (5) therein]. In these diagrams, the two-photons (particles) in each event 
(I)-(IV) are described by two lines of the same color. 
In the diagrams labeled as (I) and (II) (green and magenta color, respectively), both particles 
(photons) originate from the same well (light source). In the diagrams labeled as (III) and (IV) (blue and red 
color, respectively), each particle (photon) in the pair originates from a different well (light source). 
$L$ $(A)$ and $R$ $(B)$ denote the two trapping wells (light sources). $X_1$ and $X_2$ denote two positions in 
the time-of-flight expansion cloud (the optics far field). In the case of the optics far field, the symbols 
$x_1$ and $x_2$ (lower case) are often used \cite{mand99}, instead of the upper case $X_1$ and $X_2$.
The positions of the two particles in the time-of-flight cloud are related to the single-particle momenta $k_1$ 
and $k_2$ at the double-well trap as $X_j=\hbar k_j t_{\rm TOF})/M$, $j=1,2$ \cite{altm04}, where $M$ is
the mass of the atom and $t_{\rm TOF}$ is the experimental time of flight. Note that the use of the
term 2nd-order by us corresponds to the use of the term 4th-order in quantum optics.
}
\label{fig6}
\end{figure}

\textcolor{black}{
\section{Similarities and differences with the biphoton of quantum optics} 
\label{simi}
}

\textcolor{black}{
\subsection{Double well: A different type of source producing a larger variety of pairs of entangled particles} 
\label{ntp}
}

\textcolor{black}{
Our approach of extracting a partial coincidence-probability component $p_{\rm soc}(k_1,k_2)$ out of the total 
second-order momentum correlations ${\cal G}(k_1,k_2)$ is 
congruent to the reasoning underlying the introductory 
remarks of Mandel in Ref.\ \cite{mand99} made in the context of the case of far-field interference of two 
entangled photons originating from two separated quantal light sources $A$ and $B$. In that review, a relevant 
one-term partial joint-coincidence probability [denoted as $P_2(x_1,x_2)$, see Eq.\ (6) therein] was extracted 
from the most general, but auxiliary (as of the time of Ref.\ \cite{mand99}), multi-term interference expression 
$\widetilde{P}_2(x_1,x_2)$ given in his Eq.\ (5) [see also Eq.\ (\ref{op2}) below]. 
As explicitly shown below, Mandel's general joint-probability $\widetilde{P}_2(x_1,x_2)$ corresponds
to our total second-order momentum correlations ${\cal G}(k_1,k_2)$ \cite{note3}, whereas Mandel's partial
joint-probability $P_2(x_1,x_2)$ corresponds to our $p_{\rm soc}(k_1,k_2)$. 
}

\textcolor{black}{
To appreciate why a comparison between general and partial joint probabilities is of relevance in the context 
of the TOF physics of two ultracold atoms in a double well, it is pertinent to comment here on the type of
primary light sources in quantum optics versus that of the double well. Indeed, the primary sources
in quantum optics consist of twin pairs of spatially separated photons representing EPR-Bell-Bohm entangled 
states \cite{shih03}. These states are produced through the process of spontaneous parametric down conversion 
(SPDC). Non-entangled photon pairs from two low-density independent primary sources (usually semiconductor 
quantum dots) have also been used in quantum optics \cite{sant02}. Such separated-particles pairs 
have also been used in experiments with propagating electron \cite{liu98,bocq13,jonc12,burk07} 
(independent sources) or ultracold-atom \cite{aspe15} (entangled twin atoms) beams that aim to accurately
replicate with massive particles the quantum-optics repertoire. Below, for convenience, such pairs will be 
labeled by us using a broad brush as $(1_A,1_B)$ [or $(1_L,1_R)$], that is, this round-bracket (instead of ket) 
notation accentuates the spatial separation in these states, and omits the aspects associated with the spin 
degree of freedom and the wave function symmetrization or antisymmetrization due to quantum statistics. 
} 

\textcolor{black}{
Concerning the two-ultracold-atoms double-well primary source studied in this paper, a crucial difference from 
the primary sources described in the previous paragraph is the presence of entangled NOON($\pm$) states (broadly
labeled as $[(2_L,0_R)\pm (0_L,2_R)/\sqrt{2}]$ here) as contributing components in the two-particle wave 
function. Such NOON($\pm$) states are a direct result of the fact that double occupancy is allowed in each well 
(see Hubbard-model description in Appendix \ref{a1}); diagrams (I) and (II) in Fig.\ \ref{fig6} correspond to the 
double occupancy events. This NOON-state component enables the non-vanishing of the interference cross terms 
between {\it any two\/} of all four diagrams, (I), (II), (III), and (IV), portrayed in Fig.\ \ref{fig6}, 
and it is responsible for the full complexity of our ${\cal G}(k_1,k_2)$ or Mandel's auxiliary 
$\widetilde{P}_2(x_1,x_2)$. On the contrary, if only single-occupancy 
is allowed for each well or source [type of primary sources labeled $(1_A,1_B)$ above], only the interference
cross term between diagrams (III) and (IV) in Fig.\ \ref{fig6} survives and the quantities ${\cal G}(k_1,k_2)$ 
or $\widetilde{P}_2(x_1,x_2)$ reduce to the simpler forms, $p_{\rm soc}(k_1,k_2)$ or $P_2(x_1,x_2)$, 
respectively.} 
\textcolor{black}{Note that the light sources are designated as $A$ and $B$, whereas ``particle sources'' 
(confining wells) are denoted as $L$ (left) and $R$ (right); see Fig.\ \ref{fig6}.}

\textcolor{black}
{NOON components in the biphoton wave function were originally generated in a secondary step in quantum-optics 
experiments through the use of beam-splitters; see the seminal work of Hong-Ou-Mandel 
\cite{hong87,oubook}. For electron beams, a quantum-point 
contact is used for that purpose \cite{liu98,jonc12,bocq13}, and in ultracold-atom free-space beams, Bragg 
diffraction setups are employed \cite{aspe15}. For the case of a double-well trapped ultracold-atom pairs, 
no physical beam-splitter is required because the NOON component can be generated already in the primary 
(double-well) source. The role played by the physical beam-splitter in generating a NOON state is replicated 
in the double well by the time-evolution due to interwell tunneling \cite{kauf14,isla15} or by 
inter-particle interaction effects \cite{bran18}. However, another key aspect of the HOM experiment, i.e., the 
coalescence of the two photons on the beam-splitter, which leads to the celebrated HOM dip (typically having the 
shape of an inverted Gaussian with shoulders \cite{hong87,branc17,aspe15}), cannot be mimicked in the double-well
case because of the physical separation between the left and right wells. 
}

\textcolor{black}{
Starting with the early 2000's, NOON($+$) biphoton components have been generated using lenses in the near field 
to control the focusing of the light from a SPDC crystal on a double slit 
\cite{burl97,neve07,exte09,bobr14,wang17}. In this case the joint-coincidence probability at the far field 
resembles Mandel's $\widetilde{P}_2(x_1,x_2)$ general expression, 
and thus our 2nd-order momentum correlations ${\cal G}(x_1,x_2)$. As a result, beyond the $P_2(x_1,x_2)$ 
[$p_{\rm soc}(k_1,k_2) \propto \cos^2(d(k_1-k_2))$] pattern with fringes parallel to the main diagonal, 
some of the interferograms considered in Refs.\  \cite{burl97,neve07,exte09,bobr14,wang17}
correspond to additional terms in our Eq.\ (\ref{gsm}); namely,
they exhibit fringes along the antidiagonal or even plaid patterns. However, unlike the Hubbard-dimer double-well 
trap which naturally generates excited states with negative parity, the double-slit biphoton states generated to 
date do not contain states of the NOON($-$) variety [associated with $\sin^2(d(k_1+k_2))$ fringes], nor 
EPR-Bell-Bohm states of the $(|1_L,1_R\rangle> - |1_R,1_L\rangle>)/\sqrt{2}$ type [associated with 
$\sin^2(d(k_1-k_2))$ fringes]. In addition, a larger control of engineered two-particle entangled states in the 
double-well case is feasible via time-evolution due to tunneling and via inter-atom interactions; both of these 
tunability controls are absent in the case of the two-slit biphoton.       
}
  
Elaboration on the mathematics background associated with the quantum-optics correspondence between the 
joint-probabilities in the far field and the 2nd-order momentum correlations of two double-well trapped ultracold 
atoms is presented in the next Sec.\ \ref{deta}.\\

\subsection{Detailed mathematical analysis} 
\label{deta}

For the convenience of the reader and to facilitate comparisons, we will 
outline below relevant passages relating to the four diagrams in Fig.\ \ref{fig6} from Mandel's review 
\cite{mand99} on quantum optics, as well as from one of our previous publications on two ultracold fermionic 
atoms confined in a double-well trap \cite{bran18}.\\
~~~\\
{\it Quantum-optics view:\/} The positive- and negative-frequency parts [$E^+(x)$ and $E^-(x)$] of the optical 
field operator at a point $x$ can be used to define the 4th-order coincidence (referred to also as joint) 
probability 
\begin{align}
\widetilde{P}_2(x_1,x_2) \propto \langle E^-(x_1)E^-(x_2)E^+(x_2)E^+(x_1)\rangle,
\label{op2def}
\end{align}
where two detectors have been placed at the far-field positions $x_1$ and $x_2$. As an intermediate step, one
invokes the field decompositions ($j=1,2$)
\begin{align}
\begin{split}
E^+(x_j)& =f_A e^{i\phi_{Aj}} \hat{b}_A + f_B e^{i\phi_{Bj}} \hat{b}_B,\\
E^-(x_j)& =f^*_A e^{-i\phi_{Aj}} \hat{b}^\dagger_A + f^*_B e^{-i\phi_{Bj}} \hat{b}^\dagger_B,
\end{split}
\label{decomp}
\end{align}
where $\hat{b}_A$ and $\hat{b}_B$ are the annihilation operators for the light fields from the two sources $A$ 
and $B$, respectively, and $\phi_{Aj}$ and $\phi_{Bj}$ are corresponding accumulated phases due to differences
in the optical path lengths. Then one obtains from Eq.\ (\ref{op2def})
\begin{widetext}
\begin{align}
\begin{split}
\widetilde{P}_2(x_1,x_2) \propto &  
|f_A|^4 \langle : \hat{n}^2_A : \rangle + |f_B|^4 \langle : \hat{n}^2_B : \rangle +
2 |f_A|^2 |f_B|^2  \langle  \hat{n}_A  \rangle \langle  \hat{n}_B  \rangle
[1+\cos(\phi_{B2}-\phi_{A2}+\phi_{A1}-\phi_{B1})]+ \\
& f^{*2}_A f^2_B \langle \hat{b}^{\dagger 2}_A \hat{b}^{2}_B \rangle 
e^{i(\phi_{B2}-\phi_{A2}+\phi_{B1}-\phi_{A1})} + {\rm c.c.} +
|f_A|^2 f^*_A f_B \langle \hat{b}^{\dagger 2}_A \hat{b}_A \hat{b}_B \rangle 
[e^{i(\phi_{B1}-\phi_{A1})}+e^{i(\phi_{B2}-\phi_{A2})}] + {\rm c.c.} + \\
& |f_B|^2 f^*_B f_A \langle \hat{b}^{\dagger 2}_B \hat{b}_B \hat{b}_A \rangle 
[e^{i(\phi_{A1}-\phi_{B1})}+e^{i(\phi_{A2}-\phi_{B2})}] + {\rm c.c.}, 
\end{split}
\label{op2}      
\end{align}
\end{widetext}
where $\langle : \hat{n}^q : \rangle$ denotes the $q$th normally ordered moment of the number operator
$\hat{n}$.

The expression for $\widetilde{P}_2(x_1,x_2)$ in Eq.\ (\ref{op2}) has 16 terms, a fact that can be seen by 
considering the four amplitudes labeled (I)-(IV) in Fig.\ \ref{fig6} (this type of analysis was not presented 
in Ref.\ \cite{mand99}). Indeed one has
\begin{align}
\begin{split}
{\rm I} =& f_A^2 e^{ i(\phi_{A1}+\phi_{A2}) } \hat{b}_A^2,\\
{\rm II} =& f_B^2 e^{ i(\phi_{B1}+\phi_{B2}) } \hat{b}_B^2,\\
{\rm III} =& f_A f_B e^{ i(\phi_{A1}+\phi_{B2}) } \hat{b}_A \hat{b}_B,\\
{\rm IV} =& f_A f_B e^{ i(\phi_{A2}+\phi_{B1}) } \hat{b}_A \hat{b}_B,
\end{split}
\label{ampl}
\end{align}
and the expression in Eq.\ (\ref{op2}) can be also rewritten as
\begin{align}
\begin{split}
\widetilde{P}_2 & (x_1,x_2) = \\
& \langle (I+II+III+IV)^\dagger (I+II+III+IV) \rangle.
\end{split}
\label{op22}
\end{align}

Obviously an expansion of the r.h.s. of Eq.\ (\ref{op22}) yields 16 terms.\\
~~~\\
{\it Ultracold-atoms view:\/} As described in Appendix D of Ref.\ \cite{bran18}, the 2nd-order momentum
correlation for two atoms is defined as
\begin{align}
{\cal G} (k_1,k_2)& = \langle \hat{\psi}^\dagger(k_1) \hat{\psi}^\dagger(k_2) 
\hat{\psi}(k_1) \hat{\psi}(k_2) \rangle\\
& = \Tr [ \hat{\rho} \hat{\psi}^\dagger(k_1) \hat{\psi}^\dagger(k_2)
\hat{\psi}(k_1) \hat{\psi}(k_2)],
\end{align}
where $\hat{\rho}$ is the second-order density in the $L$, $R$ single-particle basis
\begin{align}
\hat{\rho} = \sum_{i,j,k,l=L,R}\rho_{ijkl} |i\;j\rangle \langle k\;l|,
\label{rho}
\end{align}
and 
\begin{align}
\hat{\psi}(k_i)=\sum_{j=L,R} \psi_j(k_i) \hat{c}_{ij}, 
\end{align}
where $i=1,2$. The single-particle orbitals $\psi_j(k)$ are given by Eq.\ (\ref{psik}), and $\hat{c}_{ij}$ are 
bosonic or fermionic annihilation operators.

As a result, considering only the space part of the two-atom state, one gets
\begin{align}
\begin{split}
& {\cal G}(k_1,k_2)\\
& = \sum_{i,j,k,l=L,R}\rho_{ijkl}\psi^*_i(k_1)\psi^*_j(k_2)
\psi_k(k_1)\psi_l(k_2) \\
& = \sum_{i,j,k,l=L,R}\eta_{ijkl}(k_1,k_2).
\end{split}
\label{deco}
\end{align}     

In a previous publication \cite{bran18} on two ultracold spin-1/2 fermionic atoms confined in a double-well trap 
and described by the Hubbard model, we were able to specify the 2nd-order $\hat{\rho}$ and $\hat{\eta}(k_1,k_2)$
for the Hubbard-dimer eigenstates.
 
In this section we use specifically the case of the ground-state solution $\varphi_1$ (see Appendix D.3 in
Ref.\ \cite{bran18}); for the other two-fermions Hubbard eigenstates, $\varphi_2$, $\varphi_3$, and $\varphi_4$, 
see Appendices D.4, D.5, and D.6 in Ref.\ \cite{bran18}.
Here we consider only the space part of $\varphi_1$, so our discussion applies to all three 
cases of $\varphi_1$'s associated with two spinless bosons, two spin-1/2 bosons, and two spin-1/2 fermions 
[see Eqs.\ (\ref{vphi_i})-(\ref{vphi_iii})]. Using $\mathcal{U}=U/J$ and  
$\mathcal{Q}(\mathcal{U})=\sqrt{16+\mathcal{U}^2}+\mathcal{U}$ the two matrices $\hat{\rho}$ and 
$\hat{\eta}(k_1,k_2)$
are given by
\begin{widetext}
\begin{align}
\arraycolsep=1.4pt\def\arraystretch{2.2}
\hat{\rho}&=\frac{1}{\mathcal{Q}(\mathcal{U}) \mathcal{U}+16} \;\;\;
\begin{blockarray}{ccccc}
 LL & LR & RL & RR & \\
\begin{block}{(cccc)c}
 4 & ~~\mathcal{Q}(\mathcal{U})~~ & \mathcal{Q}(\mathcal{U}) & 4 & ~~LL \\
~~ & ~~{\textstyle \frac{1}{2}}(\mathcal{Q}(\mathcal{U})\mathcal{U}+8)~~
 & ~~{\textstyle \frac{1}{2}}(\mathcal{Q}(\mathcal{U})\mathcal{U}+8)~~
 & \mathcal{Q}(\mathcal{U}) & ~~LR \\
 \BAmulticolumn{2}{c}{\multirow{2}{*}{{\LARGE h.c.}}} &
~{\textstyle \frac{1}{2}}(\mathcal{Q}(\mathcal{U})\mathcal{U}+8)~~ & \mathcal{Q}(\mathcal{U}) & ~~RL \\
 & & & 4 & ~~RR \\
\end{block}
\end{blockarray},\\
&=\;\frac{1}{2}\;\;
\begin{blockarray}{ccccc}
 LL & LR & RL & RR & \\
\begin{block}{(cccc)c}
 {\cal B}(\mathcal{U})^2 & ~{\cal A}(\mathcal{U}){\cal B}(\mathcal{U})~ 
& ~{\cal A}(\mathcal{U}){\cal B}(\mathcal{U})~ & ~{\cal B}(\mathcal{U})^2 & ~~LL \\
 & {\cal A}(\mathcal{U})^2 & {\cal A}(\mathcal{U})^2 & {\cal A}(\mathcal{U}){\cal B}(\mathcal{U}) & ~~LR \\
\BAmulticolumn{2}{c}{\multirow{2}{*}{{\LARGE h.c.}}} & {\cal A}(\mathcal{U})^2 &  
{\cal A}(\mathcal{U}){\cal B}(\mathcal{U}) & ~~RL \\
 & & & {\cal B}(\mathcal{U})^2 & ~~RR \\
\end{block}
\end{blockarray}.
\end{align}
and
{\footnotesize
\begin{align}
\hat{\eta}(k_1,k_2)=
\arraycolsep=1.4pt\def\arraystretch{2.2}
&\frac{2 s ^2 e^{-2 s ^2 \left(k_1^2+k_2^2\right)}}{\pi  (\mathcal{Q}(\mathcal{U}) \mathcal{U}+16)} \;\;\;
\begin{blockarray}{ccccc}
 LL & LR & RL & RR & \\
\begin{block}{(cccc)c}
 4 & e^{-2 i d k_2} \mathcal{Q}(\mathcal{U}) & e^{-2 i d k_1} \mathcal{Q}(\mathcal{U}) & 
4 e^{-2 i d (k_1+k_2)} & ~LL \\
 & \frac{\mathcal{Q}(\mathcal{U})\mathcal{U}}{2}+4 & 
~~\frac{1}{2} e^{-2 i d (k_1-k_2)} (\mathcal{Q}(\mathcal{U}) \
\mathcal{U}+8)~~ & e^{-2 i d k_1} \mathcal{Q}(\mathcal{U}) & ~~LR \\
\BAmulticolumn{2}{c}{\multirow{2}{*}{{\LARGE h.c.}}} & 
\frac{\mathcal{Q}(\mathcal{U})\mathcal{U}}{2}+4 & e^{-2 i d k_2} \mathcal{Q}(\mathcal{U}) & ~~RL \\
 & & & 4 & ~~RR \\
\end{block}
\end{blockarray}.
\label{eta}
\end{align}
}
\end{widetext}

Expression (\ref{deco}) has 16 terms which correspond term-by-term to the quantum-optics 
$\widetilde{P}_2(x_1,x_2)$ 
[see Eq.\ (\ref{op2})]. To facilitate this term-by-term comparison, we point out three types of identifications; 
in discussing these identifications (correspondences between the quantum optics and double-well cases), we write 
the quantum-optics quantity on the left of the arrow and the corresponding double-well quantity on the right:\\
(1) For the phases, the following correspondences apply
\begin{align}
\begin{split}
&\phi_{B1}-\phi_{A1} \rightarrow k_1 d_R-k_1 d_L=2 k_1 d,\\
&\phi_{B2}-\phi_{A2} \rightarrow k_2 d_R-k_2 d_L=2 k_2 d.
\end{split}
\label{phi_id}
\end{align} 

(2) The matrix elements $\rho_{ijkl}$ correspond to expectations values of the products of creation and
annihilation operators $b^\dagger$ and $b$, i.e.,
\begin{align}
\langle b_i^\dagger b_j^\dagger b_k b_l \rangle \rightarrow \rho_{ijkl} ,
\end{align}    
where $i,j,k,l=L,R$ when they are indices of the $\rho$ matrix and $i,j,k,l=A,B$ when they are indices for
the creation and annihilation operators; of course $L \rightarrow A$ and $R \rightarrow B$.\\
(3) The functions $f_A$ and $f_B$ correspond to
\begin{align}
f_A = f_B \rightarrow f(k)=\frac{2^{1/4}\sqrt{s}}{\pi^{1/4}}\exp (-k^2s^2).
\end{align}

As an illustration, we display the following correspondence between explicit terms in Eqs. (\ref{op2}), Eq.\
(\ref{deco}), and the $\cos^2 (d (k_1-k_2))$ term in Eq.\ (\ref{gsm}) for the total 2nd-order momentum
correlations; the case of $\varphi_1$ corresponds to having ${\cal C}_1={\cal A}({\cal U})$, ${\cal C}_2=0$, and
${\cal C}_3 = {\cal B}({\cal U})$ in Eq.\ (\ref{gsm})
\begin{widetext}
\begin{align}
\begin{split}
& 2 |f_A|^2 |f_B|^2  \langle  \hat{n}_A  \rangle \langle  \hat{n}_B  \rangle
[1+\cos(\phi_{B2}-\phi_{A2}+\phi_{A1}-\phi_{B1})] \rightarrow\\
& f^2(k_1)f^2(k_2)\big(\rho_{LRLR}+\rho_{RLRL} + \rho_{LRRL} e^{-2i(k_1-k_2)d} + 
\rho_{RLLR} e^{2i(k_1-k_2)d}\big) =\\
&2f^2(k_1)f^2(k_2) \big({\cal A}({\cal U})^2/2\big) [1+\cos\big( 2 d (k_1-k_2) \big)] =
2f^2(k_1)f^2(k_2) {\cal A}({\cal U})^2 \cos^2 \big(d (k_1-k_2) \big).
\end{split}
\end{align}

The term above is the partial joint-probability denoted as $P_2(x_1,x_2)$ by Mandel and as
$p_{\rm soc}(k_1,k_2)$ by us. 

Another example is the correspondence between explicit terms in Eqs. (\ref{op2}), Eq.\
(\ref{deco}), and the $\cos^2 (d (k_1+k_2))$ term in Eq.\ (\ref{gsm}) for the total 2nd-order momentum
correlations

\begin{align}
\begin{split}
&  f^{*2}_A f^2_B \langle \hat{b}^{\dagger 2}_A \hat{b}^{2}_B \rangle
e^{i(\phi_{B2}-\phi_{A2}+\phi_{B1}-\phi_{A1})} + {\rm c.c.} \rightarrow \\
& f^2(k_1)f^2(k_2)\big(\rho_{LLLL}+\rho_{RRRR} + \rho_{LLRR} e^{-2i(k_1+k_2)d} + 
\rho_{RRLL} e^{2i(k_1+k_2)d}\big) =\\
&2f^2(k_1)f^2(k_2) \big({\cal B}({\cal U})^2/2\big) [1+\cos\big( 2 d (k_1+k_2) \big)] =
2f^2(k_1)f^2(k_2) {\cal B}({\cal U})^2 \cos^2 \big(d (k_1+k_2) \big).
\end{split}
\end{align}

\end{widetext}

The mapping of additional terms in Eq.\ (\ref{op2}) and corresponding terms in Eq.\ (\ref{gsm}) for the ground 
state $\varphi_1$ can be established in a similar way.  

\begin{figure*}[t]
\includegraphics[width=14cm]{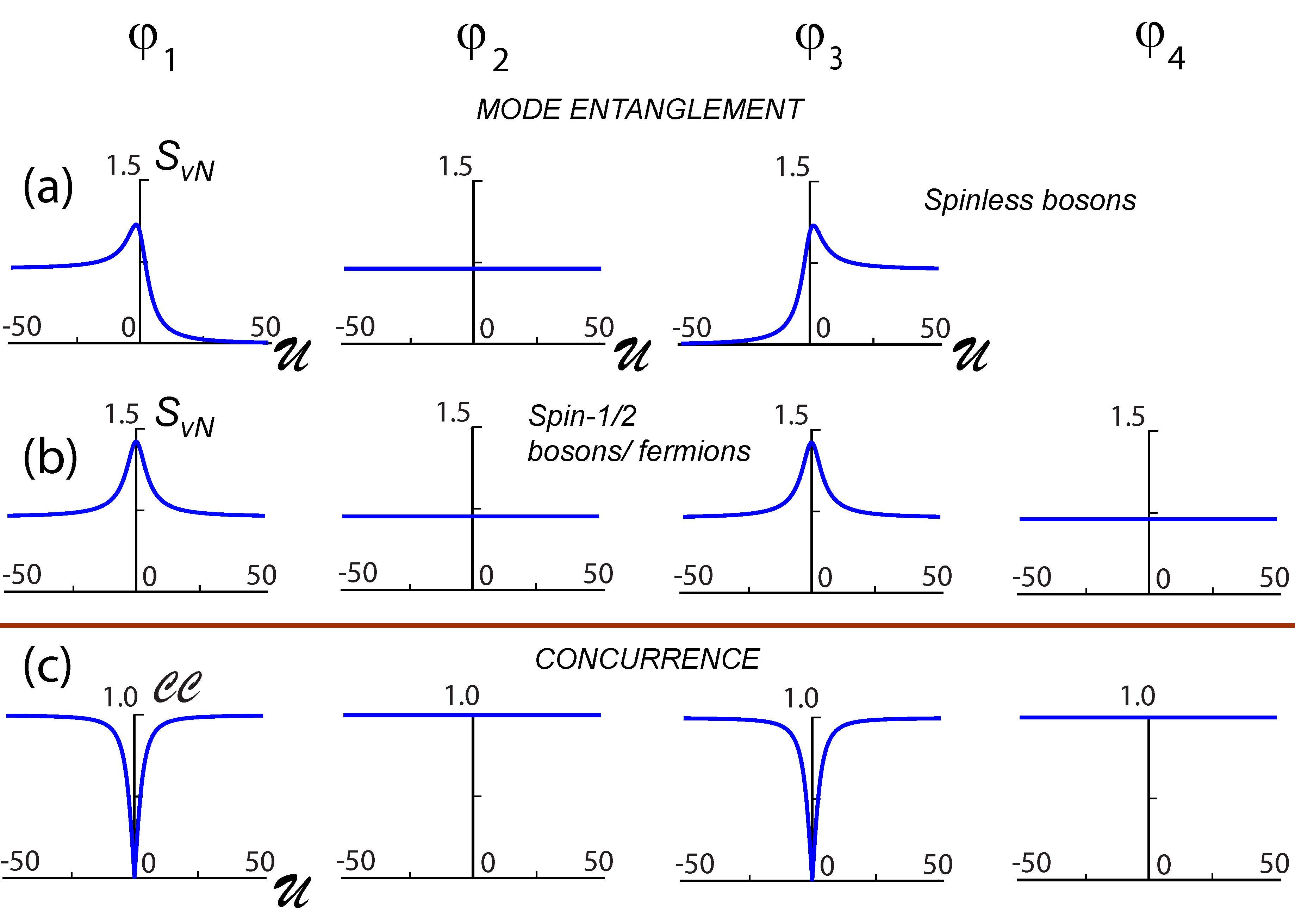}
\caption{\textcolor{black}{
(a-b) von Neumann entropy for mode entanglement and (c) concurrence for the Hubbard-dimer eigenstates 
$\varphi_i$, $i=1,\ldots,4$ as a function of ${\cal U}$. In (a) the case of two spinless bosons with only three
eigenstates is presented. The maximum value attained by the von Neumann entropy is $\ln(3)=1.09861$
at ${\cal U}=-\sqrt{2}$ for the $\varphi_1$ state or at ${\cal U}=\sqrt{2}$ for the $\varphi_3$ state. 
The asymptotic values at ${\cal U} \rightarrow \pm \infty$ are zero or $\ln(2)=0.693147$ (non-symmetric curves).
In (b), the case of two spin-1/2 bosons or fermions is presented. The maximum value attained (at ${\cal U}=0$) 
is $2\ln(2)=1.38629$, whereas the asymptotic values for ${\cal U} \rightarrow \pm \infty$ are 
$\ln(2)=0.693147$ (symmetric curves). In (c), the concurrence for all three 
cases yields the same result (of course the space antisymmetric state $\varphi_4$ is missing for the spinless 
bosons). The maximum asymptotic value attained by the concurrence (for ${\cal U} \rightarrow \pm \infty$) is 
unity, whereas the minimum value (at ${\cal U}=0$) is zero (symmetric curves).}
}
\label{fig7}
\end{figure*}

\subsection{Link to Hanbury Brown-Twiss interferometry} 
\label{hbt}

\textcolor{black}{
Of interest is the connection of our ultracold-atom interference results to the Hanbury Brown-Twiss 
interferometry. Indeed, for two thermal classical sources, Mandel's general Eq.\ (\ref{op2}) for the 
joint probability in the far-field reduces simply to
\begin{align}
\begin{split}
& \widetilde{P}_2^{\rm cl}(x_1,x_2) =(f_A^2 |I_A|^2+f_B^2 |I_B|^2)^2\\
& + 2|f_A|^2|f_B|^2 \langle I_A \rangle \langle I_B \rangle \cos(\phi_{B2}-\phi_{A2}+\phi_{A1}-\phi_{B1}),
\end{split}
\label{pcl}
\end{align} 
}

\textcolor{black}{
To derive Eq.\ (\ref{pcl}) from Eq.\ (\ref{op2}), one uses the following \cite{mand99}:
(1) the creation and annihilation operators are treated as complex $c$-number amplitudes, (2) 
$\langle \hat{n}_A (\hat{n}_B) \rangle$ becomes the mean light intensity $\langle I_A (I_B) \rangle$,
and (3) Because of phase independence, only the three terms in the first line of Eq.\ (\ref{op2}) are
non-vanishing, and they regroup to Eq.\ (\ref{pcl}). Eq.\ (\ref{pcl}) corresponds to the celebrated Hanbury 
Brown-Twiss interference \cite{baym98}, as can be seen through a direct comparison with Eq.\ (6) in 
Ref.\ \cite{baym98}.
}

\textcolor{black}{
Using the amplitude diagrams in Fig.\ \ref{fig6}, one can write
\begin{align}
\widetilde{P}_2^{\rm cl}(x_1,x_2) = \langle I^2 \rangle + \langle II^2 \rangle 
+ \langle (III+IV)^\dagger (III+IV) \rangle,
\end{align}
i.e., only the cross terms between diagrams (III) and (IV) survives.
On the other hand, as noticed (and also confirmed experimentally) by Gosh and Mandel \cite{ghos87}, the
joint probability in Eq.\ (\ref{op2}) coincides with the partial coincidence $P_2(x_1,x_2)$, i.e., our
$p_{\rm soc}^S(k_1,k_2)$ [see Eq.\ (\ref{p11s})] when exactly one photon occupies each source, i.e., 
the biphoton state is a pure EPR-Bell-Bohm state of the form 
$(|1_L,1_R\rangle + |1_R,1_L\rangle)/\sqrt{2}$. Using the diagrams of Fig.\ \ref{fig6}, one has 
\begin{align}
p_{\rm soc}^S(k_1,k_2) = \langle (III+IV)^\dagger (III+IV) \rangle.
\end{align}
}

\textcolor{black}{
Because both the classical $\widetilde{P}_2^{\rm cl}(x_1,x_2)$ and $p_{\rm soc}^S(k_1,k_2)$ [or equivalently
$P_2(x_1,x2)$] contain the same single interference cross term between (III) and (IV), some recent 
literature \cite{foel05,foel06} concerning ultracold atoms in optical lattices 
in the regime of a Mott insulator has described the 
associated time-of-flight spectra as a special version of the Hanbury Brown-Twiss effect.
We prefer, however, to follow Ref.\ \cite{ghos87}, and group the correlation maps associated with an 
EPR-Bell-Bohm state in the class of nonclassical and purely quantal biphoton interference.
}

\section{Demonstrating the violation of Bell inequalities with trapped ultracold atoms} 
\label{exp}

Thirty years after publication of the Einstein, Podolsky, and Rosen paper \cite{eins35}, presenting the EPR 
paradox and questioning the completeness of the quantum theory, a seminal proposal was put forward by John Bell 
\cite{bell64} for direct experimental testing of the local realism notion on which the local-hidden variable (LHV)
descriptions (favored by the EPR local realism) is based, versus the description advanced by quantum mechanics, 
based on nonlocal entanglement. Accordingly, experiments in which the Bell inequality is violated, reject local 
hidden variable descriptions in favor of the nonlocal quantum mechanics theory. 

Indeed using entangled pairs of (massless) photons, the violation of the Bell inequality \cite{bell64,clau69}
has been verified experimentally \cite{aspe82,aspe82.2,shih88,ou88,rari90,rari90.2,zeil98,kwia13,gius15}, 
as well as in a couple of experiments using massive particles (specifically ultracold $^9$Be$^+$ ions 
\cite{wine01} and spin-1/2 hadrons \cite{saka06}), confirming the non-local quantum-mechanical character of 
nature \cite{eins35,aspe04}.

\textcolor{black}{
In the context of two ultracold atoms confined in a double-well trap, our theoretical extraction of the partial 
coincidence probability $p_{\rm soc}(k_1,k_2)$ from the total joint coincidence probability, i.e.,
the total second-order momentum correlations ${\cal G}(k_1,k_2)$, 
allows for the use of massive trapped particles to experimentally test the Bell inequalities, in close analogy 
with previous quantum-optics experiments \cite{aspe82,aspe82.2,ou88,rari90,rari90.2} that used twin pairs of 
entangled, but separated, photons. 
}

\textcolor{black}{
Having obtained $p_{\rm soc}(k_1,k_2)$, one can then proceed to specify a Bell-Clauser-Horne-Shimony-Holt 
(Bell-CHSH) parameter \cite{aspe82,clau69,rari90,rari90.2,kher15}, ${\cal S}$, defined as
\begin{align}
{\cal S}=|Y(k_1,k_2)-Y(k_1,k_2^\prime)+Y(k_1^\prime,k_2)+Y(k_1^\prime,k_2^\prime)|.
\label{chsh}
\end{align}
}

\textcolor{black}{
The auxiliary quantity $Y(k_1,k_2)$ in Eq.\ (\ref{chsh}) is given by
\begin{widetext}
\begin{align}
\begin{split}
Y(k_1,k_2) = 
\frac{p_{\rm soc}(k_1+q, k_2+q)+p_{\rm soc}(k_1-q, k_2-q)-p_{\rm soc}(k_1+q, k_2-q)-p_{\rm soc}(k_1-q, k_2+q)}
{p_{\rm soc}(k_1+q, k_2+q)+p_{\rm soc}(k_1-q, k_2-q)+p_{\rm soc}(k_1+q, k_2-q)+p_{\rm soc}(k_1-q, k_2+q)},
\end{split}
\label{yk1k2}
\end{align}
\end{widetext}
with $\pm q= \pm \pi/(4d)$ being used to define for our 1D system the four pairs of directions along which
measurements are performed, i.e., two pairs of parallel directions [$(\leftarrow,\leftarrow)$ and 
$(\rightarrow,\rightarrow)$] and two pairs of antiparallel directions [$(\leftarrow,\rightarrow)$ and
$(\rightarrow,\leftarrow)$]. To facilitate comparisons with corresponding expressions in the Rarity and Tapster 
optics experiment \cite{rari90,rari90.2}, we note that the combinations $(\pm q, \pm q)$ in Eq.\ (\ref{yk1k2}) 
correspond to coincidences between detectors on the same side of the beamsplitter, whereas combinations 
$(\pm q, \mp q)$ correspond to coincidences between detectors on opposite sides of the beamsplitter. Moreover the 
Rarity and Tapster phases $\phi_a$ and $\phi_b$ \cite{rari90} (or $\Phi_a$ and $\Phi_b$ \cite{rari90.2}) 
correspond in our case to $2dk_1$ and $2dk_2$, respectively.  
}

\textcolor{black}{
For the Hubbard-dimer ground state, further analogies to the experiment of Refs.\ \cite{rari90,rari90.2} can be 
pointed out. In particular, in this case $p_{\rm soc}^S(k_1,k_2) \propto 1+\cos[2d(k_1-k_2)]$ [see Eq.\ 
(\ref{p11s})]. Then 
\begin{align}
p_{\rm soc}^S(k_1 \pm q, k_2 \pm q ) = 1 + \cos[2d(k_1-k_2)], 
\label{rt1}
\end{align}
and 
\begin{align}
p_{\rm soc}^S(k_1 \mp q, k_2 \pm q ) = 1 - \cos[2d(k_1-k_2)].
\label{rt2}
\end{align}
}
\textcolor{black}{
Eqs.\ (\ref{rt1}) and (\ref{rt2}) correspond to the coincidence probabilities given by Eq.\ (23) in Ref.\ 
\cite{rari90.2}. Furthermore, the quantity $Y(k_1,k_2)$ defined in Eq.\ (\ref{yk1k2}) simplifies to 
\begin{align}
Y^S(k_1,k_2) = \cos[2d(k_1-k_2)], 
\end{align}
which was also found for the case of two entangled atoms produced in collisions of Bose-Einstein condensates 
\cite{kher15}.
}

\textcolor{black}{
The local hidden-variables theory predicts an upper limit for the Bell-CHSH parameter ${\cal S}^{\rm LHV} \leq 2$
for all values of $k_1$, $k_2$, $k_1^\prime$, and $k_2^\prime$. The maximum quantal result is found directly from
Eq.\ (\ref{chsh}) for values $k_1=0$, $k_1^\prime=\pi/(4d)$, $k_2=\pi/(8d)$, and $k_2^\prime=3\pi/(8d)$; it is 
${\cal S}=2\sqrt{2}$.
}

\textcolor{black}{
Unlike Ref.\ \cite{schm17.2}, our proposal here to use the partial coincidence $p_{\rm soc}(k_1,k_2)$ instead of 
the total ${\cal G}(k_1,k_2)$ is particularly advantageous, because the maximum quantal value of 
${\cal S}=2\sqrt{2}$ decreases rapidly when higher-occupancy components are considered in the two-particle
wave function, and can even become smaller than the classical value of 2 \cite{kher15}. 
}

\textcolor{black}{
The partial coincidence $p_{\rm soc}(k_1,k_2)$ can be experimentally determined by fitting the theoretical 
expression for ${\cal G}(k_1,k_2)$ [see, e.g., Eq.\ (\ref{gsm})] to the measured 2nd-order momentum probability. 
The latter can be experimentally obtained by performing measurements on the time-of-flight (far field) expansion 
cloud \cite{berg18,schm17.2}. 
}

\section{Entanglement aspects: von Neumann entropy for mode entanglement} 
\label{vn}

In 1935 Einstein, Podolsky, and Rosen, in a celebrated paper \cite{eins35} referred to often as the EPR paper,  
announced the “EPR paradox” by calling attention to certain features of two-particle quantum theory that they 
have found most disturbing with regard to the completeness of this theory. In the same year Schr\"{o}dinger,
first in a letter to Einstein, and shortly afterwards in published papers 
\cite{schr35}, attributed these features to entanglement (translation of the German word Verschr\"{a}nkung); more 
precisely, he wrote about “entangled states” that cannot be factored into products of two single particle states 
in any representation. Schr\"{o}dinger went further to announce that: 
``I would not call [entanglement] {\it one\/},
but rather {\it the\/} characteristic trait of quantum mechanics, the one that enforces its entire departure from 
classical lines of thought.'' (p. 555) In 1951 Bohm \cite{bohmbook} used the singlet 
state of two spin-1/2 particles [see $\varphi_1$ in Eq.\ (\ref{vphi_iii})],
in a discussion of the EPR paradox, and this spawned further experimental work aiming at
preparing and exploring the properties of this, and other, entangled states (particularly in the context of the 
Bell inequality, see Sec.\ \ref{exp}). 

As already remarked in the introductory paragraph of Sec.\ \ref{exp}, principal interest 
in entanglement, aims at uncovering fundamentals of the quantum world, principally the non-intuitive notion of 
non-locality (which contradicts local realism), stating that: no physical object has distinctive individual 
properties that completely define it as an independent entity, and that the result of measurement on one system 
is not independent of measurements and/or operations performed on another, spatially separate, systems. 
Entanglement has been used in many-body theoretical and experimantal investigations of correlations, quantum 
magnetism, and quantum phase transitions in condensed-matter physics 
\cite{amic08,lida04,oste02,mint09,gu04,eise10}, many-particle ultracold atomic systems trapped in optical lattices
\cite{isla15,bloc06,monr17,luki18}, atomic and molecular systems \cite{tich11},
and even in biological structures \cite{phot10.1,phot10.2,phot13,buch11,well11,buch10,cai10,brie08}. Moreover, 
since the discovery of Shor's factoring algorithm \cite{shor99} which is anchored in entanglement, and the 
consideration of a quantum-gate mechanism based on electron spins in coupled semiconductor quantum dots, which 
can be used as a general source of spin entanglement in quantum computers \cite{burk99}, there has been a growing 
interest in entanglement in the burgeoning areas of quantum information, quantum cryptography, and quantum 
teleportation. For reviews about quantum entanglement and its applications, see Refs.\  
\cite{horo09,beng06,eise06}.

We divide our investigation of entanglement into two parts. We start our discussion here with the von Neumann
entropy as a measure of entanglement, applied to mode entanglement for the double-well-trapped dimer.  In the next
section (Sec.\ \ref{conc}) we show entanglement concurrence results for the case of two trapped fermions. 
In addition to exploring the dependence of the entanglement on the interparticle interaction strength, we study 
its time evolution. 

\textcolor{black}{
We present results for the von Neumann entropy 
associated with the mode entanglement \cite{zana02} in the case of the Hubbard-dimer eigenstates 
$\varphi_i$, $i=1,\ldots,4$ (see Sec.\ \ref{hubs2}), as well as the time-dependent wave packet $\Omega(t)$ 
described by Eq.\ (\ref{sup1}). Note that in the mode entanglement one is interested in estimating the uncertainty 
in the occupation of a given site, regardless of the identities of the particles. 
The von Neumann entropy in the case of two Hubbard sites A and B is defined as 
\begin{align}
S_{\rm vN}=-\Tr_A(\rho_A \ln(\rho_A)), 
\label{vndef}
\end{align}
where $\Tr_A$ denotes tracing with respect to indices associated with the site A, and 
$\rho_A$ is the reduced density matrix \cite{zana02,gu04}  
\begin{align}
\rho_A=\Tr_B(|\Psi\rangle \langle \Psi|)
\label{rhoa}
\end{align}
at the site A (trace taken over the site B) expressed in the local basis 
$|0\rangle_A$, $|\uparrow\rangle_A$, $|\downarrow\rangle_A$, $|\uparrow\downarrow \rangle_A$
for two spin-1/2 particles, and in the local basis $|0\rangle_A$, $|1\rangle_A$, $|2\rangle_A$ for two spinless
bosons. $|\Psi\rangle$ is any two-body Hubbard vector solution. The symbol $\ln$ here denotes natural logarithms.
}

As an example of how the calculation for $S_{\rm vN}$ proceeds, we consider the special case of the two-body
Hubbard eigenvector [see Eq.\ (\ref{a1_2}) in Appendix \ref{a1}] 
\begin{align}
{\cal V}_1({\cal U}=0)=\{1/2,\; 1/\sqrt{2},\; 1/2 \}^T
\end{align}
for the case of two noninteracting spinless bosons. In this case, one finds
\begin{align}
\begin{split}
\rho_A & = \Tr_B(|{\cal V}_1\rangle \langle {\cal V}_1|)\\
& = \frac{|2 \rangle_{A\;} {_A\langle 2|} + 2 |1 \rangle_{A\;} {_A\langle 1|} + |0 \rangle{_A\;} {_A\langle 0|}}{4},
\end{split}
\end{align}
which according to Eq.\ (\ref{vndef}) yields
\begin{align}
\begin{split}
S_{\rm vN}({\cal U}=0) & =  \ln(4)/4 + \ln(2)/2 + \ln(4)/4 \\
& = 3 \ln(2)/2 = 1.03972.
\end{split}
\end{align}

Anticipating the discussion below, we mention that this $3\ln(2)/2$ value is depicted by the point where the 
vertical axis crosses the $\varphi_1$ von Neumann-entropy curve in Fig.\ \ref{fig7}(a).

As a second example, we consider the case of the two-body Hubbard eigenvector [see Eq.\ (\ref{a1_2}) in the 
Appendix \ref{a1}]
\begin{align}
{\cal V}_1({\cal U}\rightarrow -\infty)=\{1/\sqrt{2},\; 0,\; 1/\sqrt{2} \}^T
\end{align}
for the case of two strongly attractive spinless bosons. In this case, one finds
\begin{align}
\begin{split}
\rho_A & = \Tr_B(|{\cal V}_1\rangle \langle {\cal V}_1|)\\
& = \frac{|2 \rangle_{A\;} {_A\langle 2|} + |0 \rangle{_A\;} {_A\langle 0|}}{2},
\end{split}
\end{align}
which according to Eq.\ (\ref{vndef}) yields
\begin{align}
S_{\rm vN}({\cal U} \rightarrow -\infty) = \ln(2) = 0.693147. 
\end{align}

This limit is reflected in the behavior of the $\varphi_1$ von Neumann-entropy curve in Fig.\ \ref{fig7}(a).

\subsection{Mode entanglement for the Hubbard-dimer eigenstates} 
\label{vneig}

\textcolor{black}{
Using Eqs.\ (\ref{vphi_i})-(\ref{abde}) and applying the definition (\ref{vndef}), one can find analytic 
expressions for the von Neumann entropy associated with the mode entanglement of the Hubbard-dimer
eigenstates.}\\  
~~~~\\
\noindent
{\it Two spinless bosons:\/}
In this case, one finds for the $\varphi_1$ Hubbard-dimer ground state [see Eq.\ (\ref{vphi_i})]

\begin{align}
\begin{split}
& S_{\rm vN}= \\
&\frac{8 \ln \left(\frac{1}{4} {\cal U} \left(\mathcal{W}+ {\cal U}\right)+4\right)-
\left( {\cal U} \left(\mathcal{W}+ {\cal U}\right)+
8\right) \ln \left(\frac{1}{2} \left(\frac{{\cal U}}{\mathcal{W}}+1
\right)\right)}{{\cal U} \left(\mathcal{W}+ {\cal U}\right)+16},
\end{split}
\end{align}
where
\begin{align}
{\cal W}= \sqrt{{\cal U}^2+16}.
\label{defw}
\end{align}

For the first excited state $\varphi_2$ [see Eq.\ (\ref{vphi_i})], one has simply:
\begin{align}
S_{\rm vN}=\ln(2).
\end{align}

For the second excited state $\varphi_3$ [see Eq.\ (\ref{vphi_i})], one gets:

\begin{align}
\begin{split}
& S_{\rm vN}= \\
& \frac{\left({\cal U}^2-{\cal U}\mathcal{W}+8\right) \ln \left(\frac{1}{2}-
\frac{{\cal U}}{2 \mathcal{W}}\right)-8 \ln \left(\frac{1}{4} {\cal U} 
\left({\cal U}-\mathcal{W}\right)+4\right)}{{\cal U} \left(\mathcal{W}-{\cal U}\right)-16}.
\end{split}
\end{align}
~~~~\\
\noindent
{\it Two spin-1/2 bosons or fermions:\/}
In this case, we find that for both the eigenstates $\varphi_1$ and $\varphi_3$
\begin{align}
S_{\rm vN}=\frac{8\ln({\cal R}({\cal U})/4)+({\cal R}({\cal U})-8)\ln({\cal R}(-{\cal U})/4)}
{{\cal R}({\cal U})},
\label{vneig1}
\end{align}
where 
\begin{align}
{\cal R}(\pm {\cal U}) = {\cal U}^2 \pm {\cal U} \sqrt{ {\cal U}^2 +16 } +16.
\label{ru}
\end{align}\\
~~~~\\
For the eigenstates $\varphi_2$ and $\varphi_4$, one has a constant value independent of ${\cal U}$, i.e.,
\begin{align}
S_{\rm vN}=\ln(2).
\label{vneig2}
\end{align}
We note the the result in Eq.\ (\ref{vneig1}) has been presented earlier \cite{gu04} for the case of
a two-fermion Hubbard dimer.

\textcolor{black}{
In the top two rows of Fig.\ \ref{fig7}, we compare (as a function of the Hubbard ${\cal U}$) the behavior of 
the mode-entanglement $S_{\rm vN}$ for the four possible eigenstates of the Hubbard dimer and for all three 
cases of particle pairs considered in this paper. In Fig.\ \ref{fig7}(a) the case of two spinless bosons 
[three eigenstates, see Eq.\ (\ref{vphi_i})] is presented. For the $\varphi_1$ and $\varphi_3$ states,
the $S_{\rm vN}$ curves are asymmetric about the origin. The maximum value attained by the von Neumann entropy is
$\ln(3)=1.09861$ at ${\cal U}=-\sqrt{2}$ for the $\varphi_1$ state or at ${\cal U}=\sqrt{2}$ for the 
$\varphi_3$ state, whereas the asymptotic values at ${\cal U} \rightarrow \pm \infty$ 
are zero or $\ln(2)$. A $\ln(2)$ asymptotic value indicates that the corresponding state is a pure
NOON($+$) one, whereas a vanishing asymptotic value indicates that the corresponding state is a pure
Bell($+$) one.
}
\textcolor{black}{
In Fig.\ \ref{fig7}(b), the case of two spin-1/2 bosons or fermions is presented. The $S_{\rm vN}$ curves are
now symmetric about the origin. The maximum value attained (at ${\cal U}=0$) is $2\ln(2)$, whereas the 
asymptotic values for ${\cal U} \rightarrow \pm \infty$ are $\ln(2)$.  A $\ln(2)$ asymptotic value at 
${\cal U} \rightarrow -\infty$ reflects the fact that the corresponding state is a pure NOON($+$) one, whereas
a $\ln(2)$ asymptotic value at ${\cal U} \rightarrow +\infty$ reflects the fact that the corresponding state 
is a pure Bell($+$) one.
}
\textcolor{black}{
For both the spinless and spin-1/2 pairs of particles, the maximum values of mode entanglement are attained 
at or near ${\cal U}=0$ reflecting a maximum uncertainty in the knowledge of which one of the components 
of the local basis (defined above) at a given site is present in the two-particle wave function.  
}
 
\begin{figure*}[t]
\includegraphics[width=14cm]{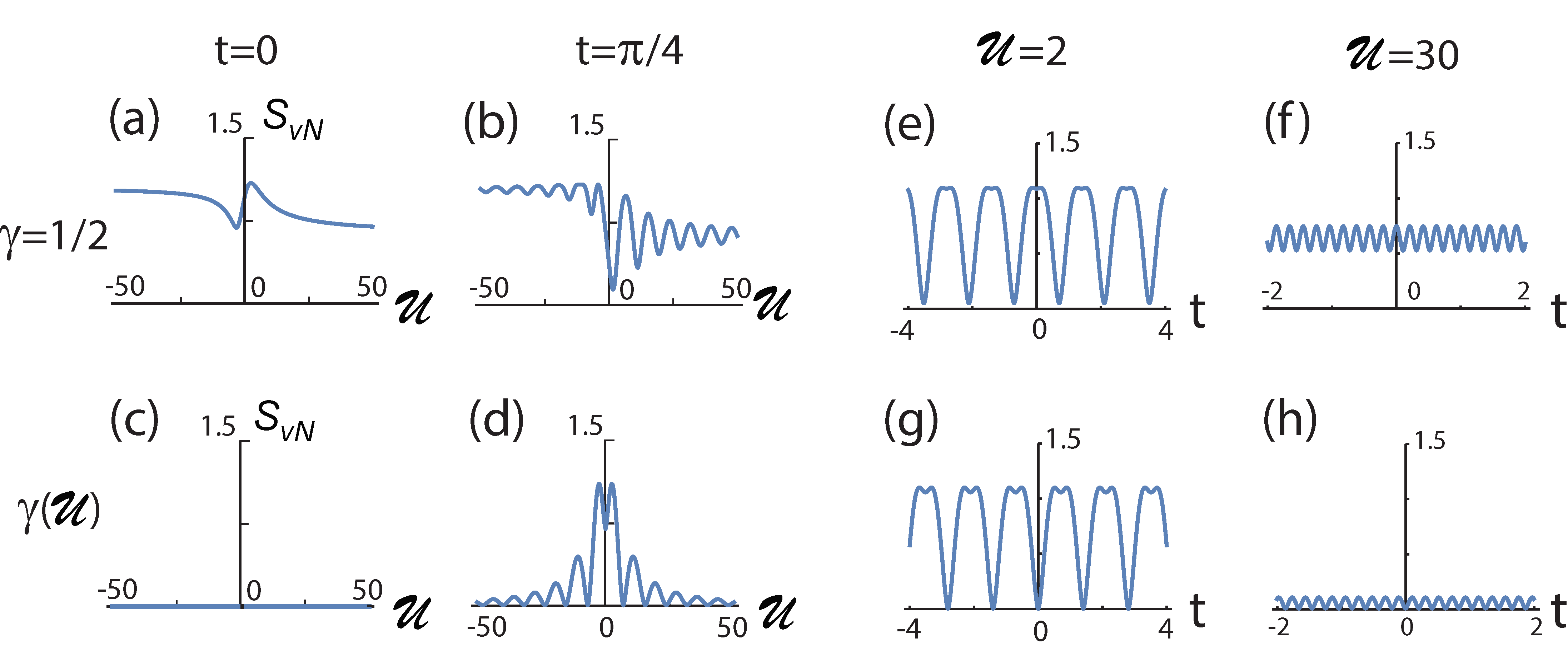}
\caption{\textcolor{black}{
von Neumann entropy for mode entanglement for two {\it spinless\/} bosons described in space by the 
time-dependent wave packet $\Omega(t)$ in Eq.\ (\ref{sup1}). The four left frames (a-d) describe variations of 
the von Neumann entropy at given times $t=0$ [(a) and (c)] and $t=\pi/4$ $\hbar/J$ [(b) and (d)] as a function 
of ${\cal U}$, and for two different cases of the mixing parameter $\gamma$ (see Sec.\ \ref{illu}), namely 
$\gamma=1/2$ [(a) and (b)] and $\gamma({\cal U}) = \frac{1}{4}({\cal U}-\sqrt{{\cal U}^2+16})$ [(c) and (d)].
The four right frames (e-h) describe variations of the von Neumann entropy at given values of ${\cal U}=2$ [(e) 
and (g)] and ${\cal U}=30$ [(f) and (h)] as a function of time $t$, and for the same two different cases of the 
mixing parameter $\gamma$, namely $\gamma=1/2$ [(e) and (f)] and $\gamma({\cal U})$ [(g) and (h)].
In agreement with the findings from the section on the von Neumann entropy for eigenstates [see Fig.\ 
\ref{fig7}(a)], the range of oscillatory variations of $S_{\rm vN}$ extends from zero to a value of 
$\ln(3)=1.09861$.} 
}
\label{fig8}
\end{figure*}

\subsection{Mode entanglement for the time-dependent wave packet $\Omega(t)$ [Eq.\ (\ref{sup1})] } 
\label{vnwav}

\textcolor{black}{
Following the same steps as in the previous section, one can also derive analytic expressions for the von 
Neumann entropy associated with the mode entanglement of the time-dependent wave packet specified in Eq.\ 
(\ref{sup1}). These expressions are listed below.}\\
~~~\\
\noindent
{\it Two spinless bosons:\/}
In this case, for a mixing parameter $\gamma=1/2$, one finds 
\begin{widetext}
\begin{align}
\begin{split}
&S_{\rm vN}(t)=\frac{1}{10 {\cal W}^2} \bigg(
\left(-16 \mathcal{W} \cos \left(tJ \mathcal{W}/\hbar\right)-5 {\cal U}^2+3 {\cal U}\mathcal{W}-80\right) 
\ln \left(\frac{1}{20} \left(\frac{16 \cos \left(tJ \mathcal{W}/\hbar\right)}{\mathcal{W}}-
\frac{3 {\cal U}}{\mathcal{W}}+5\right)\right) +\\
&\left.\left(-16 \mathcal{W} \cos \left(tJ \mathcal{W}/\hbar\right)+
5 {\cal U}^2+3 {\cal U}\mathcal{W}+80\right) \left(\ln (10)-
\ln \left(-\frac{16 \cos \left(tJ \mathcal{W}\right)}{\mathcal{W}}+
\frac{3 {\cal U}}{\mathcal{W}}+5\right)\right)\right),
\end{split}
\label{svn34_1}
\end{align}
whereas for a ${\cal U}$-dependent mixing parameter 
$\gamma({\cal U}) = \frac{1}{4}({\cal U}-\sqrt{{\cal U}^2+16})$, one gets
\begin{align}
\begin{split}
S_{\rm vN}(t)=&
\frac{1}{ {\cal W}^2 } \Big( 2 {\cal W}^2 \ln ({\cal W}) 
-\left(8 \cos \left(tJ \mathcal{W}/\hbar\right)+{\cal U}^2+8\right) 
\ln \left(8 \cos \left(tJ \mathcal{W}/\hbar\right)+{\cal U}^2+8\right)+ \\
& \left( 8 \cos \left(tJ \mathcal{W}/\hbar\right)-8\right) 
\ln \left(4-4 \cos \left(tJ \mathcal{W}/\hbar\right)\right)\Big).
\end{split}
\label{svn34_2}
\end{align}
\end{widetext}
The definition of ${\cal W}$ is given in Eq.\ (\ref{defw}).

\begin{figure*}[t]
\includegraphics[width=14cm]{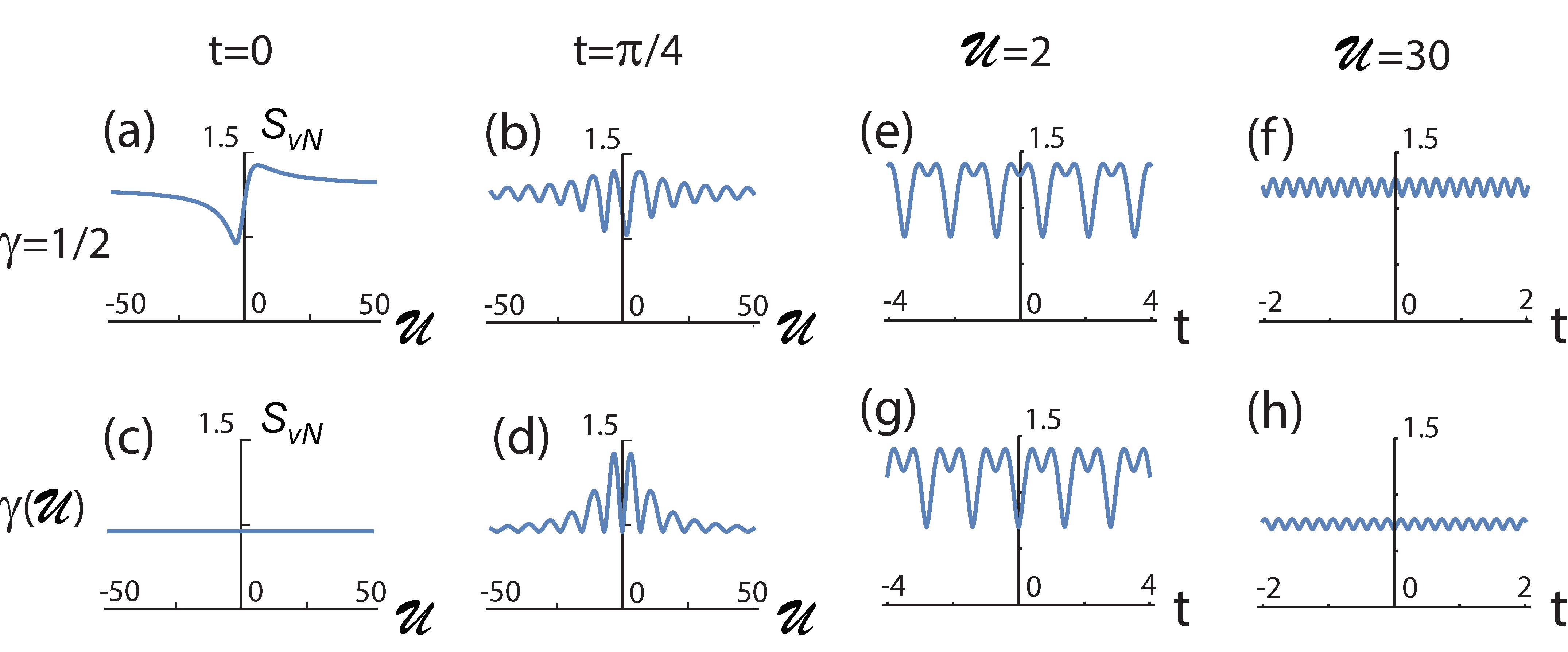}
\caption{\textcolor{black}{
von Neumann entropy for mode entanglement for two {\it spin-1/2\/} bosons or fermions described by the 
time-dependent wave packet $\Omega(t)$ in Eq.\ (\ref{sup1}). The four left frames (a-d) describe variations of 
the von Neumann entropy at given times $t=0$ [(a) and (c)] and $t=\pi/4$ $\hbar/J$ [(b) and (d)] as a function 
of ${\cal U}$ and for two different cases of the mixing parameter $\gamma$ (see Sec.\ \ref{illu}), namely 
$\gamma=1/2$ and $\gamma({\cal U}) = \frac{1}{4}({\cal U}-\sqrt{{\cal U}^2+16})$.
The four right frames (e-h) describe variations of the von Neumann entropy at given values of ${\cal U}=2$ and 
${\cal U}=30$ as a function of time $t$, and for the same two different cases of the mixing parameter $\gamma$, 
namely $\gamma=1/2$ and $\gamma({\cal U})$. In agreement with the findings from the section on the von Neumann 
entropy for eigenstates [see Fig.\ \ref{fig7}(b)], the range of oscillatory variations of $S_{\rm vN}$ extends 
from $\ln(2)=0.693147$ to a value of $2\ln(2)=1.38629$.}
}
\label{fig9}
\end{figure*}

\textcolor{black}{
Illustrations of the behavior of the $S_{\rm vN}$ in Eqs.\ (\ref{svn34_1}) and (\ref{svn34_2}) as a function of 
time or the Hubbard parameter ${\cal U}$ are presented in Fig.\ \ref{fig8}. Specifically, the four left frames
[Figs.\ \ref{fig8}(a-d)] describe variations of the von Neumann 
entropy at given times $t=0$ [Fig.\ \ref{fig8}(a) 
and Fig.\ \ref{fig8}(c)] and $t=\pi/4$ $\hbar/J$ [Fig.\ \ref{fig8}(b) and Fig.\ \ref{fig8}(d)] as a function of 
${\cal U}$, and for two different cases of the mixing parameter $\gamma$  (see Sec.\ \ref{illu}), namely 
$\gamma=1/2$ [Fig.\ \ref{fig8}(a) and Fig.\ \ref{fig8}(b)] and 
$\gamma({\cal U}) = \frac{1}{4}({\cal U}-\sqrt{{\cal U}^2+16})$ [Fig.\ \ref{fig8}(c) and Fig.\ \ref{fig8}(d)]. 
The four right frames  [Figs.\ \ref{fig8}(e-h)] describe variations of the von Neumann entropy at given 
values of ${\cal U}=2$ [Fig.\ \ref{fig8}(e) and Fig.\ \ref{fig8}(g)] and ${\cal U}=30$ [Fig.\ \ref{fig8}(f) and
Fig.\ \ref{fig8}(h)] as a function of time $t$, and for the same two different cases of the mixing parameter 
$\gamma$, namely $\gamma=1/2$ [Fig.\ \ref{fig8}(e) and Fig.\ \ref{fig8}(f)] and $\gamma({\cal U})$ [Fig.\ 
\ref{fig8}(g) and Fig.\ \ref{fig8}(h)]. The $S_{\rm vN}$ curves in Fig.\ \ref{fig8} exhibit characteristic 
oscillatory patterns as a function of $t$ and ${\cal U}$. The range of these oscillatory variations of 
$S_{\rm vN}$ extends from zero to $\ln(3)=1.09861$. This is in keeping with the minimum and 
maximum values of the corresponding $S_{\rm vN}$ associated with the Hubbard-dimer eigenstates for two spinless 
bosons [see Fig.\ \ref{fig7}(a)]. The vanishing of $S_{\rm vN}$ for all values of ${\cal U}$ in Fig.\ 
\ref{fig8}(c) reflects the fact that the initial state $\Omega(t=0)$ in this case coincides with a pure 
spinless $(1_A, 1_B)$ state for all ${\cal U}$'s as a result of the choice of $\gamma({\cal U})$ for the
mixing parameter in Eq.\ (\ref{sup1}).}\\
~~~~\\
{\it Two spin-1/2 bosons or fermions:\/}
In this case, for $\gamma=1/2$, one finds 

\begin{widetext}
\begin{align}
\begin{split}
S_{\rm vN}(t)=
\frac{1}{10 {\cal W}^2}&\left(\left(-16 {\cal W} \cos \left(tJ {\cal W}/\hbar\right)+
5 {\cal U}^2+3 {\cal W} {\cal U} + 80 \right) \ln \left(\frac{20 {\cal W}}
{-16 \cos \left(tJ {\cal W}/\hbar\right) + 5 {\cal W}+3 {\cal U}}\right)+\right.\\
&\left.\left(-16 {\cal W} \cos \left(tJ {\cal W}/\hbar\right) - 5 {\cal U}^2+
3 {\cal W} {\cal U} - 80 \right) \ln \left(\frac{16 \cos \left(tJ {\cal W}/\hbar \right)+
5 {\cal W}-3 {\cal U}}{20 {\cal W}}\right) \right),
\end{split}
\label{vn05}
\end{align}
while for $\gamma({\cal U}) = \frac{1}{4}({\cal U}-\sqrt{{\cal U}^2+16})$, one gets
\begin{align}
S_{\rm vN}(t)=
\frac{8 \left(\cos \left(tJ \mathcal{W}/\hbar\right)-1\right) \ln \left(\frac{4-
4 \cos \left(tJ \mathcal{W}/\hbar\right)}{\mathcal{W}^2}\right)+
\left(8 \cos \left(tJ \mathcal{W}/\hbar\right)+{\cal U}^2+8\right) 
\ln \left(\frac{2 \mathcal{W}^2}{8 \cos 
\left(tJ \mathcal{W}/\hbar\right)+{\cal U}^2+8}\right)}{\mathcal{W}^2}.
\label{vn06}
\end{align}
\end{widetext}

\textcolor{black}{
Illustrations of the behavior of the $S_{\rm vN}$ in Eqs.\ (\ref{vn05}) and (\ref{vn06}) as a function of time 
$t$ or the Hubbard parameter ${\cal U}$ are presented in Fig.\ \ref{fig9}. Specifically, the four left frames
[Figs.\ \ref{fig9}(a-d)] describe variations of the von Neumann entropy at given times $t=0$ [Fig.\ \ref{fig9}(a)
and Fig.\ \ref{fig9}(c)] and $t=\pi/4$ $\hbar/J$ [Fig.\ \ref{fig9}(b) and Fig.\ \ref{fig9}(d)] as a function of
${\cal U}$, and for two different cases of the mixing parameter $\gamma$  (see Sec.\ \ref{illu}), namely
$\gamma=1/2$ [Fig.\ \ref{fig9}(a) and Fig.\ \ref{fig9}(b)] and
$\gamma({\cal U}) = \frac{1}{4}({\cal U}-\sqrt{{\cal U}^2+16})$ [Fig.\ \ref{fig9}(c) and Fig.\ \ref{fig9}(d)].
The four right frames  [Figs.\ \ref{fig9}(e-h)] describe variations of the von Neumann entropy at given
values of ${\cal U}=2$ [Fig.\ \ref{fig9}(e) and Fig.\ \ref{fig9}(g)] and ${\cal U}=30$ [Fig.\ \ref{fig9}(f) and
Fig.\ \ref{fig9}(h)] as a function of time $t$, and for the same two different cases of the mixing parameter
$\gamma$, namely $\gamma=1/2$ [Fig.\ \ref{fig9}(e) and Fig.\ \ref{fig9}(f)] and $\gamma({\cal U})$ [Fig.\
\ref{fig9}(g) and Fig.\ \ref{fig9}(h)]. As was the case with the two spinless bosons [Fig.\ \ref{fig8}], the 
$S_{\rm vN}$ curves in Fig.\ \ref{fig9} exhibit characteristic oscillatory patterns as a function of $t$ and 
${\cal U}$. The range of these oscillatory variations of $S_{\rm vN}$, however, extends now from 
$\ln(2)$ to $2\ln(2)=1.38629$. This is in keeping with the minimum and maximum values of the 
corresponding $S_{\rm vN}$ associated with the Hubbard-dimer eigenstates for two spin-1/2 bosons or fermions
[see Fig.\ \ref{fig7}(b)]. The constant value of $S_{\rm vN}=\ln(2)$ for all values of ${\cal U}$ in Fig.\
\ref{fig9}(c) reflects the fact that the initial state $\Omega(t=0)$ in this case coincides with a pure Bell,
$(|1_A, 1_B\rangle + |1_B, 1_A\rangle)/\sqrt{2}$, state for all ${\cal U}$'s as a result of the choice of 
$\gamma({\cal U})$ for the mixing parameter in Eq.\ (\ref{sup1}).}

\begin{figure}[t]
\includegraphics[width=8cm]{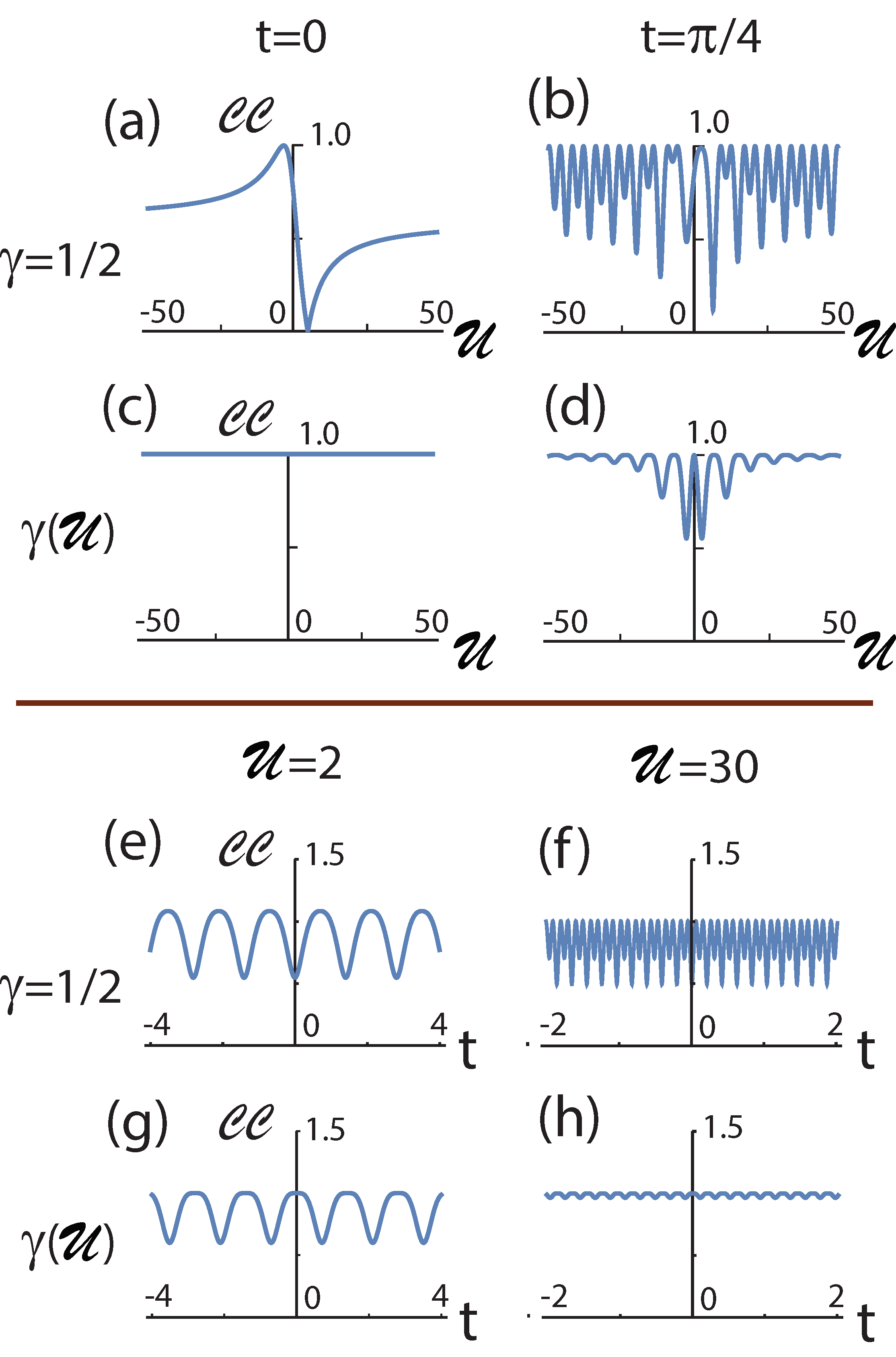}
\caption{
\textcolor{black}{
Concurrence for all three cases of a pair of particles described in space by the time-dependent wave packet 
$\Omega(t)$ in Eq.\ (\ref{sup1}). The four top frames (a-d) describe variations of the concurrence
at given times $t=0$ [(a) and (c)] and $t=\pi/4$ $\hbar/J$ [(b) and (d)] as a function
of ${\cal U}$, and for two different cases of the mixing parameter $\gamma$ (see Sec.\ \ref{illu}), namely
$\gamma=1/2$ [(a) and (b)] and $\gamma({\cal U}) = \frac{1}{4}({\cal U}-\sqrt{{\cal U}^2+16})$ [(c) and (d)].
The four bottom frames (e-h) describe variations of the concurrence at given values of ${\cal U}=2$ [(e)
and (g)] and ${\cal U}=30$ [(f) and (h)] as a function of time $t$, and for the same two different cases of the
mixing parameter $\gamma$, namely $\gamma=1/2$ [(e) and (f)] and $\gamma({\cal U})$ [(g) and (h)].
In agreement with the findings from the section on the concurrence for eigenstates [see Fig.\
\ref{fig7}(c)], the range of potential oscillatory variations of ${\cal CC}$ extends from zero to unity.}
}
\label{fig10}
\end{figure}

\section{Entanglement aspects: Concurrence for two particles} 
\label{conc}

For two fermions, one can calculate also the concurrence following the formalism of Schliemann 
{\it et al.\/} \cite{schl01} and Eckert {\it et al.\/} \cite{schl02} which generalizes for two fermions (by 
allowing for non-zero double occupancy of each Hubbard site by itinerant particles) the two-qubit concurrence 
introduced by Wootters \cite{woot98}. 
\textcolor{black}{
Due to the mapping \cite{neve07,exte09} between the Hilbert space of the
Hubbard dimer for zero total-spin projection and the two-qubit space discussed by Wootters \cite{woot98},
the concurrence results here apply to all three cases of a pair of particles considered in this paper,
that is, two spinless bosons, two spin-1/2 bosons, and two spin-1/2 fermions.} 

\subsection{Concurrence for the Hubbard-dimer eigenstates} 
\label{chde}

\textcolor{black}{
For pure states and according to Refs.\ \cite{neve07,schl01,schl02}, the concurrence ${\cal CC}$ 
for the Hubbard-dimer eigenstates can be expressed through the expansion coefficients in the $LL$, $LR$, $RL$,
and $RR$ two-particle basis. In particular, for both the Hubbard-dimer eigenstates $\varphi_1$ and $\varphi_3$ 
[see Eqs.\ (\ref{vphi_i})-(\ref{xi})] one finds 
\begin{equation}
{\cal CC} = |{\cal A}({\cal U})^2-{\cal B}({\cal U})^2|.
\label{cc}
\end{equation}
Using Eqs.\ (\ref{abde}), one obtains from the above equation
\begin{align}
{\cal CC}=\frac{| {\cal U} |} { \sqrt{ {\cal U}^2+16 } }.
\label{conc1}
\end{align}
}

On the other hand, for the eigenstates $\varphi_2$ and $\varphi_4$ [see Eqs.\ (\ref{vphi_i})-(\ref{xi})], 
one has
\begin{align}
{\cal CC}=1.
\label{conc2}
\end{align}

\textcolor{black}{
The concurrence results in Eqs.\ (\ref{conc1})-(\ref{conc2}) are plotted in Fig.\ \ref{fig7}(c).
Of course the space antisymmetric state $\varphi_4$ is not applicable for the two spinless bosons. 
In the case of the $\varphi_1$ and $\varphi_3$ eigenstates, the concurrence curves 
are symmetric about ${\cal U}=0$; the associated maximum asymptotic value attained 
(for ${\cal U} \rightarrow \pm \infty$) is unity, whereas the minimum value (at ${\cal U}=0$) is zero.
Comparing Figs.\ \ref{fig7}(b) and \ref{fig7}(c) for two spin-1/2 particles, one sees that the concurrence is 
maximum where the mode entanglement is minimum, and vice-versa. These contrasting behaviors reflect the fact 
that these two measures of entanglement probe different aspects of the two-particle wave function. Indeed, the
concurrence maximizes when the two-particle wave function exhibits a maximum uncertainty regarding the left or 
right position of each particle; or equivalently of each (up or down) spin. On the
other hand, as mentioned previously, the mode entanglement attains a maximum value when the two-particle wave 
function exhibits maximum uncertainty concerning the knowledge of which one of the components
of the local basis (defined in Sec.\ \ref{vn}) at a given site is present.
}

\subsection{Concurrence for the time-dependent wave packet $\Omega(t)$ [Eq.\ (\ref{sup1})]} 
\label{ctdp}

Following same steps as in the previous section, one finds for the time-dependent concurrence of $\Omega(t)$ 
[Eq.\ (\ref{sup1})] when $\gamma=1/2$
\begin{widetext}
\begin{align}
{\cal CC}(t)=
\frac{ \sqrt{25 {\cal U}^2 \sin^2 \left(tJ {\cal W}/\hbar \right)+
\left(16-3 {\cal U} \cos \left(tJ {\cal W}/\hbar \right)\right)^2 } } {5{\cal W}},
\label{cct1}
\end{align}
\noindent
whereas for $\gamma({\cal U}) = \frac{1}{4}({\cal U}-\sqrt{{\cal U}^2+16})$, one gets
\begin{align}
{\cal CC}(t)=
\frac{\sqrt{-8 {\cal U}^2 \left(\cos \left(2 tJ {\cal W}/\hbar \right)-
4 \cos \left(tJ {\cal W}/\hbar \right)\right)+{\cal U}^4+8 {\cal U}^2+256}}{{\cal W}^2}.
\label{cct2}
\end{align}
\end{widetext}
As aforementioned, the definition of ${\cal W}$ is given in Eq.\ (\ref{defw}).

\textcolor{black}{
Illustrations of the behavior of ${\cal CC}$ in Eqs.\ (\ref{cct1}) and (\ref{cct2}) as a function of time
$t$ or the Hubbard parameter ${\cal U}$ are presented in Fig.\ \ref{fig10}. 
Specifically, the four top frames (a-d) describe variations of the concurrence
at given times $t=0$ [(a) and (c)] and $t=\pi/4$ $\hbar/J$ [(b) and (d)] as a function
of ${\cal U}$, and for two different cases of the mixing parameter $\gamma$ (see Sec.\ \ref{illu}), namely
$\gamma=1/2$ [(a) and (b)] and $\gamma({\cal U}) = \frac{1}{4}({\cal U}-\sqrt{{\cal U}^2+16})$ [(c) and (d)].
The four bottom frames (e-h) describe variations of the concurrence at given values of ${\cal U}=2$ [(e)
and (g)] and ${\cal U}=30$ [(f) and (h)] as a function of time $t$, and for the same two different cases of the
mixing parameter $\gamma$, namely $\gamma=1/2$ [(e) and (f)] and $\gamma({\cal U})$ [(g) and (h)].
In agreement with the findings from the section on the concurrence for eigenstates [see Fig.\
\ref{fig7}(c)], the range of potential oscillatory variations of ${\cal CC}$ extends from zero to unity.
In Fig.\ \ref{fig10}(c), the constant value of unity reflects the fact that the initial state at $t=0$ remains
an EPR-Bell-Bohm state for all values of ${\cal U}$. When the mixing parameter is given by $\gamma({\cal U})$,
a ${\cal CC}$ value of unity is reached in the limit of strong interaction ${\cal U} \rightarrow \pm \infty$,
and for any given value of $t$ [see, e.g., Fig.\ \ref{fig10}(d)].
However, note that this limiting concurrence value does not characterize uniquely the associated two-particle
wave function. Indeed, in this case, an EPR-Bell-Bohm state emerges for repulsive interaction,
whereas a NOON(+) state emerges for attractive interaction.}

\section{Summary} 
\label{summ}

Over the past few years we have witnessed significant gains in development and implementation of experimental 
programs that exploit atom cooling techniques and laser-generated microtraps \cite{joch15,berg18} or optical 
tweezers \cite{kauf14,rega18} in conjunction with time-of-flight and {\it in situ\/} measurements. Paralleling this 
experimental progress, advances were made in the formulation, implementation, testing, and employment of 
theoretical methodologies, including microscopic Hamiltonian exact-diagonalization and Hubbard-Hamiltonian  
modeling \cite{bran17,bran18,bran15,bran16,bran17.2}. Motivated by these developments, we focused in this paper on 
the quantum mechanical physical nature of two ultracold atoms confined by a one-dimensional double-well potential 
and on analogies, similarities, and differences between this system and the physics of biphotons, studied for over 
three decades \cite{mand99,shihbook,oubook} in the field of quantum optics with the use of photon sources (such as 
the spontaneous parametric down conversion), mirrors, beam splitters and photon detectors. 

Aiming at enabling and aiding investigation of the building blocks of quantum simulators, which in the spirit of 
Feynman's inspiration \cite{feyn82} ``will do exactly the same as nature,'' (p. 468) we addressed here theoretically 
issues and methodologies of joint theoretical and experimental relevance for investigations of the fundamentals of 
quantum mechanics. In particular, we explored numerically and analytically quantum entanglement, the properties of 
EPR-Bell-Bohm states, and entanglement measures (the von Neumann mode-entropy 
in Sec.\ \ref{vn} and two-particle concurrence in Sec.\ \ref{conc}) 
of quantum eigenstates of double-well trapped ultracold atom-dimer systems and the 
time-evolution of prepared wave packets, as well as Bell-Clauser-Horne-Shimony-Holt testing of the nonlocal nature 
of the quantum world (Sec.\ \ref{exp}). 
 
It is pertinent to emphasize here the goal of studying foundational questions of the quantum world using a small 
confined ultracold atom dimer, particularly in light of a comment made by one of the fathers of the quantum theory
\cite{gerrybook}, Erwin Schr\"{o}dinger, who in 1952 wrote \cite{schr52} 
``... we never experiment with just one electron or atom or 
(small) molecule. In thought experiments, we sometimes assume that we do; this invariably entails ridiculous 
consequences ... in the first place it is fair to state that we are not experimenting with single particles, any 
more than we raise ichthyosauria in the zoo.'' (p. 239) Indeed, it took more than 20 years for this opinion to be 
challenged by Dehmelt \cite{dehm75,wine86}. Moreover, employing cooling and trapping techniques developed and 
refined since, it has become now possible to experimentally trap, manipulate, and measure a precise number of 
ultracold neutral atoms \cite{joch15,berg18,kauf14,rega18}, providing impetus for the developments that led to our 
studies. 

Drawing on the many years of fruitful explorations in quantum optics, we demonstrated here an extensive 
correspondence between the dynamical evolution of two (massive and interacting) ultracold fermionic or bosonic 
atoms trapped in a double well with the physics underlying the nonlocal quantum interference exhibited by  
(massless and noninteracting) biphotons \cite{mand99,shihbook,oubook}; see Sec.\ \ref{simi}. 
We find this correspondence to extend beyond
the sinusoidal pattern of the integrated coincidence probability, encompassing in detail the underlying frequency 
interferograms (spectral frequency correlation maps) and their fringes \cite{gerr15.1,gerr15.2,kwia18}. Throughout 
we illustrated our results for case studies relating to the following double-well-trapped ultracold atom dimer 
systems: (i) two spinless bosons, (ii) two spin-1/2 bosons, and (iii) two spin-1/2 fermions
(for details of these states, see Sec.\ \ref{hubb} and Appendix \ref{a1}).

\textcolor{black}{
Importantly, we showed that the optical (with massless noninteracting photons) frequency-frequency correlations 
associated with binary single-occupancy {\it sources\/} correspond to a distinct contribution in the total 
second-order momentum correlation maps of the two trapped massive and interacting particles. This contribution is 
associated with the single-occupancy component (EPR-Bell-Bohm component) of the two-atom wave function and exhibits
a general form of a cosine-square quantum beat on the momenta difference.} This finding will enable 
experimental extraction of the massive-particle soc coincidence interference from time-of-flight measurements which 
mirror \cite{altm04} the total second-order momentum correlations of the trapped ultracold particles. In this 
context, we noted ongoing efforts in the experimental community to explicitly measure \cite{berg18} the total 
second-order momentum correlations of two interacting double-well trapped fermions or to devise protocols based on 
such correlations for the characterization of entanglement of two noninteracting distinguishable bosons 
\cite{schm17.2}. 

We specifically investigated the two-particle coincidence interferogram, where a special role is played in the 
interpretation of the time-of-fight experiments by using the extracted 
partial joint-coincidence probability spectrum 
$p_{\rm soc}(k_1, k_2)$ for detecting a pair of particles in the time-of-flight expansion image (far field), with 
the double-well-trapped particles belonging to the single-occupancy component of the two-atom wave function; 
see Sec.\ \ref{simi}. This partial joint-coincidence 
probability is of particular significance here, because unlike the primary photon optics sources used in the 
biphoton HOM experiments (where the twin-pair of photons are generated in an EPR-Bell-Bohm entangled state), our 
source, namely an ultracold atom dimer trapped in a double well, contains an entangled double-occupancy 
[NOON($\pm$)] component. This partial joint coincidence probability is related to the part of the momentum total 
wave function that involves exclusively the symmetrized or antisymmetrized cross products of both the left and 
right single-particle orbitals, respectively. The extracted single-occupancy component 
$p_{\rm soc}(k_1, k_2)$, is also key to the above-noted evaluation 
of the Bell-CHSH inequalities; see Sec.\ \ref{exp}. 

Finally, we drew attention to analogies of the second-order momentum correlations for two ultracold atoms trapped 
in a double well with the total coincidence measurements performed in most recent double-slit biphoton 
quantum-optics experiments \cite{neve07,bobr14,exte09,wang17} which, in addition to the EPR-Bell-Bohm component, 
include also a double-occupancy NOON component in the prepared biphoton state.


With the framework developed here, we close with some remarks on recent studies 
of double photoionization of molecules in high laser fields \cite{akou07,krei08,wait16}, following early 
proposals by Cohen and Fano in 1966 \cite{fano66} and Kaplan and Markin in 1969 \cite{kapl69}. 
These authors predicted that photo-electrons emitted from diatomic molecules (e.g., N$_2$, 
O$_2$, H$_2$) will exhibit two-center interference patterns in the angular distribution of the emitted electrons 
with respect to the molecular axis. In that early work, which was inspired by undulations observed in measured 
records of photoabsorption cross sections versus photon energy for a few simple molecules \cite{sams66}, 
Huygen’s point of view has been invoked, regarding the absorption of a photon by the initial coherent state of 
the homonuclear diatomic molecule. The absorption process was taken as causing the launch of two coherent 
electron waves from each of the molecule’s nuclei (protons in the case of H$_2$). In this single photoionization 
case, as well as in the later extensions to double photoionization \cite{akou07,krei08,wait16}, the superposition
of the emitted waves from the two sources (atoms) generates an interference pattern, akin to that produced by the
interference of a photon (two-photons, or matter-waves) in a double-slit experiment, where the interference 
patterns show a periodicity that depends on the initial internuclear distance and the momenta of the emitted 
electrons. These molecular-level results associated with double photoionization are in the spirit of our findings
for the case of a double-well-trapped ultracold atom dimer (see Sects.\ III and IV).
 
In the case of H$_2$, the initial two-electron electronic ground state can be well described dominantly by a 
Heitler-London wave function (an entangled, EPR-Bell-Bohm, state, which has the two electrons located at the 
two different protons), and by an added smaller contribution from a double-occupancy component (with both 
electrons localized on one of the protons); the latter state is sometime termed, “the ionic contribution”, and it 
corresponds to what we have referred to above as a NOON, $(|2,0\rangle \pm |0,2\rangle)/\sqrt{2}$, 
state. 

Early analysis of the experimental data showed that out of the four different interfering breakup 
double-ionization channels contributing to the double electron ejection from H$_2$ (see Fig.\ 2 in 
Ref.\ \cite{krei08}) only the two channels corresponding to the small (double-occupancy) NOON component of the 
ground state contribute to the observed data. This result has been confirmed in more recent experiments and 
ab-initio calculations \cite{wait16}, where the measured and ab-initio-calculated momentum-momentum, 
(${\bf k}_1$,${\bf k}_2$), correlation map (constructed for the ejected electron pair) shows diffraction fringes 
along the antidiagonal in the (${\bf k}_1$,${\bf k}_2$) map (see Fig.\ 3 in  Ref.\ \cite{wait16}), 
corresponding to a two-electron emission probability 
proportional to $\cos^2 [({\bf k}_1+{\bf k}_2) \cdot {\bf R}/2]$, with no EPR component, 
$\cos^2[({\bf k}_1- {\bf k}_2) \cdot {\bf R}/2]$; see Eq.\ (1) in Ref.\ \cite{wait16}; the  latter exhibiting  
diffraction fringes along the main diagonal of the momentum-momentum interferogram, see  Secs.\ III.C and IV.B. 
Interestingly, two-point (second-order) momentum correlation functions for the electronic ground state of H$_2$  
(described with the use of the Heitler-London wave function),  exhibiting a diffraction behavior similar to
an EPR-Bell-Bohm state, have been derived some 80 years ago by Coulson \cite{coul41} in early studies aiming at 
gaining momentum-space insights into molecular bonding.  

\begin{acknowledgments}
This work has been supported by a grant from the Air Force Office of Scientific Research (AFOSR) under
Award No. FA9550-15-1-0519. Calculations were carried out at the GATECH Center for Computational Materials
Science.
\end{acknowledgments}

\appendix

\section{Diagonalization of the Hubbard-dimer Hamiltonians}
\label{a1}

\noindent 
{\it Two spinless bosons:\/} Using the basis kets
\begin{align}
\ket{2,0},\;\; \ket{1,1},\;\;\ket{0,2},
\label{a1_1}
\end{align}
where $\ket{n_L,n_R}$ (with $n_L+n_R=2$) corresponds to a permanent with $n_L$ ($n_R$) particles in the $L$ ($R$)
site, one derives the following matrix Hamiltonian associated with the spinless bosonic Hubbard Hamiltonian in 
Eq.\ (\ref{hbspl})
\begin{align}
H=\left(
\begin{array}{cccc}
 U & -\sqrt{2}J & 0 \\
 -\sqrt{2}J & 0 & -\sqrt{2}J  \\
 0 & -\sqrt{2}J & U  \\
\end{array}
\right).
\label{a11}
\end{align}

The three eigenenergies of the matrix (\ref{a11}) are given by the quantities $E_1$, $E_2$ and $E_3$
in Eq.\ (\ref{ener}) of the main text. The corresponding three normalized eigenvectors are
\begin{align}
\begin{split}
{\cal V}_1&=\{{\cal B}({\cal U})/\sqrt{2},\; {\cal A}({\cal U}),\; {\cal B}({\cal U})/\sqrt{2} \}^T\\
{\cal V}_2&=\{1/\sqrt{2},\; 0,\; -1/\sqrt{2} \}^T\\
{\cal V}_3&=\{{\cal E}({\cal U})/\sqrt{2},\; {\cal D}({\cal U}),\; {\cal E}({\cal U})/\sqrt{2} \}^T,
\end{split}
\label{a1_2}
\end{align}  
where the coefficients ${\cal A}({\cal U})$, ${\cal B}({\cal U})$, ${\cal D}({\cal U})$, and ${\cal E}({\cal U})$ 
are given in Eq.\ (\ref{abde}).

To generate the space-dependent expressions in Sect.\ \ref{hubs2}, one uses the following mappings:
\begin{align}
\begin{split}
&\ket{1,1} \rightarrow \Phi_{S1}(x_1,x_2)\\
&\ket{2,0}-\ket{0,2} \rightarrow \sqrt{2} \Phi_{S2}(x_1,x_2)\\
& \ket{2,0}+\ket{0,2} \rightarrow \sqrt{2} \Phi_{S3}(x_1,x_2).
\end{split}
\label{a1_3}
\end{align} 
~~~~\\
\noindent
{\it Two spin-1/2 bosons:\/} We seek solutions for states with $S_z=0$. Using the basis set
\begin{align}
\ket{\uparrow\downarrow,0},\;\; \ket{\uparrow,\downarrow},\;\;\ket{\downarrow,\uparrow}
\;\;\ket{0,\uparrow\downarrow},
\label{a1_4}
\end{align}
where the kets correspond to permanents in first quantization, one derives the following matrix Hamiltonian 
with the spin-1/2 bosonic Hubbard Hamiltonian in Eq.\ (\ref{hbspf})
\begin{align}
H=\left(
\begin{array}{cccc}
 U & -J & -J & 0 \\
 -J & 0 & 0 & -J \\
 -J & 0 & 0 & -J \\
 0 & -J & -J & U 
\end{array}
\right).
\label{a12}
\end{align}

The four eigenenergies of the matrix (\ref{a12}) are given by the quantities $E_i$, $i=1,\ldots,4$, 
in Eq.\ (\ref{ener}) of the main text. The corresponding four normalized eigenvectors are
\begin{align}
\begin{split}
{\cal V}_1&=\{{\cal B}({\cal U})/\sqrt{2},\; {\cal A}({\cal U})/\sqrt{2},\; {\cal A}({\cal U})/\sqrt{2},\; 
{\cal B}({\cal U})/\sqrt{2} \}^T\\
{\cal V}_2&=\{1/\sqrt{2},\; 0,\; 0,\; -1/\sqrt{2} \}^T\\
{\cal V}_3&=\{{\cal E}({\cal U})/\sqrt{2},\; {\cal D}({\cal U})/\sqrt{2},\; {\cal D}({\cal U})/\sqrt{2},\; 
{\cal E}({\cal U})/\sqrt{2} \}^T\\
{\cal V}_4&=\{0,\; 1/\sqrt{2},\; -1/\sqrt{2},\; 0 \}^T,
\end{split}
\label{a1_5}
\end{align}  
where the coefficients ${\cal A}({\cal U})$, ${\cal B}({\cal U})$, ${\cal D}({\cal U})$, and ${\cal E}({\cal U})$ 
are given in Eq.\ (\ref{abde}).

To generate the space-dependent expressions in Sect.\ \ref{hubs2}, one uses the following mappings:
\begin{align}
\begin{split}
&\ket{\uparrow,\downarrow}+\ket{\downarrow,\uparrow} \rightarrow \sqrt{2} \Phi_{S1}(x_1,x_2) \chi(1,0)\\
&\ket{\uparrow\downarrow,0}-\ket{0,\uparrow\downarrow} \rightarrow \sqrt{2} \Phi_{S2}(x_1,x_2) \chi(1,0)\\
& \ket{\uparrow\downarrow,0}+\ket{0,\uparrow\downarrow} \rightarrow \sqrt{2} \Phi_{S3}(x_1,x_2) \chi(1,0)\\
&\ket{\uparrow,\downarrow}-\ket{\downarrow,\uparrow} \rightarrow \sqrt{2} \Phi_{A}(x_1,x_2) \chi(0,0).
\end{split}
\label{a1_6}
\end{align}
~~~~\\
\noindent
{\it Two spin-1/2 fermions:\/} We again seek solutions for states with $S_z=0$. Using the basis set
\begin{align}
\ket{\uparrow\downarrow,0},\;\; \ket{\downarrow,\uparrow},\;\;\ket{\uparrow,\downarrow}
\;\;\ket{0,\uparrow\downarrow},
\label{a1_7}
\end{align}
where the kets correspond to determinants in first quantization, one derives the following matrix Hamiltonian 
associated with the spin-1/2 fermionic Hubbard Hamiltonian in Eq.\ (\ref{hf}) 
\begin{align}
H=\left(
\begin{array}{cccc}
 U & J & -J & 0 \\
 J & 0 & 0 & J \\
 -J & 0 & 0 & -J \\
 0 & J & -J & U \\
\end{array}
\right)
\label{a13}
\end{align}

The four eigenenergies of the matrix (\ref{a13}) are given again by the quantities $E_i$, $i=1,\ldots,4$, 
in Eq.\ (\ref{ener}) of the main text. The corresponding four normalized eigenvectors are
\begin{align}
\begin{split}
{\cal V}_1&=\{{\cal B}({\cal U})/\sqrt{2},\; -{\cal A}({\cal U})/\sqrt{2},\; {\cal A}({\cal U})/\sqrt{2},\; 
{\cal B}({\cal U})/\sqrt{2} \}^T\\
{\cal V}_2&=\{1/\sqrt{2},\; 0,\; 0,\; -1/\sqrt{2} \}^T\\
{\cal V}_3&=\{{\cal E}({\cal U})/\sqrt{2},\; -{\cal D}({\cal U})/\sqrt{2},\; {\cal D}({\cal U})/\sqrt{2},\; 
{\cal E}({\cal U})/\sqrt{2} \}^T\\
{\cal V}_4&=\{0,\; 1/\sqrt{2},\; 1/\sqrt{2},\; 0 \}^T,
\end{split}
\label{a1_8}
\end{align}  
where the coefficients ${\cal A}({\cal U})$, ${\cal B}({\cal U})$, ${\cal D}({\cal U})$, and ${\cal E}({\cal U})$ 
are given in Eq.\ (\ref{abde}).

To generate the space-dependent expressions in Sect.\ \ref{hubs2}, one uses the following mappings:
\begin{align}
\begin{split}
&\ket{\uparrow,\downarrow}-\ket{\downarrow,\uparrow} \rightarrow \sqrt{2} \Phi_{S1}(x_1,x_2) \chi(0,0)\\
&\ket{\uparrow\downarrow,0}-\ket{0,\uparrow\downarrow} \rightarrow \sqrt{2} \Phi_{S2}(x_1,x_2) \chi(0,0)\\
& \ket{\uparrow\downarrow,0}+\ket{0,\uparrow\downarrow} \rightarrow \sqrt{2} \Phi_{S3}(x_1,x_2) \chi(0,0)\\
&\ket{\uparrow,\downarrow}+\ket{\downarrow,\uparrow} \rightarrow \sqrt{2} \Phi_{A}(x_1,x_2) \chi(1,0).
\end{split}
\label{a1_9}
\end{align}

Note that the difference between the matrices in Eqs.\ (\ref{a12}) and (\ref{a13}) is an opposite
sign in specific matrix elements; this is due to the commutation versus anticommutation property between 
bosonic and fermionic creation and annihilation operators.  

\section{Glossary}
\label{a2}

${\cal CC}$: the concurrence, a measure of entanglement, see Eqs.\ (\ref{cc})-(\ref{cct2}).

EPR:  Einstein, Podolsky, and Rosen, see Ref.\ \cite{eins35}.

EPR-Bell-Bohm state:  an entangled state associated with single occupancy of the left $(L)$ and right $(R)$  wells,
that is, $(|1_L,1_R\rangle \pm |1_R,1_L\rangle)/\sqrt{2}$. See Sec.\ \ref{qomip}.

${\cal G}(k_1, k_2)$:   The second-order total momentum correlation function.

$\gamma$: mixing coefficient in Eq.\  (\ref{sup1}) for $\Omega (t)$.

HOM: Hong-Ou-Mandel, see Ref.\ \cite{hong87}.

$J$: the Hubbard-Hamiltonian tunneling parameter.

soc: single-occupancy component.

NOON state: a two particle entangled state corresponding to double-occupancy of either the left $(L)$ or right 
$(R)$  well, that is NOON($\pm$) $= (|2,0\rangle \pm |0,2\rangle)/\sqrt{2}$. See Sec.\ \ref{qomip}.

$\Omega (t)$: the (time-dependent) initial state [in Eq.\ (\ref{sup1})], that is a superposition of the lowest and 
highest in energy pair of eigenstates.

$P_{11}$ : the {\it in-situ\/} (integrated) joint single-occupancy probability, associated with destructive 
interference when $P_{11}=0$.

$P_{20+02}$ :  the {\it in-situ\/} (integrated) double-occupancy probability. See Eq.\ (\ref{pp11s}).

$p_{\rm soc}(k_1, k_2)$: the partial joint-coincidence probability spectrum for detecting a pair of particles  
in the time-of-flight expansion image (far field) with the double-well-trapped particles belonging to 
the single-occupancy component of the two-atom wave function.

$p_{20+02} (k_1, k_2)$:  probability  for detecting a pair of particles in the time-of-fight expansion image
with both particles originating from the same well (either the left or right one). See Eqs.\ (\ref{p20s}) and
(\ref{p20a}).

${\cal S}$: Bell-Clauser-Horne-Shimony-Holt (Bell-CHSH) parameter, see Eq.\ (\ref{chsh}).

soc: single-occupancy component.

$S_{\rm vN}$ : the von Neumann entropy, see Eq.\ (\ref{vndef}).

$\rho_A$: the reduced density matrix at site $A$, see Eq.\ (\ref{rhoa}).

$U$: the Hubbard-Hamiltonian onsite interparticle interaction parameter.

${\cal U}$: the reduced onsite interaction parameter, $U/t$.

$L, A, 1$: equivalent indices for the left well.

$R, B, 2$: equivalent indices for the right well.


\begin{thebibliography}{99}
\bibitem{geme09}
N. Gemelke, X. Zhang, C.-L. Hung, and C. Chin,
{\it In situ\/} observation of incompressible Mott-insulating domains in ultracold atomic gases,
Nature {\bf 460}, 995 (2009).
\bibitem{grei09}
W. S. Bakr, J. I. Gillen, A. Peng, S. F\"{o}lling, and M. Greiner,
A quantum gas microscope for detecting single atoms in a Hubbard-regime optical lattice,
Nature {\bf 462}, 74 (2009).
\bibitem{grei15}
M. F. Parsons, F. Huber, A. Mazurenko, Ch. S. Chiu, W. Setiawan, K. Wooley-Brown, 
S. Blatt, and M. Greiner,
Site-resolved imaging of fermionic $^6$Li in an optical lattice,
Phys. Rev. Lett. {\bf 114}, 213002 (2015).
\bibitem{hule15}
P.M. Duarte, R.A. Hart, T.-L. Yang, X. Liu, Th. Paiva, E. Khatami, R.T. Scalettar, 
N. Trivedi, and R.G. Hulet,
Compressibility of a fermionic Mott insulator of ultracold atoms,
Phys. Rev. Lett. {\bf 114}, 070403 (2015).
\bibitem{joch15}
S. Murmann, A. Bergschneider, V.M. Klinkhamer, G. Z\"{u}rn, T. Lompe, and S. Jochim,
Two fermions in a double well: Exploring a fundamental building block of the Hubbard model,
Phys. Rev. Lett. {\bf 114}, 080402 (2015).
\bibitem{foel05}
S. F\"{o}lling, F. Gerbier, A. Widera, O. Mandel, T. Gericke and I. Bloch,
Spatial quantum noise interferometry in expanding ultracold atom clouds,
Nature {\bf 434}, 491 (2005).
\bibitem{foel06}
T. Rom, Th. Best, D. van Oosten, U. Schneider, S. F\"{o}lling, B. Paredes, and I. Bloch,
Free fermion antibunching in a degenerate atomic Fermi gas released from an optical lattice,
Nature {\bf 444}, 733 (2006).
\bibitem{bouc16}
B. Fang, A. Johnson, T. Roscilde, and I. Bouchoule,
Momentum-space correlations of a one-dimensional Bose gas,
Phys. Rev. Lett. {\bf 116}, 050402 (2016).
\bibitem{hodg17}
S.S. Hodgman, R.I. Khakimov, R.J. Lewis-Swan, A.G. Truscott, and K.V. Kheruntsyan,
Solving the quantum many-body problem via correlations measured with a momentum microscope,
Phys. Rev. Lett. {\bf 118}, 240402 (2017).
\bibitem{schm17.1}
Th. Schweigler, V. Kasper, S. Erne, I. Mazets, B. Rauer, F. Cataldini, T. Langen, 
Th. Gasenzer, J. Berges, and J. Schmiedmayer,
Experimental characterization of a quantum many-body system via higher-order correlations,
Nature {\bf 545}, 323 (2017).
\bibitem{schm17.2}
M. Bonneau, W.J. Munro, K. Nemoto, and J. Schmiedmayer,
Characterizing two-particle entanglement in a double-well potential,
arXiv:1711.08977.
\bibitem{berg18}
A. Bergschneider, V.M. Klinkhamer, J.H. Becher, R. Klemt, L. Palm, G. Z\"{u}rn, S. Jochim, Ph.M. Preiss,
Correlations and entanglement in an itinerant quantum system,
arXiv:1807.06405v1.
\bibitem{roma04}
I. Romanovsky, C. Yannouleas, and U. Landman, 
Crystalline boson phases in harmonic traps: Beyond the Gross-Pitaevskii mean field, 
Phys. Rev. Lett. {\bf 93}, 230405 (2004).
\bibitem{baks07}
L.O. Baksmaty, C. Yannouleas, and U. Landman,
Rapidly rotating boson molecules with long- or short-range repulsion: An exact diagonalization study,
Phys. Rev. A {\bf 75}, 023620 (2007).
\bibitem{pfan07}
F. Deuretzbacher, K. Bongs, K. Sengstock, and D. Pfannkuche,
Evolution from a Bose-Einstein condensate to a Tonks-Girardeau gas: An exact diagonalization study,
Phys. Rev. A {\bf 75}, 013614 (2007).
\bibitem{zinn14}
A.G. Volosniev, D.V. Fedorov, A.S. Jensen, M. Valiente, N.T. Zinner,
Strongly interacting confined quantum systems in one dimension,
Nature Commun. {\bf 5}, 5300 (2014). 
\bibitem{poll18}
P. Mujal, A. Polls, and B. Juli\'{a}-D\'{i}az,
Fermionic properties of two interacting bosons in a two-dimensional harmonic trap,
Condens. Matter {\bf 3}, 9 (2018).
\bibitem{bran17}
B.B. Brandt, C. Yannouleas, and U. Landman,
Two-point momentum correlations of few ultracold quasi-one-dimensional trapped fermions: Diffraction patterns
Phys. Rev. A {\bf 96}, 053632 (2017).
\bibitem{bran18}
B.B. Brandt, C. Yannouleas, and U. Landman,
Interatomic interaction effects on second-order momentum correlations and Hong-Ou-Mandel interference of 
double-well-trapped ultracold fermionic atoms,
Phys. Rev. A {\bf 97}, 053601 (2018).
\bibitem{ming02}
A. Minguzzi, P.Vignolo, and M.P.Tosi,
High-momentum tail in the Tonks gas under harmonic confinement,
Phys. Lett. A {\bf 294}, 222 (2002).
\bibitem{olsh03}
M. Olshanii and V. Dunjko,
Short-distance correlation properties of the Lieb-Liniger system and momentum distributions of trapped 
one-dimensional atomic gases,
Phys. Rev. Lett. {\bf 91}, 090401 (2003).
\bibitem{kauf14}
A.M. Kaufman, B.J. Lester, C.M. Reynolds, M.L. Wall, M. Foss-Feig,
K.R.A. Hazzard, A.M. Rey, and C.A. Regal,
Two-particle quantum interference in tunnel-coupled optical tweezers,
Science {\bf 345}, 306 (2014).
\bibitem{isla15}
R. Islam, R. Ma, Ph. M. Preiss, M.E. Tai, A. Lukin, M. Rispoli, and M. Greiner,
Measuring entanglement entropy in a quantum many-body system,
Nature {\bf 528}, 77 (2015).
\bibitem{bloc06}
P. Treutlein, T. Steinmetz, Y. Colombe, B. Lev, P. Hommelhoff, J. Reichel, M. Greiner, O. Mandel,
A. Widera, T. Rom, I. Bloch, and Th. W. H\"{a}nsch,
Quantum information processing in optical lattices and magnetic microtraps,
Fortschr. Phys. {\bf 54}, 702 (2006).
\bibitem{shih03}
Y.H. Shih,
Entangled biphoton source - property and preparation,
Rep. Prog. Phys. {\bf 66}, 1009 (2003).
\bibitem{sant02}
Ch. Santori, D. Fattal, J. Vu\v{c}kovi\'{c}, G.S. Solomon, and Y. Yamamoto,
Indistinguishable photons from a single-photon device,
Nature {\bf 419}, 594 (2002).
\bibitem{hong87}
C. K. Hong, Z. Y. Ou, and L. Mandel,
Measurement of subpicosecond time intervals between two photons by interference,
Phys. Rev. Lett. {\bf 59}, 2044 (1987).
\bibitem{mand99}
L. Mandel,
Quantum effects in one-photon and two-photon interference,
Rev. Mod. Phys. {\bf 71}, S274 (1999).
\bibitem{shihbook}
Y.H. Shih,
{\it An Introduction to Quantum Optics: Photon and Biphoton Physics\/} 
(CRC Press, Boca Raton, Florida, 2011)
\bibitem{oubook}
Z.Y. Ou,
{\it Multi-photon Quantum Interference\/}
(Springer, New York, 2007).
\bibitem{altm04}
E. Altman, E. Demler, and M.D. Lukin,
Probing many-body states of ultracold atoms via noise correlations,
Phys. Rev. A {\bf 70}, 013603 (2004).
\textcolor{black}{
\bibitem{note2}
Unlike the use of upper-case $X_1$ and $X_2$, the far-field space coordinates are denoted usually with 
lower-case letters $x_1$ and $x_2$ in the literature of quantum optics (see, e.g., Ref.\ \cite{mand99}).
For our purposes here, the lower-case $x_1$ and $x_2$ will be used in conjunction with the two-particle wave 
function of the trapped atoms when expressed in space coordinates.}
\bibitem{branc17}
A.M. Bra\'{n}czyk,
Hong-Ou-Mandel interference,
arXiv:1711.00080.
\bibitem{aspe15}
R. Lopes, A. Imanaliev, A. Aspect, M. Cheneau, D. Boiron, and C.I. Westbrook,
Atomic Hong-Ou-Mandel experiment,
Nature {\bf 520}, 66 (2015).
\bibitem{liu98}
R.C. Liu, B. Odom, Y. Yamamoto, and S. Tarucha,
Quantum interference in electron collision,
Nature {\bf 391}, 263 (1998).
\textcolor{black}{
\bibitem{jonc12}
Electron and hole Hong-Ou-Mandel interferometry,
T. Jonckheere, J. Rech, C. Wahl, and T. Martin
Phys. Rev. B {\bf 86}, 125425 (2012). 
\bibitem{bocq13}
E. Bocquillon, V. Freulon, J.-M. Berroir, P. Degiovanni, B. Pla\c{c}ais,
A. Cavanna, Y. Jin, and G. F\`{e}ve,
Coherence and indistinguishability of single electrons emitted by independent sources,
Science {\bf 339}, 1054 (2013).
\bibitem{burk07}
G. Burkard,
Spin-entangled electrons in solid-state systems,
J. Phys.: Condens. Matter {\bf 19}, 233202 (2007).
}
\bibitem{kher14}
R.J. Lewis-Swan and K.V. Kheruntsyan,
Proposal for demonstrating the Hong-Ou-Mandel effect with matter waves,
Nature Commun. {\bf 5}, 3752 (2014).
\bibitem{ghos87}
R. Ghosh and L. Mandel,
Observation of nonclassical effects in the interference of two photons,
Phys. Rev. Lett. {\bf 59}, 1903 (1987).
\bibitem{mand88}
Z.Y. Ou and L. Mandel,
Observation of spatial quantum beating with separated photodetectors,
Phys. Rev. Lett. {\bf 61}, 54 (1988).
\bibitem{shih88}
Y.H. Shih and C.O. Alley,
New type of Einstein-Podolsky-Rosen-Bohm experiment using pairs of light quanta produced by 
optical parametric down conversion,
Phys. Rev. Lett. {\bf 61}, 2921 (1988).
\bibitem{fran89}
J.D. Franson,
Bell inequality for position and time,
Phys. Rev. Lett. {\bf 62}, 2205 (1989).
\bibitem{mand90}
Z.Y. Ou, X.Y. Zou, L.J. Wang, and L. Mandel,
Observation of nonlocal interference in separated photon channels,
Phys. Rev. Lett. {\bf 65}, 321 (1990).
\bibitem{kwia90}
P.G. Kwiat, W.A. Vareka, C.K. Hong, H. Nathel, and R.Y. Chiao,
Correlated two-photon interference in a dual-beam Michelson interferometer,
Phys. Rev. A {\bf 41}, 2910(R) (1990).
\bibitem{rari90}
J.G. Rarity and P.R. Tapster,
Experimental violation of Bell's inequality based on phase and momentum,
Phys. Rev. Lett. {\bf 64}, 2495 (1990).
\bibitem{shih96}
T.B. Pittman, D.V. Strekalov, A. Migdall, M.H. Rubin, A.V. Sergienko, and Y.H. Shih,
Can two-photon interference be considered the interference of two photons?
Phys. Rev. Lett. {\bf 77}, 1917 (1996).
\bibitem{remp04}
Th. Legero, T. Wilk, M. Hennrich, G. Rempe, and A. Kuhn,
Quantum beat of two single photons,
Phys. Rev. Lett. {\bf 93}, 070503 (2004).
\bibitem{neve07}
L. Neves, G. Lima, E.J.S. Fonseca, L. Davidovich, and S. P\'{a}dua,
Characterizing entanglement in qubits created with spatially correlated twin photons,
Phys. Rev. A {\bf 76}, 032314 (2007).
\textcolor{black}{
\bibitem{bobr14}
I.B. Bobrov, D.A. Kalashnikov, and L.A. Krivitsky,
Imaging of spatial correlations of two-photon states,
Phys. Rev. A {\bf 89}, 043814 (2014).
\bibitem{exte09}
W.H. Peeters, J.J. Renema, and M.P. van Exter,
Engineering of two-photon spatial quantum correlations behind a double slit,
Phys. Rev. A {\bf 79}, 043817 (2009).
\bibitem{wang17}
D.-J. Zhang, S. Wu, H.-G. Li, H.-B. Wang, J. Xiong, and K. Wang,
Young's double-slit interference with two-color biphotons,
Sci. Rep. {\bf 7}, 17372 (2017).
\bibitem{gerr15.1}
T. Gerrits, F. Marsili, V. B. Verma, L. K. Shalm, M. Shaw, R. P. Mirin, and S. W. Nam,
Spectral correlation measurements at the Hong-Ou-Mandel interference dip,
Phys. Rev. A {\bf 91}, 013830 (2015).
\bibitem{gerr15.2}
R.-B. Jin, Th. Gerrits, M. Fujiwara, R. Wakabayashi, T. Yamashita, S. Miki, H. Terai, R. Shimizu, 
M. Takeoka, and M. Sasaki,
Spectrally resolved Hong-Ou-Mandel interference between independent photon sources,
Opt. Express {\bf 23}, 28836 (2015).
\bibitem{kwia18}
K. Zielnicki, K. Garay-Palmett, D. Cruz-Delgado, H. Cruz-Ramirez, M. F. O'Boyle, B. Fang, 
V. O. Lorenz, A. B. U'Ren, and P. G. Kwiat,
Joint spectral characterization of photon-pair sources,
J. Mod. Opt. {\bf 65}, 1141 (2018).
\bibitem{bran15}
B.B. Brandt, C. Yannouleas, and U. Landman, 
Double-well ultracold–fermions computational microscopy: Wave function anatomy of attractive pairing   
and Wigner molecule entanglement and natural orbitals, 
Nano Lett. {\bf 15}, 7105 (2015). 
\bibitem{bran16}
C. Yannouleas, B.B. Brandt, and U. Landman, 
Ultracold few fermionic atoms in needle-shaped double wells: 
Spin chains and resonating spin clusters from microscopic Hamiltonians emulated via 
antiferromagnetic Heisenberg and $t$-$J$ models, 
New J. Phys. {\bf 18}, 073018 (2016).  
\bibitem{bran17.2}
B.B. Brandt, C. Yannouleas, and U. Landman, 
Bottom-up configuration-interaction emulations of ultracold fermions in entangled and tunnel-coupled
two-dimensional optical plaquettes: Building blocks of unconventional superconductivity, 
Phys. Rev. A {\bf 95}, 043617 (2017).
\bibitem{aspe82}
A. Aspect, Ph. Grangier, and G. Roger,
Experimental realization of Einstein-Podolsky-Rosen-Bohm Gedankenexperiment: 
A new violation of Bell's inequalities,
Phys. Rev. Lett. {\bf 49}, 91 (1982).
\bibitem{aspe82.2}
A. Aspect, J. Dalibard, and G. Roger,
Experimental test of Bell's inequalities using time-varying analyzers,
Phys. Rev. Lett. {\bf 49}, 1804 (1982).
\bibitem{ou88}
Z.Y. Ou and L. Mandel,
Violation of Bell's inequality and classical probability in a two-photon correlation experiment,
Phys. Rev. Lett. {\bf 61}, 50 (1988).
\bibitem{hubb13}
Editorial,
The Hubbard model at half century, Editorial in: 
Nat. Phys. {\bf 9}, 523 (2013).
\bibitem{hubb63}
J. Hubbard, 
Electron correlations in narrow energy bands,
Proc. R. Soc. Lond. A {\bf 276}, 238 (1963). 
\bibitem{hubb64}
J. Hubbard, 
Electron correlations in narrow energy bands III. An improved solution,
Proc. R. Soc. Lond. A {\bf 281}, 401 (1964). 
\bibitem{jaks98}
D. Jaksch, C. Bruder, J.I. Cirac, C.W. Gardiner, and P. Zoller,
Cold bosonic atoms in optical lattices, 
Phys. Rev. Lett. {\bf 81}, 3108 (1998). 
\bibitem{jaks05}
D. Jaksch and P. Zoller,  
The cold atom Hubbard tool box, 
Ann. Phys. (N.Y.) {\bf 315}, 52 (2005).
\bibitem{essl10}
T. Esslinger, 
Fermi-Hubbard physics with atoms in an optical lattice, 
Ann. Rev. Cond. Matt. Phys. {\bf 1}, 129 (2010).
\bibitem{dutt15}
O. Dutta, M. Gajda, Ph. Hauke, M. Lewenstein, Dirk-S\"{o}ren L\"{u}hmann, B.A. Malomed, 
T. Sowi\'{n}ski, and J. Zakrzewski,
Non-standard Hubbard models in optical lattices: a review,
Rep. Prog. Phys. {\bf 78}, 066001 (2015).
\bibitem{fish89}
M.P.A. Fisher, P.B. Weichman, G. Grinstein, and D.S. Fisher,
Boson localization and the superfluid-insulator transition,
Phys. Rev. B {\bf 40}, 546 (1989). 
\bibitem{rama93}
K. Sheshadri, H.R. Krishnamurthy, R. Pandit, and T.V. Ramakrishnan,
Superfluid and insulating phases in an interacting-boson model: Mean-Field theory and the RPA,
EPL {\bf 22}, 257 (1993).
\bibitem{grei02}
M. Greiner, O. Mandel, T. Esslinger, T.W. H\"{a}nsch and I. Bloch, 
Quantum phase transition from a superfluid to a Mott insulator in a gas of ultracold atoms, 
Nature {\bf 415}, 39 (2002). 
\bibitem{joer08}
R. J\"{o}rdens, N. Strohmaier,  K. G\"{u}nter, H. Moritz and T. Esslinger,  
A Mott insulator of fermionic atoms in an optical lattice, 
Nature {\bf 455}, 204 (2008).
\bibitem{poll10}
B. Juli\'{a}-D\'{i}az, D. Dagnino, M. Lewenstein, J. Martorell, and A. Polls,
Macroscopic self-trapping in Bose-Einstein condensates: Analysis of a dynamical quantum phase transition,
Phys. Rev. A {\bf 81}, 023615 (2010).
\bibitem{penn17}
F. Lingua and V. Penna,
Continuous-variable approach to the spectral properties and quantum states of the two-component 
Bose-Hubbard dimer,
Phys. Rev. E {\bf 95}, 062142 (2017).
\bibitem{ande04}
P.W. Anderson, P.A. Lee, M. Randeria, T.M. Rice, N. Trivedi, and F.C. Zhang,
The physics behind high-temperature superconducting cuprates: the ``plain vanilla'' version of RVB, 
J. Phys.: Condens. Matter {\bf 16}, R755 (2004).
\bibitem{wang06}
K. Wang,
Quantum theory of two-photon wavepacket interference in a beamsplitter,
J. Phys. B: At. Mol. Opt. Phys. {\bf 39}, R293 (2006).
\bibitem{rega18}
A.M. Kaufman, M.C. Tichy, F. Mintert, A.M. Rey, C.A. Regal, 
The Hong-Ou-Mandel effect with atoms, 
Adv. Atom. Mol. Opt. Phys. {\bf 67}, 377 (2018).
\bibitem{note3}
We remind here that after the time-of-flight expansion, the particle momenta $k_1$ and $k_2$ 
of the two trapped ultracold atoms correspond to their measured far-field space coordinates $X_1$ 
and $X_2$.
\textcolor{black}{
\bibitem{burl97}
A.V. Burlakov, M.V. Chekhova, D.N. Klyshko, S.P. Kulik, A.N. Penin, Y.H. Shih, and D. V. Strekalov,
Interference effects in spontaneous two-photon parametric scattering from two macroscopic regions,
Phys. Rev. A {\bf 56}, 3214 (1997).
}
\bibitem{baym98}
G. Baym,
The physics of Hanbury Brown-Twiss intensity interferometry: from stars to nuclear collisions,
Acta Phys. Polon. B {\bf 29}, 1839 (1998).
}
\bibitem{eins35}
A. Einstein, B. Podolsky, and N. Rosen,
Can quantum-mechanical description of physical reality be considered complete?,
Phys. Rev. {\bf 47}, 777 (1935).
\textcolor{black}{
\bibitem{bell64}
J.S. Bell,
On the Einstein-Podolsky-Rosen paradox,
Physics Physique Fizika {\bf 1}, 195 (1964).
\bibitem{clau69}
J.F. Clauser, M.A. Horne, A. Shimony, and R.A. Holt,
Proposed experiment to test local hidden-variable theories,
Phys. Rev. Lett. {\bf 23}, 880 (1969).
Erratum: Phys. Rev. Lett. {\bf 24}, 549 (1970).
\bibitem{zeil98}
G. Weihs, Th. Jennewein, Ch. Simon, H. Weinfurter, and A. Zeilinger,
Violation of Bell's inequality under strict Einstein locality conditions,
Phys. Rev. Lett. {\bf 81}, 5039 (1998).
\bibitem{rari90.2}
J.G. Rarity and P.R. Tapster,
Two-color photons and nonlocality in fourth-order interference,
Phys. Rev. A {\bf 41}, 5139 (1990).
\bibitem{kwia13}
B.G. Christensen, K.T. McCusker, J.B. Altepeter, {\it et al.\/},
Detection-loophole-free test of quantum nonlocality, and applications,
Phys. Rev. Lett. {\bf 111}, 130406 (2013).
\bibitem{gius15}
M. Giustina, M.A.M. Versteegh, S. Wengerowsky, {\it et al.\/},
Significant-loophole-free test of Bell’s theorem with entangled photons,
Phys. Rev. Lett. {\bf 115}, 250401 (2015).
\bibitem{wine01}
M.A. Rowe, D. Kielpinski, V. Meyer, C.A. Sackett, W.M. Itano, C. Monroe, and D.J. Wineland,
Experimental violation of a Bell's inequality with efficient detection,
Nature {\bf 409}, 791 (2001).
\bibitem{saka06}
H. Sakai, T. Saito, T. Ikeda, K. Itoh, T. Kawabata, H. Kuboki, Y. Maeda, N. Matsui, C. Rangacharyulu, M. Sasano, 
Y. Satou, K. Sekiguchi, K. Suda, A. Tamii, T. Uesaka, and K. Yako,
Spin correlations of strongly interacting massive fermion pairs as a test of Bell’s inequality,
Phys. Rev. Lett. {\bf 97}, 150405 (2006); Erratum Phys. Rev. Lett. {\bf 97}, 179901 (2006).
\bibitem{aspe04}
A. Aspect,
John Bell and the second quantum revolution, foreword of {\it Speakable and unspeakable 
in quantum mechanics: J.S. Bell papers on quantum mechanics,\/}
(Cambridge University Press, New York, 2004).
\bibitem{kher15}
R.J. Lewis-Swan and K.V. Kheruntsyan,
Proposal for a motional-state Bell inequality test with ultracold atoms,
Phys. Rev. A {\bf 91}, 052114 (2015).
}
\bibitem{schr35}
E. Schr\"{o}dinger, Die gegenw\"{o}rtigsituation der quantenmechanik, 
Naturwissenschaften {\bf 23}, 807 (1935)
[ John D. Trimmer, The Present Situation in Quantum Mechanics: A Translation 
of Schr\"{o}dinger's "Cat Paradox" Paper, Proc. Am. Phil. Soc. {\bf 124}, 323 (1980)];   
E. Schr\"{o}dinger, 
Discussion of probability relations between separated systems, 
Math. Proc. Camb. Phil. Soc. {\bf 31}, 555 (1935).
\bibitem{bohmbook}
D. Bohm, 
{\it Quantum Theory\/} (Prentice –Hall, New York, 1951).
\bibitem{amic08}
L. Amico, R. Fazio, A. Osterloh, and V. Vedral,
Entanglement in many-body systems,
Rev. Mod. Phys. {\bf 80}, 517 (2008).
\bibitem{lida04}
L.A. Wu, M.S. Sarandy, and D.A. Lidar,
Quantum phase transitions and bipartite entanglement,
Phys. Rev. Lett. {\bf 93}, 250404 (2004).
\bibitem{oste02}
A. Osterloh, L. Amico, G. Falci, and R. Fazio,
Scaling of entanglement close to a quantum phase transition,
Nature {\bf 416}, 608 (2002).
\bibitem{mint09}
F. Mintert, A.M. Rey, I.I. Satija, and C.W. Clark, 
Phase transitions, entanglement and quantum noise interferometry in cold atoms,
Europhys. Lett. {\bf 86}, 17003 (2009).
\bibitem{gu04}
S.-J. Gu, S.-S. Deng, Y.-Q. Li, and H.-Q. Lin,
Entanglement and quantum phase transition in the extended Hubbard model,
Phys. Rev. Lett. {\bf 93}, 086402 (2004).
\bibitem{eise10}
J. Eisert, M. Cramer, and M.B. Plenio,
Area laws for the entanglement entropy,
Rev. Mod. Phys. {\it 82}, 277 (2010).
\bibitem{monr17}
N.M. Linke, S. Johri, C. Figgatt, K.A. Landsman, A.Y. Matsuura, and Ch. Monroe,
Measuring the Renyi entropy of a two-site Fermi-Hubbard model on a trapped ion quantum computer,
arXiv:1712.08581v1.
\bibitem{luki18}
A. Lukin, M. Rispoli, R. Schittko, M.E. Tai, A.M. Kaufman, 
S. Choi, V. Khemani, J. L\'{e}onard, and M. Greiner,
Probing entanglement in a many-body-localized system,
arXiv:1805.09819v2.
\bibitem{tich11}
M.C. Tichy, F. Mintert, and A. Buchleitner, 
Essential entanglement for atomic and molecular physics, 
J. Phys. B: At. Mol. Opt. Phys. {\bf 44}, 192001 (2011).
\bibitem{phot10.1}
M. Sarovar, A. Ishikazi, G.R. Fleming, K.B. Whaley, 
Quantum entanglement in photosynthetic light-harvesting complexes,
Nat Phys. {\bf 6}, 462 (2010).
\bibitem{phot10.2}
P. Nalbach and  M. Thorwart, 
Quantum coherence and entanglement in photosynthetic light-harvesting complexes, 
Quantum efficiency in complex systems, Pt. 1: Biomolecular systems,  
Semiconductors and Semimetals {\bf 83}, 39 (2010). 
\bibitem{phot13}
Y. Kurashige, G.K.-L. Chan,  and T. Yanai,  
Entangled quantum electronic wavefunctions of the Mn$_4$CaO$_5$ cluster in photosystem II, 
Nat. Chem. {\bf 5}, 660 (2013).
\bibitem{buch11}
T. Scholak, T. Wellens, and A. Buchleitner, 
Optimal networks for excitonic energy transport,
J. Phys. B: At. Mol. Opt. Phys. {\bf 44}, 184012 (2011).
\bibitem{well11}
T. Scholak, F. de Melo, T. Wellens, F. Mintert, and A. Buchleitner,
Efficient and coherent excitation transfer across disordered molecular networks,
Phys. Rev. E {\bf 83}, 021912 (2011).
\bibitem{buch10}
T. Scholak, F. Mintert, T. Wellens, and A. Buchleitner, 
Transport and Entanglement,
Semiconductors and Semimetals {\bf 83}, 1 (2010).
\bibitem{cai10}
J. Cai, S. Popescu, and H.J. Briegel, 
Dynamic entanglement in oscillating molecules and potential biological implications,
Phys. Rev. E {\bf 82}, 021921 (2010).
\bibitem{brie08}
H.J. Briegel and S. Popescu, 
Entanglement and intra-molecular cooling in biological systems? -
A quantum thermodynamic perspective, arXiv:0806.4552
\bibitem{shor99} 
P.W. Shor, 
Polynomial-time algorithms for prime factorization and discrete logarithms on a quantum computer,
SIAM Rev. {\bf 41}, 303 (1999).
\bibitem{burk99}
G. Burkard, D. Loss, and D.P. DiVincenzo, 
Coupled quantum dots as quantum gates, 
Phys. Rev. B {\bf 59}, 2070 (1999).
\bibitem{beng06}
I. Bengtsson and K. Zyczkowski,  
{\it Geometry of quantum states: An introduction to quantum entanglement\/}
(Cambridge University Press, Cambridge, 2006)
\bibitem{eise06}
J. Eisert and D. Gross,
{\it Lectures on quantum information\/},
Eds. D. Bruss and G. Leuchs (Wiley-VCH, Weinheim, 2006)
\bibitem{horo09}
R. Horodecki, P. Horodecki, M. Horodecki and K. Horodecki,
Rev. Mod. Phys. {\bf 81}, 865 (2009).
\bibitem{zana02}
P. Zanardi,
Quantum entanglement in fermionic lattices,
Phys. Rev. A {\bf 65}, 042101 (2002).
\bibitem{schl01}
J. Schliemann, D. Loss, and A.H. MacDonald,
Double-occupancy errors, adiabaticity, and entanglement of spin qubits in quantum dots,
Phys. Rev. B {\bf 63}, 085311 (2001).
\bibitem{schl02}
K. Eckert, J. Schliemann, D. Bru\ss, and M. Lewenstein,
Quantum correlations in systems of indistinguishable particles,
Ann. Phys. (New York) {\bf 299}, 88 (2002).
\bibitem{woot98}
W.K. Wootters,
Entanglement of formation of an arbitrary state of two qubits,
Phys. Rev. Lett. {\bf 80}, 2245 (1998).
\bibitem{feyn82}
R.P. Feynman,  
Simulating physics with computers,  
Int. J. Theor. Phys. {\bf 21}, 467 (1982).
\bibitem{gerrybook}
C.C. Gerry and P.L. Knight,
{\it Introductory Quantum Optics\/}
(Cambridge University Press, Cambridge, 2004).
\bibitem{schr52}
E. Schr\"{o}dinger, 
Are there quantum jumps?: Part I, 
Br. J. Philos. Sci. {\bf 3}, 109 (1952); 
Are there quantum jumps?: Part II,
{\it ibid.\/} {\bf 3}, 233 (1952). 
\bibitem{dehm75}
H. Dehmelt,
Proposed 10 nu greater than nu laser fluorescence spectroscopy on T1 mono-ion oscillator II,  
Bull. Am. Phys. Soc. {\bf 20}, 60 (1975); Coherent spectroscopy on single atomic system at rest in free space. II,
J. Phys. (Paris) Colloq. {\bf 42}, C8-299 (1981).
\bibitem{wine86}
J.C. Bergquist, R.G. Hulet, W.M. Itano, and D.J. Wineland, 
Observation of quantum jumps in a single atom, 
Phys. Rev. Lett. {\bf 57}, 1699 (1986). 
\bibitem{akou07}
D. Akoury et al, 
The simplest double slit: Interference and entanglement in double photoionization of H$_2$,
Science {\bf 318}, 949 (2007).
\bibitem{krei08} 
K. Kreidi et al., 
Interference in the collective electron momentum in double photoionization, 
Phys. Rev. Lett. {\bf 100}, 133005 (2008).
\bibitem{wait16}
M. Waitz, et al., 
Two-particle interference of electron pairs on a molecular level,
Phys. Rev. Lett. {\bf 117}, 083002 (2016).
\bibitem{fano66}
H.D. Cohen and U. Fano, 
Interference in photo-ionization of molecules,  
Phys. Rev. {\bf 150}, 30 (1966).
\bibitem{kapl69}
I.G. Kaplan and A.P. Markin, 
Interference phenomena in photoionization of molecules,
Sov. Phys. Dokl. {\bf 14}, 36 (1969).
\bibitem{sams66}
J.A.R. Samson and R.B. Cairns,
A study of kinetic energies of electrons produced by photoionization,
J. Opt. Soc. Am. {\bf 56}, 552 (1966).
\bibitem{coul41}
C.A. Coulson, 
Momentum distribution in molecular systems. Part I. The single bond,
Math. Proc. Cambridge Phil. Soc. {\bf 37}, 55 (1941).
\end{thebibliography}
\end{document}